\documentclass[pof,a4paper,graphicx,numbers]{revtex4-1}
\pdfoutput=1

\usepackage[table]{xcolor}
\usepackage{tikz}
\usepackage[english]{babel}
\usepackage{mdframed}
\usepackage{physics}
\usepackage{amssymb}
\usepackage{amsmath}
\usepackage{url}
\usepackage{mathtools}
\usepackage[]{geometry}
\usepackage{tikz}
\usepackage{tikz-3dplot}
\usepackage{microtype}
\usepackage{soul}
\usepackage{hyperref}  
\usepackage{graphicx}
\usepackage{epstopdf}
\usepackage[utf8]{inputenc}
\usepackage[T1]{fontenc}
\usepackage{etoolbox}
\usepackage{caption}
\usepackage[list=false]{subcaption}
\usepackage{adjustbox}
\usepackage{rotating}
\usepackage[section]{placeins}

\usetikzlibrary{decorations,decorations.pathreplacing,calc, arrows.meta,arrows,quantikz}
\DeclarePairedDelimiter\ceil{\lceil}{\rceil}

\draft 

\begin{document}

\title{Quantum Algorithm for Lattice Boltzmann (QALB) Simulation of Incompressible Fluids with a Nonlinear Collision Term} 

\author{Wael Itani}
\email{itani@nyu.edu}
\affiliation{Tandon School of Engineering, New York University, New York, NY 11201 , United States of America}

\author{Katepalli R. Sreenivasan}
\email{katepalli.sreenivasan@nyu.edu}
\affiliation{Tandon School of Engineering, New York University, New York, NY 11201 , United States of America}
\affiliation{Courant Institute of Mathematical Sciences, New York University, New York, NY 10012, United States of America}
\affiliation{Department of Physics, New York University, New York, NY 10003, United States of America}

\author{Sauro Succi}
\thanks{Corresponding Author}
\email{sauro.succi@gmail.com}
\affiliation{Fondazione Istituto Italiano di Tecnologia,
Center for Life Nano-Neuroscience at la Sapienza, 00161 Roma, Italy}
\affiliation{Physics Department, Harvard University, Cambridge Massachusetts, USA}

\date{\today}

\begin{abstract}
  We propose a quantum algorithm for solving physical problems represented by the lattice Boltzmann formulation. Specifically, we deal with the case of a single phase, incompressible fluid obeying the Bhatnagar-Gross-Krook model. We use the framework introduced by Kowalski that links the nonlinear dynamics of a system to the evolution of bosonic modes, assigning a Carleman linearization order to the truncation in the bosonic Fock space of the bosons.  The streaming and collision steps are both achieved via unitary operators. A quantized version of the nonlinear collision term has been implemented, without introducing variables of discrete densities coupled from neighbouring sites, unlike the classical Carleman technique. We use the compact mapping of the bosonic modes to qubits that uses a number of qubits which scales logarithmically with the size of truncated bosonic Fock space. The work can be readily extended to the multitude of multiphysics problems which could adapt the lattice Boltzmann formulation. 
\end{abstract}


\maketitle 

\newpage
\tableofcontents
\newpage

\section{Introduction}
\subsection{The Case for a Quantum Computer in Fluid Dynamics Research}

Feynmann popularized the concept of “quantum computing” in his lecture “Simulating Physics with Computers” \cite{feynmanSimulatingPhysicsComputers1982}. He was focused more on computing quantum problems which are very hard to compute on digital computers. Coincidentally, he is also the one behind labeling turbulence a central unsolved problem of classical physics \cite{feynmanFeynmanLecturesPhysics2011}. Peter Shor's remarks that the Church-Turing thesis -  which states that anything computable by any means is computable on a Turing machine - is in fact a statement about the physical world. Conveniently, Shor, whose algorithm is a landmark in the development of quantum computing, when mentioning physical systems current digital computers struggle to simulate, even before quantum mechanics, cites turbulence, and adds “which I unfortunately have nothing further to say” \cite{shorIntroductionQuantumAlgorithms2001}. Quantum computing algorithms that deal with non-linear systems are not easy to design, given the linearity of quantum mechanics. They are, however, at the center of many industrial and scientific problems, and the interest in harnessing the power of quantum computers to solve them has drastically increased with the recent advances in quantum computing technology.

The prospects of quantum computing are tantalizing
indeed, offering as they do, the potential chance of putting the quantum
superposition principle at use to explore and simulate in polynomial time problems that
present exponential barriers to classical algorithms \cite{groverQuantumMechanicsHelps1997}.
On general grounds, computer simulations
occupy the following four-quadrants of the physics-computing plane:
\begin{itemize}
\item[] CC: Classical computing for Classical physics;
\item[] CQ: Classical computing for Quantum physics;
\item[] QC: Quantum computing for Classical physics;
\item[] QQ: Quantum computing for Quantum physics.
\end{itemize}

To date, CC and CQ are by far the main and vastly more populated
quadrants: fluid dynamics and molecular dynamics belong in CC,
while electronic structure simulations are a prototypical CQ case. CC is often presented with polynomial computational complexity
(for instance, turbulence requires computing times that vary with the Reynolds number as its third or larger power)
and the CQ case often faces exponential complexity, a typical
case in point being the quantum many-body problem.

For nearly five decades Moore's law and parallel computing have 
successfully sustained an exponential growth of both CC and CQ 
quadrants, but for a few years now it is apparent that
the trend has slowed down, mostly on account of power consumption issues. In contrast, QQ offers, {\it in principle}, a natural escape from the
barrier of exponential complexity.

It is well known, however, that turning this potential into a concrete tool faces
daunting problems: for one, not all problems can be expressed
in terms of quantum computing algorithms; and, even when it is possible to do so,
their practical execution on actual quantum hardware
(QPU's) is confronted with efficiency issues and
the severe problem of decoherence. The so-called Quantum Advantage (QA), namely the expectation that
a quantum algorithm cannot be beaten by any classical one,
remains to be realized at present.
Not surprisingly, the most promising candidates are problems
in quantum chemistry and materials---namely, those centered
around the quantum many-body problem \cite{tacchinoQuantumComputersUniversal2020a}---although many other
applications (involving a search in high-dimensional spaces \cite{rogetGroverSearchNaturally2020}) hold promises
as well. In contrast, the QC quadrant remains largely unpopulated \cite{josephKoopmanvonNeumannApproach2020}.

This situation is hardly surprising because many classical problems
feature two major hurdles for quantum computing: 
nonlinearity and non-unitarity (dissipation). Nonlinearity implies dephasing \cite{hippertUniversalManybodyDiffusion2021}, hence loss of orthogonality,
because the rotation in the Hilbert space depends on the initial 
state vector: two different state vectors acted upon by the
same Hamiltonian rotate by different angles. 
Loss of orthogonality means loss of
information \cite{pokharelBetterthanclassicalGroverSearch2022a} and the time to tell apart two overlapping states scales
as $1/O$ where $O$ is the degree of orthogonality (overlap), with 
$O=0$ for the parallel case and $O=1$ for the orthogonal.
This is an inevitable problem at high Reynolds numbers, where loss of orthogonality
occurs very quickly.
Dissipation, on the other hand, cannot be dealt exactly by deterministic 
unitaries but typically requires a probabilistic implementation which comes with
a corresponding non-zero failure rate. Yet, given the paramount relevance of classical physics 
to science and engineering, an increasing group of quantum computing 
researchers is turning attention to this major challenge \cite{lopezDerivationMathematicalAnalysis2013,tennieQuantumComputersWeather2022}.

This paper occupies the QC quadrant, with specific focus
on the physics of fluids and, more precisely, the formulation of a quantum 
algorithm for fluids based on lattice kinetic theory. The motivations are straightforward: fluid turbulence faces a complexity of $Re^3$
or higher, $Re$ being the Reynolds number, basically the strength
of nonlinearity over viscous effects.
Most real life problems feature Reynolds numbers in the many-billions
(for an airplane $Re \sim 10^8$),  placing them well beyond the
reach of the best electronic supercomputers, now in
the exascale range \cite{slotnickCFDVision2030}.
In contrast, a $Q$ qubit computer (IBM is currently at Q=433, nearing
500 in next two years), offers $2^{Q} \sim 10^{3Q/10}$ binary
degrees of freedom. A turbulent flow at a given Reynolds number
contains $Re^{9/4}$ degrees of freedom (or more), hence the (minimum) 
number of qubits required to represent such a turbulent flow is given by
\begin{equation}
Q \sim \frac{15}{2} log(Re).
\end{equation} 
This shows that $Q=60$ is already matching the exascale capacity \cite{succiExascaleLatticeBoltzmann2019}, while the Reynolds number would skyrocket to about $Re \sim 10^{20}$ with
$Q=120$,  
far beyond the capacity of any foreseeable classical computer.
Hence, the potential definitely exists, although its actual realization
faces mounting difficulties with increasing Reynolds numbers.

Despite the promise, it is important to note that the current error rate of quantum computing hardware increases the computational complexity exponentially in the circuit area \cite{eddinsExperimentallyCharacterizingError2023}, thereby effectively effacing any quantum advantage \cite{aaronsonReadFinePrint2015}.
 
The specific focus on lattice kinetic theory is motivated by the hope
that the substantial advantages offered by such formulation in the case
of classical fluids can somehow be transferred to the quantum realm.
In particular, we refer to the fact that nonlinearity and nonlocality are
disentangled, i.e., streaming is  nonlocal but linear, while collisions
are nonlinear but local. Since the first simulation based on the lattice Boltzmann model in 1989 \cite{higueraSimulatingFlowCircular1989}, the adoption of the model has grown, especially with the advent of GPUs which could parallelize the model's algorithms thanks to the aforementioned separation. Lattice models could, similarly, utilize the parallelism afforded by quantum computers \cite{bharadwajQuantumComputationFluid2020a}. Similar to the case with classical computers \cite{higueraSimulatingFlowCircular1989}, algorithms for the simulation of classical fluids on a quantum computer using lattice kinetic theory started with focus on lattice gas \cite{yepezLatticeGasQuantumComputation1998,yepezQuantumComputationFluid1999,vahalaQuantumLatticeGas2008} to lattice Boltzmann \cite{yepezOpenQuantumSystem2006,yepezEfficientQuantumAlgorithm2002,vahalaQuantumLatticeGas2008}. However, we make a distinction between those that attempt to achieve a quantum analogue \cite{mezzacapoQuantumSimulatorTransport2015} versus those which use of a quantum computer is due to the advantage it affors when carrying out the arithmetic \cite{moawadInvestigatingHardwareAcceleration2022a,steijlQuantumAlgorithmsFluid2020,steijlQuantumAlgorithmsNonlinear2020,steijlQuantumCircuitImplementation2023}

In contrast, the Navier-Stokes fluid self-advection
operator which is non-local and non-linear at once. The result is that information
travels along space-time material lines defined by the flow velocity
$\frac{dx}{dt}=u(x,t)$, while in lattice kinetic theory information moves along
straight lines defined by discrete velocities $v$, namely $\frac{dx}{dt}=v$,
where $v$ is a constant. In addition, thanks to the extra dimensions inherent to the phase space $\Gamma = (x,v)$, both the pressure and the strain-rate are locally available online with
no need for solving the Poisson equation requiring second order spatial derivatives. 
This is expected to simplify the structure of the Carleman linearization in comparison to the
fluid case, which involves complex pressure-velocity and pressure-strain-rate correlations. Furthermore, the Carleman-linearized formulation would benefit from quantum algorithms designed for a high-dimensional phase-space \cite{pfefferHybridQuantumclassicalReservoir2022a}.

\subsection{Early Attempts at Quantum Simulation of Fluids}

The early efforts moved from cellular automata, to quantum lattice gas algorithms, to reach the first quantum lattice Boltzmann scheme in the 90s \cite{benziLatticeBoltzmannEquation1992,succiLatticeBoltzmannEquation1993}. Shortly after, the approaches have been further refined for implementation on the nuclear magnetic resoance  (NMR) quantum platforms \cite{yepezQuantumComputationFluid1999,vahalaQuantumLatticeGas2008,yepezLatticeGasQuantumComputation1998,yepezEfficientQuantumAlgorithm2002,yepezOpenQuantumSystem2006,boghosianEntropicLatticeBoltzmann2001} of interest at the time \cite{maheshImplementingLogicGates2001,  priceConstructionImplementationNMR1999,  priceMultiqubitLogicGates2000,  somarooExpressingOperationsQuantum1998,  williamsQuantumComputingQuantum1999}.

\subsection{Recent Attempts}

Recent attempts at quantum algorithms for fluid simulation generally fall into three categories: variational approaches, extensions of quantum linear system algorithms (QLSA) to deal with the nonlinearities, and the physically-motivated. The latter could in fact overlap with the former, as with \cite{lloydQuantumAlgorithmNonlinear2020} which physical basis has been revisited in \cite{gellerUniverseNonlinearQuantum2021}.

Variational methods exist in both paradigms, the adiabatic-type, and the gate-based quantum computer. The quantum annealing paradigm of the former casts the system as an energy minimization problem performed iteratively on a graph \cite{srivastavaBoxAlgorithmSolution2019}. While the state-of-art of quantum annealers also presents them as a more plausible near-term platform \cite{srivastavaBoxAlgorithmSolution2019}, their binary optimization approach has produced unsatisfactory results for Navier-Stokes equations despite grid independence \cite{raySolvingNavierStokesEquation2019}. Variational gate-based algorithms could handle nonlinearities more promptly than linear system methods with their extensions, as the former are logarithmic in the condition number, compared to the square root scaling of the latter \cite{lubaschVariationalQuantumAlgorithms2020}. Variational approaches typically involve coupling between classical and quantum algorithms, such that the terms are varied classically based on the quantum results, as~\cite{lubaschVariationalQuantumAlgorithms2020} have done to produce velocity fields from layered networks. However, moving back and forth between the classical and the quantum is an approach sufficient to kill the quantum speedup in the absence of quantum random access memory (QRAM).

Hybrid, classical-quantum, algorithms form a larger class of which variational algorithms are a subclass. Typically, a quantum step is used to accelerate a classical algorithm, but a classical step could also be used to complement a quantum algorithm \cite{steijlQuantumAlgorithmsFluid2020}. The former is demonstrated in \cite{gaitanSupplementaryInformationFinding} which exhibited the use of the Quantum Amplitude Estimation Algorithm (QAEA) to evaluate the mean-value of the integral of Navier-Stokes equations. 

Solving systems of linear equations independent of time has been shown to be polynomial in the condition number and logarithmic in dimension \cite{schleichHowSolveLinear}. Papers tackling nonlinearity generally attempt to extend a class of quantum algorithms for linear systems, largely based on the HHL algorithm \cite{harrowQuantumAlgorithmSolving2009}. The algorithm, dubbed "quantum" matrix inversion, provides the solution of a linear system of equations . It has been used as a basis for algorithm with exponential runtime for systems in which nonlinearities are of the polynomial type, where the "duplicate" registers are created to match the polynomial order, has been proposed in~\cite{leytonQuantumAlgorithmSolve2008}. 

We have seen a divergence towards citing Navier-Stokes as a future direction for papers discussing solving nonlinear differential equations on quantum computers \cite{liuEfficientQuantumAlgorithm2020,anEfficientQuantumAlgorithm2022}, away from the early physically-motivated algorithms for fluid simulation, as the quantum lattice gas one mentioned above \cite{yepezEfficientQuantumAlgorithm2002,  yepezLatticeGasQuantumComputation1998,  yepezOpenQuantumSystem2006,  yepezQuantumComputationFluid1999}. The is an improvement over using linear governing equations \cite{budinskiQuantumAlgorithmAdvection2021,mezzacapoQuantumSimulatorTransport2015}, or assuming the existence of nonlinear oracles \cite{gaitanFindingFlowsNavier2020}. However, there might be advantages in the physical nature of the nonlinearity, and a physically-motivated representation which would readily extensible for multiscale, multiphysics, simulations.

On the other hand, the attempts to revive the physically-motivated algorithms beyond quantum systems, by \cite{steijlQuantumAlgorithmsFluid2020,mezzacapoQuantumSimulatorTransport2015,budinskiQuantumAlgorithmAdvection2021,lloydQuantumAlgorithmNonlinear2020} and others, are promising. The work of \cite{mezzacapoQuantumSimulatorTransport2015} stands out as it presents itself as a method not of quantum computation per se, but of quantum simulation. The latter leverages the correspondence between the Dirac and lattice Boltzmann equations. We note that the streaming step, specular reflection, and periodic boundary conditions for a quantum lattice Boltzmann algorithm have been developed for a more generic representation of variables \cite{todorovaQuantumAlgorithmCollisionless2020}.

\section{Lattice Boltzmann}
The lattice Boltzmann (LB) formulation uses neither the continuum-scale velocity fields typical of Navier-Stokes solvers, nor the miniscule description of molecular dynamics, but discrete probablity densities $f$ of fluids parcels moving on a $D$-dimensional lattice in $Q$ discrete directions. This makes it inherently suitable for use as the basis of a quantum algorithm. It stems from solving the Boltmzmann kinetic equation over a lattice, restricting the velocity, and thus stepsize of particles in a single timestep, to $\vec{c}_i$, $\vec{c}_i=c_i\vec{e}_i$, defined in the directions of the lattice vectors $\vec{e}_i$s. Similar to how it is derived from the Boltzmann equation, it could be adapted for a range of partial-differential equations \cite{succiLatticeBoltzmannEquation2018}. For a single, not mixed, phase fluid, it takes the form:

\begin{equation}
\begin{aligned}
\label{lBolEq}
    \frac{1}{\Delta t}(f_i(\vec{x}+\vec{c}_i\Delta t, t + \Delta t) - f_i(\vec{x},t)) = {\Omega_i}(\vec{f}(\vec{x},t)) 
\end{aligned}
\end{equation}

where under the BGK approximation, the collision term, $\Omega_i$, which is a relaxation towards equilibrium, is simply the finite difference between the value of the discrete probability density $f_i(\vec{x},t)$ and a corresponding equilibrium function $f_i^{eq}(\vec{x},t)$, over relaxation time $\tau$:
\begin{equation}
    \Omega_i = -\frac{1}{\tau} (f_i -f_i^{eq})
\end{equation}
We note that for a multiphase flow, the model could be extended to include multiple relaxation processes towards the equilibrium distribution of each phase.

Evolving towards the solutions involves two operations at each timestep, collision and advection (or streaming). During collision, the discrete densities are updated based on the values of the local densities in the same cell according to the collision term seen on the right hand side of the discretized equation Eq.~(\ref{lBolEq}):

\begin{equation}
\label{colBolEq}
    f_i(\vec{x}, t + \Delta t) = f_i(\vec{x},t) +\Delta t\Omega_i(\vec{f})
\end{equation}

From the right hand side of Eq.~(\ref{lBolEq}), one can see that the collision process is a relaxation towards an equilibrium distribution. The BGK term for a single=phase fluid utilizes an equilibrium function:

\begin{equation}
    \begin{aligned}
        f_i^{eq} = w_i\rho (1+3\frac{\vec{e}_i\cdot\vec{u}}{c_s^2}+\frac{9}{2}(\frac{\vec{e}_i\cdot\vec{u}}{c_s^2})^2-\frac{3}{2}\frac{\vec{u}^2}{c_s^2})
    \end{aligned}
\end{equation}

Under the incompressible assumption, the equilibrium function becomes:

\begin{equation}
\label{approxO}
\begin{aligned}
    f_i^{eq}(\vec{x}, t) ={w_i}(1+3\vec{e}_i\cdot f_j\vec{e}_j+\frac{9}{2}(\vec{e}_i\cdot f_j\vec{e}_j)^2-\frac{3}{2}f_jf_k\vec{e}_j\cdot\vec{e}_k))
    \end{aligned}
\end{equation}

For example, for D1Q3, we have:
\begin{equation}
{\Omega}(\vec{f}(\vec{x},t)) = - \frac{1}{\tau}
\begin{pmatrix}
\frac{1}{2}
(f_1 + f_3 - (f_1 - f_3)^2 - \frac{1}{3})\\
(f_2 + (f_1 - f_3)^2 -\frac{2}{3})\\
\frac{1}{2}(f_1 + f_3 - (f_1 - f_3)^2 - \frac{1}{3})
\end{pmatrix}
\end{equation}

Streaming proceeds by propagating each discrete density along its respective lattice direction to the neighboring cell. While collision is a local nonlinear operation, streaming is a nonlocal linear one with is said to be exact in reference to the zero round-off error when performed on a classical computer.
\begin{equation}
\label{strBolEq}
    f_i(\vec{x},t + \Delta t) \xrightarrow{} f_i(\vec{x}+\vec{c}_i\Delta t, t + \Delta t)
\end{equation}
A quantum algorithm for lattice Boltzmann, then, requires an implementation of the two, collision and streaming, steps.

\section{The Lattice Boltzmann Equation in the Mode-Coupling Form}
\begin{figure*}
\centering
\includegraphics[scale=1]{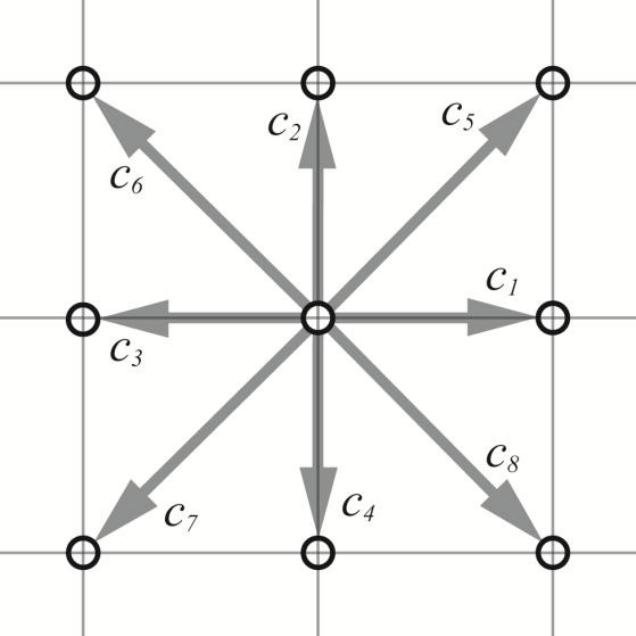}
\caption{A single site in the standard D2Q9 lattice\label{fig:d2q9fivefam}}
\end{figure*}
We consider the lattice Boltzmann (LB) equation
in single-time relaxation form and one spatial dimension, without loss
of generality:
\begin{widetext}
\begin{equation}
\label{LBC1}
f_i(x+c_i \Delta t,~t+\Delta t) -f_i(x,t)=(1-\omega)f_i(x,t)+\omega f_i^{eq}(x,t).
\end{equation}
\end{widetext}
Here $f_i(x,t)=f(x,v,t) \delta(v-c_i)$ is a set of discrete Boltzmann
distributions moving with discrete velocity  
$c_i, i=1,b$, chosen according to a suitable symmetry group; $\omega = \Omega \Delta t$.
The LHS is the free streaming along direction $c_i$ while the RHS stands for
collisional relaxation towards the local equilibrium $f_i^{eq}$ on a typical
timescale $\tau=1/\Omega$.
The local equilibrium compatible with the Navier-Stokes equation of 
incompressible fluid dynamics reads as \cite{succiLatticeBoltzmannEquation2001,succiLatticeBoltzmannEquation2018}:

\begin{equation}
\label{LEQ}
f_i^{eq} = w_i \rho(1+\frac{c_iu}{c_s^2}+ \frac{Q_iu^2}{2c_s^4}), 
\end{equation}
where $w_i$ is a suitable set of weights normalized to unity.
In the above, $\rho=\sum_i f_i$  is the fluid density, $u= \sum_i f_i c_i$ is the fluid 
current and $Q_i = c_i^2 -c_s^2$, $c_s^2$ being the lattice sound speed,
is a constant $O(1)$ in lattice units.

In $d=2$, the standard D2Q9 model, namely the second-rank tensor, can be obtained through taking the tensor product of the first-rank vector $(-1,0,1)$ for D1Q3.
Likewise, the $D3Q27$ lattice directions
are defined by the third-rank tensor product of D1Q3 = $(-1,0,1)$ (see Fig.2).

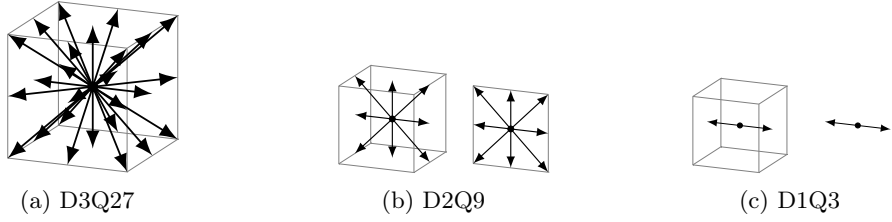
\begin{figure}[ht]
\centering
\captionsetup[subfigure]{justification=centering}
\subcaptionbox[D3Q27]{D3Q27\label{subfig:d3q27}}
[3cm]
{\resizebox{3cm}{!}{
\tdplotsetmaincoords{74}{113}
         \begin{tikzpicture}[tdplot_main_coords,
axis/.style={thick, ->, >=stealth'}]

\coordinate (O) at (0.5,0.5,0.5);

\def \a {1}       
\def \b {1}       
\def \c {1}       

 \foreach \u in {0,1,...,\a}
    \foreach \v in {0,1,...,\b}
      \foreach \w in {0,1,...,\c}
        \draw[very thin,gray] (\u,\v,0) -- (\u,\v,\w); 
 \foreach \u in {0,1,...,\a}
    \foreach \v in {0,1,...,\b}
      \foreach \w in {0,1,...,\c}
        \draw[very thin,gray] (\u,0,\w) -- (\u,\v,\w); 
 \foreach \u in {0,1,...,\a}
    \foreach \v in {0,1,...,\b}
      \foreach \w in {0,1,...,\c}
        \draw[very thin, gray] (0,\v,\w) -- (\u,\v,\w);

\draw plot [mark=*, mark size=1] coordinates{(O)};

 \foreach \u in {0,1,...,\a}
    \foreach \v in {0,1,...,\b}
      \foreach \w in {0,1,...,\c}
        \draw[thin,-latex,black](O) -- (\u,\v,\w);
        
    \foreach \v in {0,1,...,\b}
      \foreach \w in {0,1,...,\c}
        \draw[thin,-latex,black](O) -- (0.5,\v,\w);
 \foreach \u in {0,1,...,\a}
      \foreach \w in {0,1,...,\c}
        \draw[thin,-latex,black](O) -- (\u,0.5,\w);
 \foreach \u in {0,1,...,\a}
    \foreach \v in {0,1,...,\b}
        \draw[thin,-latex,black](O) -- (\u,\v,0.5);

 \foreach \u in {0,1,...,\a}
        \draw[thin,-latex,black](O) -- (\u,0.5,0.5);
    \foreach \v in {0,1,...,\b}
        \draw[thin,-latex,black](O) -- (0.5,\v,0.5); 
      \foreach \w in {0,1,...,\c}
        \draw[thin,-latex,black](O) -- (0.5,0.5,\w); 

\end{tikzpicture}
}
}
\hspace{0.1\textwidth}
\subcaptionbox[D2Q9]{D2Q9\label{subfig:d2q9}}
[3cm]
{\resizebox{3cm}{!}{
\tdplotsetmaincoords{74}{113}
         \begin{tikzpicture}[tdplot_main_coords,
axis/.style={thick, ->, >=stealth'}]

\coordinate (O) at (0.5,0.5,0.5);

\def \a {1}       
\def \b {1}       
\def \c {1}       

 \foreach \u in {0,1,...,\a}
    \foreach \v in {0,1,...,\b}
      \foreach \w in {0,1,...,\c}
        \draw[very thin,gray] (\u,\v,0) -- (\u,\v,\w); 
 \foreach \u in {0,1,...,\a}
    \foreach \v in {0,1,...,\b}
      \foreach \w in {0,1,...,\c}
        \draw[very thin,gray] (\u,0,\w) -- (\u,\v,\w); 
 \foreach \u in {0,1,...,\a}
    \foreach \v in {0,1,...,\b}
      \foreach \w in {0,1,...,\c}
        \draw[very thin, gray] (0,\v,\w) -- (\u,\v,\w);

\draw plot [mark=*, mark size=1] coordinates{(O)};

    \foreach \v in {0,1,...,\b}
      \foreach \w in {0,1,...,\c}
        \draw[thin,-latex,black](O) -- (0.5,\v,\w);

    \foreach \v in {0,1,...,\b}
        \draw[thin,-latex,black](O) -- (0.5,\v,0.5); 
      \foreach \w in {0,1,...,\c}
        \draw[thin,-latex,black](O) -- (0.5,0.5,\w); 

\end{tikzpicture}

\tdplotsetmaincoords{74}{113}
         \begin{tikzpicture}[tdplot_main_coords,
axis/.style={thick, ->, >=stealth'}]

\coordinate (O) at (0.5,0.5,0.5);

\def \a {1}       
\def \b {1}       
\def \c {1}       

 \foreach \v in {0,1,...,\b}
      \foreach \w in {0,1,...,\c}
        \draw[very thin,gray] (0.5,\v,0) -- (0.5,\v,\w); 
 \foreach \v in {0,1,...,\b}
      \foreach \w in {0,1,...,\c}
        \draw[very thin,gray] (0.5,0,\w) -- (0.5,\v,\w);

\draw plot [mark=*, mark size=1] coordinates{(O)};

    \foreach \v in {0,1,...,\b}
      \foreach \w in {0,1,...,\c}
        \draw[thin,-latex,black](O) -- (0.5,\v,\w);

    \foreach \v in {0,1,...,\b}
        \draw[thin,-latex,black](O) -- (0.5,\v,0.5); 
      \foreach \w in {0,1,...,\c}
        \draw[thin,-latex,black](O) -- (0.5,0.5,\w); 

\end{tikzpicture}}}
\hspace{0.1\textwidth}
\subcaptionbox[D1Q3]{D1Q3\label{subfig:d1q3}}
[3cm]
{\resizebox{3cm}{!}{
\tdplotsetmaincoords{74}{113}
         \begin{tikzpicture}[tdplot_main_coords,
axis/.style={thick, ->, >=stealth'}]

\coordinate (O) at (0.5,0.5,0.5);

\def \a {1}       
\def \b {1}       
\def \c {1}       

 \foreach \u in {0,1,...,\a}
    \foreach \v in {0,1,...,\b}
      \foreach \w in {0,1,...,\c}
        \draw[very thin,gray] (\u,\v,0) -- (\u,\v,\w); 
 \foreach \u in {0,1,...,\a}
    \foreach \v in {0,1,...,\b}
      \foreach \w in {0,1,...,\c}
        \draw[very thin,gray] (\u,0,\w) -- (\u,\v,\w); 
 \foreach \u in {0,1,...,\a}
    \foreach \v in {0,1,...,\b}
      \foreach \w in {0,1,...,\c}
        \draw[very thin, gray] (0,\v,\w) -- (\u,\v,\w);

\draw plot [mark=*, mark size=1] coordinates{(O)};

    \foreach \v in {0,1,...,\b}
        \draw[thin,-latex,black](O) -- (0.5,\v,0.5); 
\end{tikzpicture}

\tdplotsetmaincoords{74}{113}
         \begin{tikzpicture}[tdplot_main_coords,
axis/.style={thick, ->, >=stealth'}]

\coordinate (O) at (0.5,0.5,0.5);

\def \a {1}       
\def \b {1}       
\def \c {1}       

 \foreach \u in {0,1,...,\a}
    \foreach \v in {0,1,...,\b}
      \foreach \w in {0,1,...,\c}
        \draw[very thin,white] (\u,\v,0) -- (\u,\v,\w); 
 \foreach \u in {0,1,...,\a}
    \foreach \v in {0,1,...,\b}
      \foreach \w in {0,1,...,\c}
        \draw[very thin,white] (\u,0,\w) -- (\u,\v,\w); 
 \foreach \u in {0,1,...,\a}
    \foreach \v in {0,1,...,\b}
      \foreach \w in {0,1,...,\c}
        \draw[very thin,white] (0,\v,\w) -- (\u,\v,\w);

\draw plot [mark=*, mark size=1] coordinates{(O)};

    \foreach \v in {0,1,...,\b}
        \draw[thin,-latex,black](O) -- (0.5,\v,0.5); 
\end{tikzpicture}}}
\caption{Different lattice configurations in three, two and one dimensions}
\label{fig:lattices}
\end{figure}

\subsection{Mode Coupling Form}

By the definitions of $\rho$ and $u$, the local equilibrium can be written 
in the mode-coupling form as
\begin{equation}
f_i^{eq} =  L_{ij}f_j+Q_{ijk} f_j f_k, 
\end{equation}
where  
\begin{eqnarray}
L_{ij}=w_i(1+c_i c_j/c_s^2)\\
Q_{ijk} = w_i Q_i c_j c_k/2c_s^4.
\end{eqnarray}
As a result the LBE takes the form
\begin{equation}
f_i(x+c_i, t+1) = A_{ij}f_j + B_{ijk}f_j f_k, 
\end{equation}
where we have set $A_{ij}=\delta_{ij}-\omega L_{ij}$ and $B_{ijk}=\omega Q_{ijk}$.
Note that the RHS has two fixed points, an unstable trivial vacuum $f_i=0$
and a non-trivial stable one, $f_i = f_i^{eq}$.
This is the desired mode-coupling form which proves expedient 
to the Carleman formulation, to be discussed next.
Before doing so, let us note that, owing to conservation laws, the above matrices
display these sum-rules: 
\begin{eqnarray}
\label{SUMRULES}
\sum_i w_i = 1,\;
\\\sum_i L_{ij}=\sum_j L_{ij}=1,\;
\\\sum_i Q_{ijk} = 0.
\end{eqnarray}

As we shall show, these conservation laws lay at the ground of the 
exact closure at the second order of the Carleman procedure
for the homogeneous kinetic equation.

\section{Classical Carleman Linearization}

Carleman linearization works by introducing additional variables that are monomials up to a chosen order, making it ideal for linearizing polynomial expressions or nonlinear functions which are well-expressed as power-series. The evolution of the additional variables could be simply derived from that of the original variables by using the original system of differential equations along with the product rule from calculus. Together with the original variables $\vec{f}(t)$, the additional variables are called the Carleman variables, $\vec{V}(t)$, and the chosen order of truncation is called the Carleman order $O_c$. Starting from a system of differential equations with a nonlinear driving function $\Omega_i$:
\begin{equation}
\label{Eq::density_time_derivative}
    \frac{\partial f_i(t)}{\partial t} = \Omega_i(\vec{f}(t)),
\end{equation}
one can linearize it into
\begin{equation}
\label{Carl_Equation}
    \frac{\partial \vec{V}(t)}{\partial t} = C\vec{V}(t).
\end{equation}
where $C$ is a matrix of constant coefficients derived from:
\begin{equation}
\label{Eq::Carleman_var_def}
    \frac{\partial V_n(t)}{\partial t}  = \Sigma_{i=1}^Q\frac{\partial V_n(t)}{\partial f_i}\Omega_i(\vec{f}(t)), 
\end{equation}
by identifying each monomial that appears with its corresponding variable $V_i$ on the right hand side of Eq.~(\ref{Eq::Carleman_var_def}), and removing all monomials beyond the chosen order $O_c$.

Performing the Carleman linearization for the lattice Boltzmann method with classical numerical techniques gives rise to some issues~\cite{itaniAnalysisCarlemanLinearization2022}, such as a combinatorial increase in the number of variables, as well as the loss of the the exactness of streaming, a major advantage of the lattice Boltzmann formulation.

Once Eq.(\ref{Eq::Carleman_var_def}) is solved for $\vec{V}(t)$, an approximate solution $\vec{f}_C$ to the original variables $\vec{f}(t)$ can be found, by looking at the subset of Carleman variables $\vec{V}$ of first-order, which coincide with $\vec{f}$ by construction. Because we are truncating $\vec{V}$ at order $O_c$, the solution $f_c$ will be approximate.

Let us illustrate the idea for the simple case of 
the logistic equation. 

\subsection{Carleman Treatment of the Logistic Equation}

Consider the logistic equation
\begin{equation}
\partial_t f = -af+bf^2;\;\;\;f(0)=f_0
\end{equation}
with $a,b$ both positive.
This can be seen as the homogeneous version $(\partial_x f =0)$ of a corresponding
kinetic equation with two fixed points, a stable one, $f=0$, and
an unstable one, $f=a/b=K$, where $K$ is the so-called carrying-capacity.
Note that $R=b/a=1/K$ measures the strength of the nonlinearity (and is akin to the Reynolds number).
The exact solution reads 
\begin{equation}
f(t) = \frac{f_0 e^{-at}}{1-\frac{f_0}{K}(1-e^{-at})}.
\end{equation}
For $0<f_0<K$ this decays to zero, while for $f_0>K$ it develops
a finite-time singularity in a time lapse $at_{sing} \sim K/f_0$.  

The Carleman procedure is readily shown to lead to the
infinite chain of ODE's
\begin{equation}
\frac{df_k}{dt} = - k(a f_k +  b f_{k+1}),\;\;\; k=1, k_{max}
\end{equation}
with initial conditions $f_k(0)= f_0^k$.
The first order truncation $f_2=0$ yields a pure exponential
decay $f_1^{(0)}(t)= f_0 e^{-at}$. Besides missing the slow
transient, this also provides an incorrect asymptotic amplitude.
However, as long as $f_0/K \ll 1$, it is readily shown
that a simple Euler time marching with sufficiently small time step 
$a \Delta t \ll 1$, delivers a pretty accurate solution
with just a few Carleman iterates.
However, the convergence deteriorates rapidly as $f_0/K \to 1$.

\begin{figure*}[!htbp]
\centering
\input{Figures/3}
\caption{Absolute Carleman error with overlap $Rf_0=0.5$ (\ref{subfig:error1}) 
and $Rf_0=0.01$ (\ref{subfig:error2}), with time
step $\Delta t=0.01$, with truncation from the first (orange) to the fourth (violet) order.
(\ref{subfig:error3}) is the absolute Carleman error with overlap $Rf_0=0.1$ and time
step $\Delta t = 0.001$, with truncation from the first to the fourth order.
}
\end{figure*}
\subsection{Carleman Linearization for Kinetic Theory versus Fluid Dynamics} 

In the spirit of developing a Carleman-based quantum algorithm for fluids, it is natural
to focus the attention on the Navier-Stokes equations. 
A simple inspection shows that the Carleman linearization of the Navier-Stokes 
equations meets  with a number of complications due to correlations between the fluid flow
$u_\mu$, $\mu=x,y,z$, the strain sensor $D_{\mu \nu} = \partial_\mu u_\nu$ and the fluid
pressure $p$. In particular, a dynamic equation for the fluid pressure is required
to generate the pressure-velocity and pressure-strain correlators, and
the dynamic equations for the corresponding correlators are quite cumbersome.

On the other hand, the Carleman linearization of the kinetic equation gives rise to
a hierarchy of multiscalars, $f_i \to f_i f_j \to f_i f_j f_k \to ...$ and 
the corresponding dynamic equations remain first order in space and time, which
we expect to provide a significant advantage for the formulation of a
quantum computing algorithm.  
More importantly,  thanks to the basic mass-momentum conservation laws, there
is no need to track all the multiscalars above, but only a limited 
subset of linear combinations therefrom. 

Either way, the key question is the convergence as a function of the Reynolds number.
Based on the logistic results, one would expect Carleman convergence to occur 
under the constraint 
\begin{equation}
\label{CCO}
|\frac{f-f^{eq}}{f^{eq}}| \ll 1/Re.
\end{equation}
This looks quite demanding at high Reynolds number, although 
a moment's thought reveals that such a constraint is 
fully in line with the hydrodynamic limit of kinetic theory.

To this end, let us remind that hydrodynamics emerges from the kinetic theory
in the limit of weak departure from local equilibrium, or, differently 
restated, for small Knudsen numbers $Kn = \lambda/L \sim |f-f^{eq}|/f^{eq}\ll 1$, 
where $\lambda$ is the molecular mean free path and $L$ a characteristic hydrodynamic scale.
The next observation is that the Knudsen number scales inversely with
the Reynolds number $Re=UL/\nu$, according to the so-called von K\'arm\'n relation
\begin{equation}
Kn = \frac{Ma}{Re},
\end{equation}
where $Ma$ is the Mach number.
Taking $Ma \sim O(1)$, and $f^{eq} \sim 1$, the above relation coincides
with the constraint Eq.~(\ref{CCO}). 
For a typical car, we have $Re \sim 10^7$, indicating that the departure from
local equilibrium is on the order of the seventh digit.
This is unquestionably a stringent request,  mitigated however by the hydrodynamic
conservation laws, as we shall discuss shortly.

\subsection{Carleman Lattice Boltzmann (CLB) Scheme}

We will now discuss this last point by providing the explicit form of
the Carleman Lattice Boltzmann (CLB) scheme to first and second orders.

\subsubsection{First-order CLB}

To the first order the standard Lattice Boltzmann (LB) equation for $f_i$ is 
\begin{equation}
\label{LBC1firstorder}
f_i(x+c_i \Delta t, t+\Delta t) = (1-\omega) f_i(x,t) + \omega g_i(x,t),
\end{equation}
where $\omega = \Omega \Delta t$, as before, and we have set $g_i \equiv f_i^{eq}$ to 
denote a generic collisional attractor.
At this stage, the Carleman array of variables is just $F_1=[f_i]$,

It is to be noted further that the negative lattice viscosity, also known as propagation
viscosity $\nu_P = -\frac{1}{2}$ (in lattice units $\Delta x = \Delta t =1$),
can be incorporated within an effective physical viscosity, $\nu = c_s^2 (1/\omega-1/2)$,
thereby permitting one to march in large steps $\Delta t \sim O(\tau)$ without losing stability.  
This stands in contrast to Euler marching for ODEs, which requires $\Delta t \ll \tau$.

\subsubsection{Second-order CLB}

We write the LB equation at two different locations $x_i=x+c_i$ and $x_j=x+c_j$, as
\begin{equation}
\label{LBC121}
f_i(x_i,t+1) = (1-\omega) f_i(x,t) + \omega g_i(x,t)
\end{equation}
\begin{equation}
\label{LBC122}
f_j(x_j,t+1) = (1-\omega) f_j(x,t) + \omega g_j(x,t)
\end{equation}
where we have $\Delta t=1$ for convenience.

Multiplying one equation by the other, we obtain

\begin{widetext}
\begin{equation}
\label{}
f_{ij}(x_i,x_j,t+1) 
= (1-\omega)^2 f_{ij}(x,x,t)  
+ 2 \omega (1-\omega) h_{ij}(x,x;t)
+ \omega^2 g_{ij}(x,x,t),
\end{equation}
\end{widetext}
where we have set
\begin{equation}
g_{ij}= g_i g_j
\end{equation}
for the double-equilibrium and
\begin{equation}
h_{ij} = \frac{1}{2}(f_ig_j+g_if_j)
\end{equation}
for the semi-equilibrium.

Several comments are in order. First, we note that for $\omega=0$, we still have an exact free-streaming 
formulation $f_{ij}(x_i,x_j,t+1)= f_{ij}(x,x,t)$. Second, the second term on the right-hand-side takes the symbolic form
form $h=fLf+fQff$, while the third one gives $g=LfLf+LfQff+QffLf+QffQff$, indicating
coupling with third and fourth order Carleman variables.
Clearly, Carleman truncation at order two retains only $fLf$ and $LfLf$, although
a better approximation might be obtained by replacing $f$ in the higher order terms
with a zero-velocity equilibrium, $f_i \sim \rho w_i$.
We also observe that the above notation invites a natural analogy with tensor 
networks which might be worth exploring for the future \cite{gourianovQuantuminspiredApproachExploit2022}. 

Most important of all, as a consequence of streaming, the second order 
Carleman array involves a pair (Carleman pairs) of locations $x_i$
and $x_j$, typical of a local two-body problem.
The number of Carleman variables at this stage is thus
$bL$ for $f_i(x)$, and $b(b+1)L/2$ for $f_{ij}(x_i,x_j)$.
\begin{figure*}
\label{CarlePairs}
\centering
\includegraphics[scale=0.5]{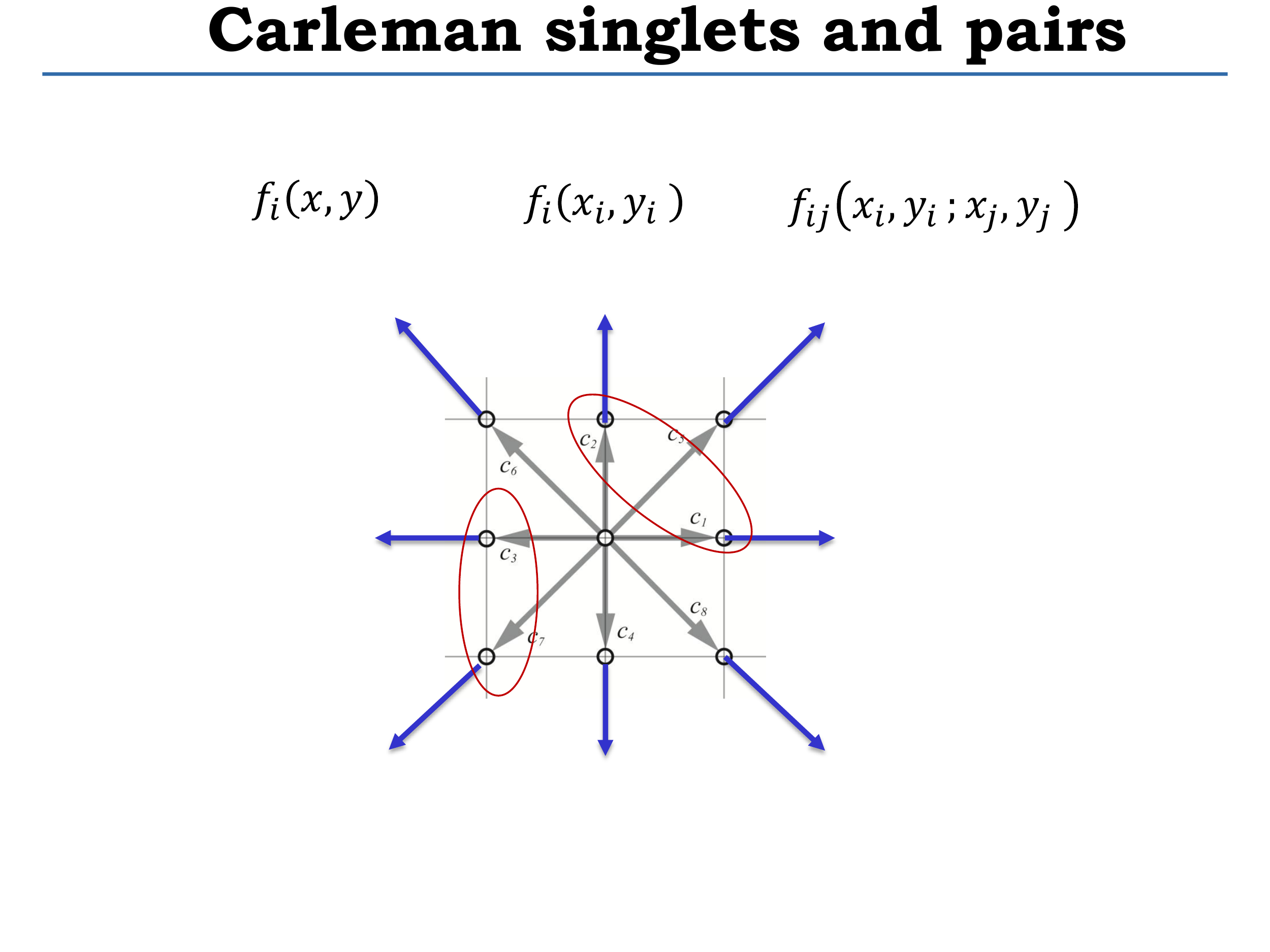}
\caption{The first three Carleman structures: local singlet $f_i(x,y)$ (central grey), 
non-local singlet $f_i(x_i,y_i)$ (blue arrows) and nonlocal pair $f_{ij}(x_i,y_i,x_j,y_j)$ (red circled).
The encircled pairs correspond to $f_{12}$ and $f_{37}$ respectively. 
The figure refers to a two-dimensional D2Q9 for better visualization purposes. 
}
\end{figure*}
\subsection{Carleman Structure for Kinetic Theory versus Fluid Dynamics} 

The above consideration signals a combinatorial many-body proliferation of Carleman 
variables, scaling like the size of the grid to some power 
associated with the Carleman truncation order, $b^2$
at order 2, $b^3$ at order 3, and so on.

In a nutshell, {\it Carleman linearization at order $k$ transforms 
a nonlinear one-body problem $d$ into a linear k-body problem.} This is the principal effect of nonlocality.
 
But let us for the moment suspend the issue of nonlocality 
and focus on local nonlinearity, namely the streaming-free 
homogeneous LB.

\subsection{Homogeneous Case}
Equation \ref{LBC1} takes the form:
\begin{equation}
\label{LBC1homo}
f_i(x,t+1) = (1-\omega) f_i(x,t) + \omega f_i^{eq}(x,t)
\end{equation}
This is a set of $b$ ODE's, whose local attractor is $f_i^{eq}$.
The equation for $f_{ij}(x,t)$ now reads simply as:
\begin{widetext}
\begin{equation}
\label{LBC2}
f_{ij}(x,t+1) 
= (1-\omega)^2 f_{ij}(x,t)  
+ 2 \omega h_{ij}(x,t) 
+ \omega^2 g_{ij}(x,t)
\end{equation}
\end{widetext}

The second order CLB reads formally as 
\begin{widetext}
\begin{eqnarray}
\label{LBC2T}
f_{i}(t+1) = [(1-\omega) \delta_{ij} + \omega L_{ij}] f_j 
+\omega  Q_{ijk} f_{jk}\\
f_{ij}(t+1) = (1-\omega)^2 f_{ij}  
+ \omega (1-\omega)[L_{jk} f_{ki} + L_{il} f_{lj}]
+ \omega^2 L_{ik} L_{jl} f_{kl}.
\end{eqnarray}
\end{widetext}

This is elegant and can be readily generalized to higher orders.
However, it shows a steep power-law growth, $N^k$ at the $k^{th}$
Carleman order, which becomes rapidly unsustainable on classical
computers.

\subsection{From Truncation to Exact Closure}
\label{subsec:truncexact}
The above treatment does not make use of the conservation laws and resulting
sum rules, relations in Eq.~(\ref{SUMRULES}), which, for the homogeneous case,
permit one to close the Carleman hierarchy exactly at the second order.

The main observation is that local equilibria depend parametrically only
on the fluid density $\rho$ and the flow field $u$.
For the case of incompressible flows, we can set $\rho=1$, so that the local
equilibria, hence the quadratic term in the Carleman procedure, depend only
on the quadratic term $u^2=\sum_{ij} f_{ij} c_i c_j$ and not on 
the single components $f_{ij}$.

For the sake of concreteness, let us report the explicit calculation
for the D1Q3 case, with the following basic parameters:
$c_0=0,c_1=-1,c_2=+1$,
$w_0=4/6,w_1=1/6,w_2=1/6$,
$Q_0=-1/3,Q_1=2/3,Q_2=2/3$. The corresponding local equilibria are
\begin{eqnarray}
f_0^{eq} = \frac{2}{3}(1-\frac{u^2}{3})\\
f_1^{eq} = \frac{1}{6}(1-3u+3u^2)\\
f_2^{eq} = \frac{1}{6}(1+3u+3u^2).
\end{eqnarray}

The equation of motion for the first Carleman level 
$F_1= [f_0,f_1,f_2]$ is
$
\frac{d f_i}{dt} = -\omega (f_i-f_i^{eq}).
$
From the expression for the local equilibria, it is clear that
the quadratic coupling is confined to the $u^2=(f_2-f_1)^2$ term;
consequently, it is sufficient to define just {\it one} second-level 
Carleman variable, $F_2 = (f_2-f_1)^2$.
On the other hand, since $u$ is left unchanged by the collision operator,
we clearly have $du^2/dt=0$.
This shows that the four-component Carleman system, given by 
$F_{12}= [f_0,f_1,f_2;(f_2-f_1)^2]$, can be truncated
exactly to the second order \cite{itaniAnalysisCarlemanLinearization2022}.
This is an example of a simple but very effective dimensional reduction
via nonlinear mapping: instead of tracking $f_1^2,f_2^2, f_1 f_2$ separately, it is
sufficient to track the single invariant $(f_1-f_2)^2$.
The same technique is readily extended to two and three-dimensions, with three 
and six extra Carleman variables, $u_x^2,u_x u_y,u_y^2$ 
and $u_x^2,u_xu_y,u_xu_z,u_y^2,u_yu_z, u_z^2$, respectively. 

Unfortunately, this wonderful property is impaired by the streaming step, 
since $u(x)$ is no longer a dynamic invariant. 
Hence, the issue of nonlocality cannot be sidestepped; due to
the combinatorial growth of the degrees of freedom for increasing Carleman 
orders, it is clear that, on classical computers, trading nonlinearity 
for higher dimensions and non-locality is a self-inflicted exertion.

Differently restated, the CLB scheme ought to be run 
on quantum computers. 

\section{Motivation}
\subsection{Carleman Embedding}

The Carleman embedding technique introduces an extended set of variables \cite{itaniAnalysisCarlemanLinearization2022}, defining higher powers of the original ones, such that the system is linear in the extended set

\begin{figure}
    \centering
    \input{Figures/5}
    \caption{Illustration of a set of original and extended variables used to linearize a system under the Carleman embedding approach}
    \label{fig:carlvars}
\end{figure}

The number of Carleman variables $N$ grows combinatorially in the linearization order $O_c$ and the number of original variables $Q$. Its correspondence to mapping the variables to eigenvalues of operators in an infinite Hilbert space, thereby to a linear embedding in the quantum mechanical framework, has long been know \cite{kowalskiNonlinearDynamicalSystems1997}. There lies an opportunity for a more efficient encoding of the linearization technique in the latter framework, but a mapping to native qubit gates awaits. Moreover, despite the necessary mathematical foundation available, an analytical study of the effects of such truncation is lacking to the best of our knowledge \cite{alpayIndefiniteInnerProduct2018}. Consider a state encoding three variable $x$,$y$ and $z$ along with their higher powers, up to a normalization:
\begin{equation}
\begin{aligned}
    \ket{x}\ket{y}\ket{z} &= (\Sigma_{i,j,k=0}^{O_c} x^i)\ket{i})(\Sigma_{j=0}^{O_c}y^j\ket{j})(\Sigma_{k=0}^{O_c}z^k\ket{k})
    \\&= \Sigma_{i,j,k=0}^{O_c} x^i y^j z^k \ket{i}\ket{j}\ket{k}
    \end{aligned}
    \label{eq:basicidea}
\end{equation}
where $\ket{i}$, $\ket{j}$ and $\ket{k}$ refer to the states encoding the binary representation of the values of the respective indices $\ket{i=3} = \ket{011}$. As such, in the quantum Carleman, as opposed to the classical Carleman technique, we no longer talk about the number of Carleman variables $N$ and the linearization order separately $O_c$, as they are easily identified for each variable as seen, and we define $N=O_c$. Moreover, instead of simple monomials of the variables, we can talk about linear combinations of such monomials, including the special case where these linear combinations correspond to the Hermite polynomials. 

\section{Quantum Linear Embedding}
\subsection{Quantized Classical Variables}

The standard method to move from classical physics to quantum physics is to map the classical variable(s) to a quantum variable(s).
\begin{equation}
u \rightarrow \hat{u}
\end{equation}
To store the variable as it is evolved, we must utilize the quantum computer equivalent of a (\textit{classical}) bit, a qubit. A register (\textit{set}) of qubits could be thought of as in some state:
\begin{equation}
\ket{\psi}
\end{equation}
Then, for our mapped classical variables, we must consider a state:
\begin{equation}
\ket{u}
\end{equation}
which is an eigenstate of the quantum operator corresponding to the classical variable, shown here to depend on a single independent variable, time:
\begin{equation}
\hat{u} \ket{u(t)} = u(t) \ket{u}
\end{equation}
If the evolution:
\begin{equation}
\ket{u(t = 0)} \rightarrow \ket{u(t = T)}
\end{equation}
abides by the restriction of being unitary (norm-preserving), then the results can be retrieved as:
\begin{equation}
\bra{u(t)}\hat{u}\ket{u(t)} = \bra{u}u(t)\ket{u(t)} = u(t) \braket{u(t)}{u(t)} = u(t)
\end{equation}
It is worthy to note here that quantum computers in reality are open quantum systems, but this nonunitary evolution is commonly dismissed, when discussing quantum algorithms theoretically, as noise. In practice, this noise might be advantageous in simulating dissipative dynamics, but this discussion is beyond the scope of this work.
\subsection{Higher-Order Terms}

Before we proceed, recall that the Schr\"odinger equation is linear. The right-hand side of the equation is factored into a Hamiltonian (\textit{matrix}) acting upon the state. When we are discussing retrieving higher-order terms of the quantized variables, we must then first think in terms of these higher orders being stored and updated. Considering the classical variable $u$:
\begin{equation}
u^0 = 1, u^1 = u, u^2, u^3, \dots
\end{equation}
are its orders. As such $N$ degrees of freedom are required to encode $N$ orders of $u$. Conveniently, a quantum computer offers a computational space that increases exponentially with the number of qubits. As such, we expect to be able to encode $N+1$ values, counting the zeroth order, in $\log_2{(N+1)}$ qubits. We also note that we need not encode the values as is, but linear combinations would also do well in encoding the information. One such linear combination is rescaling each value representing an order $n$ by $\sqrt{(n!)}$:
\begin{equation}
\frac{1}{\sqrt{0!}} = 1, \frac{u}{\sqrt{1!}} = u, \frac{u^2}{\sqrt{2!}}, \frac{u^3}{\sqrt{3!}}
\end{equation}
This encoding is known in quantum jargon, up to a normalization factor, as a coherent state:
\begin{equation}
\ket{\alpha} = \Sigma_{n=0}^\infty \frac{\alpha^n}{\sqrt{n!}} \ket{n}
\end{equation}
where $\ket{n}$ is a computational (binary) basis representation of $n$:
\begin{equation}
\begin{aligned}
\ket{n = 0} =& \ket{0\dots0\dots00}
\\\ket{n = 1} =& \ket{0\dots0\dots01}
\\\vdots
\end{aligned}
\end{equation}
Encoding a variable in a coherent state corresponds to choosing to quantize a classical variable to a bosonic lowering operator, to which the coherent state is an eigenstate (eigenvector), with the computational basis corresponding to an occupation number basis, upon which the lowering operator acts as:
\begin{equation}
\hat{a} \ket{n} = \sqrt{n} \ket{n-1}
\end{equation}
We note that the Hilbert space used to study a quantum system allows us to generalize linear algebra to an infinite dimensional space. The language of eigenstates and operators is then an efficient way to discuss the concept of Carleman linearization, which is mathematically equivalent, also casting a finite system into an infinite system to linearize it. The powers of the variable, then, form an orthogonal polynomial basis, with choosing a different basis through orthogonal linear combinations corresponding to the choice of a different quantizing operator. For further details, the reader is referred to \cite{kowalskiNonlinearDynamicalSystems1997}.
\subsection{Canonical Conjugate Operator}

In classical physics, we sepak of a pair of conjugate variables where one is the Fourier transform of the other. In quantum physics, as variables are quantized to operators, we speak of conjugate operators as those having nonzero commutation relation i.e. they do not commute:
\begin{equation}
[\hat{A},\hat{B}] = \hat{A}\hat{B}-\hat{B}\hat{A} \neq 0
\end{equation}
such that the uncertainity principle applies to them, and they can only be simultaneously observed up to a (\textit{fundamental}) error. Of special importance are canonical conjugate operators. Consider the raising operator $\hat{a}^\dagger = ((\hat{a})^{*})^{T}$
\begin{equation}
\hat{a}^\dagger \ket{n} = \sqrt{n+1} \ket{n}
\end{equation},
it is a canonical conjugate of the lowering operator $\hat{a}$, which together satisfy:
\begin{equation}
[\hat{a},\hat{a}^\dagger] = \hat{I}
\end{equation}
Similarly, the position and momentum operators, which return the mean position and momentum of bosons distributed across energy levels when observed, and which are a linear combination of the lowering and raising operators: 
\begin{align}
\hat{q} =& \frac{1}{\sqrt{2}} (\hat{a}+\hat{a}^\dagger)
\\\hat{p} =& -\frac{i}{\sqrt{2}} (\hat{a}-\hat{a}^\dagger)
\end{align},
are canonical conjugate operators:
\begin{align}
[\hat{q},\hat{p}] =& i\hat{I}
\end{align}
such that:
\begin{equation}
[\hat{q}^n,-i\hat{p}] = n\hat{q}^{n-1}
\end{equation}
From their definitions, it is seen that the operator $\hat{p}_i$ is Hermitian, $\hat{p}^\dagger = \hat{p}$, and so is the operator $\hat{q}$. 
\subsection{Bosonic Fock Space}
\subsubsection{Coherent State}

The representation of coherent states in the occupation number basis of the bosonic Fock space is one of the commonly encountered physical statevectors which comes arguably close to the desired statevector encoding powers of a value.
\begin{equation}
    \ket{\alpha} = e^{-\frac{1}{2}|\alpha|}e^{\alpha \hat{a}^\dagger-\alpha^*\hat{a}} \ket{0}= e^{-\frac{1}{2}|\alpha|^2} \Sigma_{n=0}^\infty \frac{\alpha^n}{\sqrt{n!}} \ket{n}
\end{equation}
for the case of $\alpha$ encoding a real value, $\alpha=\alpha^*$.
\subsubsection{Bosonic Operators}

The lowering used above in the definition of a displacement operator are defined as:
\begin{equation}
    \hat{a} = \begin{pmatrix}
        0 & 0 & 0 & 0 & \dots & 0 & \dots
        \\ \sqrt{1} & 0 & 0 & 0 & \dots & 0 & \dots
        \\ 0 & \sqrt{2} & 0 & 0 & \dots & 0 & \dots
        \\ 0 & 0 & \sqrt{3} & 0 & \dots & 0 & \dots
        \\ \vdots & \vdots & \vdots & \ddots & \ddots & \dots & \dots
        \\ 0 & 0 & 0 & \dots & \sqrt{n} & 0 & \dots
        \\ \vdots & \vdots & \vdots & \vdots & \vdots & \ddots & \ddots
    \end{pmatrix}
\end{equation}
as to act on the state $\ket{n}$ in the following manner:
\begin{equation}
\begin{aligned}
    \hat{a} \ket{n} &= \sqrt{n} \ket{n-1}
    \\ \hat{a}^\dagger \ket{n} &= \sqrt{n+1}\ket{n+1}
    \\ [\hat{a},\hat{a}^\dagger] &= \hat{I}
\end{aligned}
\end{equation}
Furthermore, the number operator is associated with the observable the number of particles in the Fock space:
\begin{equation}
    \begin{aligned}
            \hat{n}\ket{n} &= \hat{a}^\dagger\hat{a} \ket{n} = \hat{n}\ket{n}
    \end{aligned}
\end{equation}
One may also talk of the coherent state being an approximate eigenstate of the position and momentum operators:
\begin{equation}
\begin{aligned}
    \hat{q} &= i\partial_{\hat{p}} = \frac{1}{\sqrt{2}}(\hat{a}+\hat{a}^\dagger)
    \\ \hat{p} &= -i\partial_{\hat{q}} = \frac{i}{\sqrt{2}}(\hat{a}^\dagger-\hat{a})
    \\ [\hat{q},\hat{p}] &= i[\hat{a},\hat{a}^\dagger] = i\hat{I}
    \end{aligned}
\end{equation}
Clear from the above definition, the position and momentum operators are Hermitian, unlike the raising and lowering operators. For a system of multiple bosonic modes, we denote the operators acting on each by a subindex, e.g. $\hat{a}_k,\hat{q}_k,\dots$. 
\subsection{Mapping Differential Equation Driving-Functions to Bosonic Hamiltonians}

Considering a position eigenstate:
A consequence of the canonical commutation relation is that a position eigenstate could be defined in terms of a "translation" operator acting on a zero-position eigenstate:
\begin{align}
\ket{f} =& \ket{q = f}
\\\hat{q}\ket{f} =& f\ket{f}
\\\ket{f} =& e^{-i\hat{p}f}\ket{q = 0}
\end{align}
and for more than one variable dependent on time $t$ and position $\vec{x}$ as independent variables:
\begin{equation}
\ket{\vec{f}(\vec{x},t)} = e^{-i\vec{\hat{p}}\cdot\vec{f}(\vec{x},t)}\ket{\vec{q} = \vec{0}}
\label{eq:mapping}
\end{equation}
where the operator acting on $\ket{\vec{q} = \vec{0}}$ is the translation operator, which is a unitary operator:
\begin{equation}
(e^{-i\vec{\hat{p}}\cdot\vec{f}(\vec{x},t)})(e^{-i\vec{\hat{p}}\cdot\vec{f}(\vec{x},t)})^\dagger = (e^{-i\vec{\hat{p}}\cdot\vec{f}(\vec{x},t)})(e^{i\vec{\hat{p}}^\dagger\cdot\vec{f}(\vec{x},t)}) = \hat{I}
\end{equation}
since $\hat{p}$ is Hermitian. That is to say the position eigenstate could be normalized for each variable separately rather than the overall tensored state, independent of the value it encodes, given that the value itself is less than unity.

If $\vec{f}$ evovles according to the differential equation:
\begin{equation}
\partial_t\vec{f} = \Omega(\vec{f})
\end{equation},
then differentiating Eq.~(\ref{eq:mapping}) with respect to time gives:
\begin{equation}
     \partial_t\ket{\vec{f}(\vec{x},t)} = -i\vec{\hat{p}}\cdot \Omega(\vec{f}(\vec{x},t)) \ket{\vec{f}(\vec{x},t)}
\end{equation}
such that we can bring it down to the Schr\"odinger form by defining the Hamiltonian:
\begin{equation}
\label{orgexp}
    \hat{H} = \vec{\hat{p}}\cdot \Omega(\vec{f}(\vec{x},t)) = \vec{\hat{p}} \cdot \Omega(\vec{\hat{q}}).
\end{equation}
where the rightmost equality holds due to the choice of encoding the vector of variables of $\vec{f}$ into a position eigenstate, i.e. the classical variables are quantized to a bosonic position operator, Eq.~(\ref{eq:mapping}). Notice that the derived Hamiltonian is not Hermitian:
\begin{equation}
\hat{H} \neq \hat{H}^\dagger
\end{equation}
since the position and momentum operators acting on the system subspace of the system i.e. correpsonding to the same variable, do not commute
\begin{equation}
[\hat{q}_i,\hat{p}_j] = i\delta_{ij}\hat{I}
\end{equation}
\subsection{Choice of Quantized Operators}

In what follows, the choice of quantizing the classical discrete densities into position operators is not arbitrary. $\hat{q}$ and $\hat{p}$ are Hermitian operators, such that when we want to implement them we should expect better scaling in the number of terms of Pauli words (tensored Pauli operators) than the ladder operators. Moreover, when truncated, we can think of the lowering operator $\hat{a}$ relaying dependence upon unknown, truncated, terms, and $\hat{a}^\dagger$ taking providing an information loss mechanism, as such we may expect better error scaling for $f_i \rightarrow \hat{q} = \frac{1}{\sqrt{2}}(\hat{a}+\hat{a}^\dagger)$ than $f_i \rightarrow \hat{a}$. Apart from the above practical considerations, note that the generator of Hermite polynomials appears naturally in the Boltzmann (equilibrium) distribution.
\subsubsection{Boltzmann Distribution as a Generator of Rescaled Hermite Polynomials}
The equilibrium function used in lattice Boltzmann scheme is an approximation for the Boltzmann equilibrium distribution, obtained by truncating the power series of the exponential, as:
\begin{eqnarray}
    f^{eq} = \frac{\rho}{(2\pi R T)^{\frac{d}{2}}} e^{-\frac{(\Vec{c}-\Vec{u})^2}{2RT}} 
\end{eqnarray}
where $\rho$ is the fluid density, $T$ the thermodynamic temperature, and $R$ the gas constant. 

\begin{eqnarray}
\label{eq:fiequ}
    f_i^{eq} =  \frac{\rho}{(2\pi R T)^{\frac{d}{2}}} e^{-\frac{(\Vec{c}_i-\Vec{u})^2}{2RT}} =&  \frac{\rho}{(2\pi R T)^{\frac{d}{2}}} e^{-\frac{\Vec{c}_i^2}{2RT}} e^{\frac{(2\Vec{c}_i\cdot\Vec{u}-\Vec{u}^2)}{2RT}}
    \\ =& \frac{\rho}{(2\pi R T)^{\frac{d}{2}}} e^{-\frac{\Vec{c}_i^2}{2RT}} e^{\frac{\Sigma_{\mu=1}^d(2c_{i,\mu}u_\mu-{u}_\mu^2)}{2RT}}
    \\ =& \frac{\rho}{(2\pi R T)^{\frac{d}{2}}} e^{-\frac{\Vec{c}_i^2}{2RT}} \Pi_{\mu=1}^d \Sigma_{k=0}^\infty \frac{1}{k!} (\frac{c_{i,\mu}}{\sqrt{2RT}})^k H_k(\frac{u_\mu}{\sqrt{2RT}}),
\end{eqnarray}
where we have identified the generating function for the Hermite polynomials,
\begin{equation}
    e^{2xy-y^2} = \Sigma_{n=0}^\infty H_n(x) \frac{y^n}{n!},
\end{equation}
where $H_n(x)$ is the $n^{th}$ Hermite polynomial, with $x = \frac{u_\mu}{\sqrt{2RT}}$ and $y = \frac{c_{i,\mu}}{2RT}$.
\section{Unitary Evolution Substeps}
\subsection{Unitary Collision}
\subsubsection{Hermitization Approach}

\paragraph{Constant Divergence}
Following \cite{kowalskiNonlinearDynamicalSystems1997}, we first manipulate Eq.~(\ref{orgexp}) to arrive at 
\begin{equation}
\label{Eq::H_terms}
    \hat{H} = \frac{1}{2} \vec{\hat{p}} \cdot \Omega(\vec{\hat{q}}) +\frac{1}{2} \Omega(\vec{\hat{q}}) \cdot \vec{\hat{p}}+ (\frac{1}{2} \vec{\hat{p}} \cdot \Omega(\vec{\hat{q}})  - \frac{1}{2} \Omega(\vec{\hat{q}}) \cdot \vec{\hat{p}})
\end{equation}
We can write the Hamiltonian as:
    \begin{equation}
\label{Eq::Hreplaced}
\begin{aligned}
    \hat{H} =& \frac{1}{2} \vec{\hat{p}} \cdot \Omega(\vec{\hat{q}}) +\frac{1}{2} \Omega(\vec{\hat{q}}) \cdot \vec{\hat{p}}+ \frac{1}{2} [\hat{{p}}_i,\Omega_i(\vec{\hat{q}})]
    \end{aligned}
\end{equation}
where we can recognize a Hermitian part $\frac{1}{2} \vec{\hat{p}} \cdot \Omega(\hat{q}) + \frac{1}{2} \Omega(\hat{q}) \cdot \vec{\hat{p}} $ and anti-Hermitian term:
\begin{equation}
    \frac{1}{2}\sum_{i=1}^{Q} [\hat{{p}}_i,\Omega_i(\hat{q})] 
\end{equation}
Since $\Omega$ is assumed to be analytic, and $[\hat{q},\hat{p}]=i\hat{I}$, we have:
\begin{equation}
\label{Eq::p_q_commutators}
    [\hat{{p}}_i,\Omega_i(\hat{q})] = -i\frac{\partial \Omega_i(\vec{\hat{q}})}{\partial \hat{q}_i}
\end{equation}

We calculate the divergence with respect to position of the collision operator for the incompressible case.
\begin{equation}
\begin{aligned}
{\Sigma}_{i=1}^{Q} \frac{\partial}{\partial \hat{q}_i}\Omega_i(\hat{q}) = {\Sigma}_{i=1}^{Q} \frac{\partial}{\partial f_i}\Omega_i(\vec{f}) =& -\frac{1}{\tau}(Q-D)
\\=& \begin{cases}
    -\frac{2}{\tau}\hat{I} & D1Q3
    \\-\frac{7}{\tau} \hat{I} & D2Q9
    \\-\frac{24}{\tau}\hat{I} & D3Q27
\end{cases}
\end{aligned}
\end{equation}
which depend on the relaxation factor of the LB formulation, $\tau$, dictated by the viscosity of the fluid being simulated, is seen.

We define a normalized state:
\begin{equation}
    \begin{aligned}
        \ket{\vec{f}'(\Vec{x},t)} = e^{\frac{1}{2}\int_0^t \nabla \cdot \vec{\Omega}(\vec{f})dr } \ket{\vec{f}(\Vec{x},t)}
        \label{eq:unitarystate}
    \end{aligned}
\end{equation}

$\ket{\vec{f}'(\Vec{x},t)}$ evolves under the Hermitian Hamiltonian:
\begin{equation}
    \hat{H}'= \frac{1}{2} \Sigma_{i=1}^Q \hat{p}_i\Omega_i(\vec{\hat{q}})+\Omega_i(\vec{\hat{q}})\hat{p}_i = \hat{H}^{'\dagger}
\end{equation}

At $t = 0$, the two states overlap:
\begin{equation}
\ket{\vec{f}'(\Vec{x},t = 0)} = \ket{\vec{f}(\Vec{x},t = 0)}
\end{equation}
but the normalized state undergoes unitary collision with the operator $e^{-i\Delta t \hat{H}'}$, and the results consistent $\ket{\vec{f}(\vec{x},t)}$ can be recovered by multiplying by $e^{\frac{T\Delta t}{2\tau}(Q-D)}$ with $T$ total number of timesteps taken. 
\subsubsection{Hydrodynamic Variables Approach}

\label{sec:hydrovar}
\paragraph{Partial Linearity}
We see that collision term is linear in a rescaling factor. This is readily seen from:

\begin{equation}
    \vec{f}'(\Vec{x},t) = \alpha \Vec{f}(\Vec{x},t)
\end{equation}
which gives
\begin{equation}
    \rho'(\vec{x},t) = \alpha \rho(\vec{x},t)
\end{equation}

\begin{equation}
    \vec{u}'(\vec{x},t) = \vec{u}(\vec{x},t)
\end{equation}
such that $f_i^{'eq}(\vec{x},t) = f_i^{eq}(\vec{u}'(\vec{x},t)) = f_i^{eq}(\vec{u}(\vec{x},t)) = f_i^{eq}(\vec{x},t)$, giving:
\begin{equation}
    \Omega_i(\vec{f}) = \Omega_i(\frac{1}{\alpha}\vec{f}') = -\frac{1}{\tau} \frac{1}{\alpha} (f'_i(\vec{x},t)-f_i^{'eq}(\vec{x},t)) = \frac{1}{\alpha} \Omega_i(\vec{f}')
\end{equation}
Therefore:
\begin{equation}
\label{eq:normalizedomega}
    \Omega_i(\alpha \vec{f}) = \alpha \Omega_i(\vec{f})
\end{equation}
This result is expected as the collision function is nonlinear only in $\vec{u}$ in which the discrete densities appear as a linear combination normalized by the density:
\begin{equation}
\vec{u}(\vec{x},t) = \frac{1}{\rho} \Sigma_{i=1}^Q f_i(\vec{x},t)\vec{e}_i
\end{equation}

\paragraph{Time-Depedent Rescaling}
We revisit the first-order in time expression for the collision:
\begin{align}
f_i(\vec{x},t+1) =& f_i(\vec{x},t)-\frac{\Delta t}{\tau}(f_i(\vec{x},t)-f_i^{eq}[\rho,\vec{u}(\vec{x},t)])
\\f_i(\vec{x},t+1) =& (1-\frac{\Delta t }{\tau})f_i(\vec{x},t)+\frac{\Delta t}{\tau}(f_i^{eq}[\rho,\vec{u}(\vec{x},t)])
\\\frac{1}{1-\frac{\Delta t }{\tau}}f_i(\vec{x},t+1) =& f_i(\vec{x},t)+\frac{\frac{\Delta t}{\tau}}{1-\frac{\Delta t }{\tau}}(f_i^{eq}[\rho,\vec{u}(\vec{x},t)])
\\\frac{1}{1-\frac{\Delta t }{\tau}}f_i(\vec{x},t+1) =& f_i(\vec{x},t)+\frac{1}{\frac{\tau}{\Delta t}-1}(f_i^{eq}[\rho,\vec{u}(\vec{x},t)])
\end{align}
We see that the discrete densities,$f_i \; \forall \; i \in [1,Q]$, and therefore the density $\rho$, carry a normalization factor $(\frac{1}{1-\frac{\Delta t}{\tau}})^t$ where $t$ is the number of timesteps taken. For clarity, let us then define:
\begin{equation}
\Tilde{f}_i(\vec{x},t) = (\frac{1}{1-\frac{\Delta t}{\tau}})^t f_i(\vec{x},t)
\end{equation}
and:
\begin{equation}
\Tilde{\rho}(t) = (\frac{1}{1-\frac{\Delta t}{\tau}})^t
\end{equation}
which does not depend on the location $\vec{x}$ since we have the flow to be incompressible.
We do not define $\vec{\Tilde{u}}$ since:
\begin{equation}
\vec{u} = \frac{1}{\rho} \Sigma_{i=1}^Q f_i \vec{e}_i = \frac{1}{\Tilde{\rho}} \Sigma_{i=1}^Q \Tilde{f}_i\vec{e}_i
\end{equation}
The evolution of the rescaled discrete density is then written as:
\begin{equation}
\Tilde{f}_i(\vec{x},t+1) = \Tilde{f}_i(\vec{x},t)+w_i\frac{\Delta t}{\tau}(\frac{1}{1-\frac{\Delta t}{\tau}})^{t+1}(1+ 3\frac{\Vec{e}_i\cdot\Vec{u}}{c^2}+9\frac{(\Vec{e}_i\cdot\Vec{u})^2}{2c^4}-3\frac{\Vec{u}^2}{2c^2})
\end{equation}
\paragraph{Implementation with Rescaling}
To implement a unitary collisoin term through this approach, instead of the position eigenstates encoding the discrete densities introduced above $\ket{\vec{f}(\vec{x},t)}$ we need to encode the rescaled density $\ket{\vec{\Tilde{f}}(\vec{x},t)}$, as well as the velocity $\ket{\vec{u}(\vec{x},t)}$, such that the variables at each lattice cite become endcoded in:
\begin{equation}
\ket{\vec{\Tilde{f}}(\vec{x},t)}\ket{\vec{u}(\vec{x},t)}
\end{equation}
The position and momentum operators corresponding to the $i^{th}$ encoded discrete density are denoted $\hat{q}_i$ and $\hat{p}_i$ respectively, for $i \in [1,Q]$. For the velocity components, they are $\hat{u}_d$ and $\hat{\mu}_d$ respectively for $d\in[1,D]$.
The collision step would need to be represented by the actoin of three consecutive unitaries achieving:
\begin{itemize}
\item Computing $\vec{u}(\vec{x},t)$: $\ket{\vec{0}}\rightarrow\ket{\vec{u}(\vec{x},t)}$
\item Relaxing the discrete densities: $\ket{\vec{\Tilde{f}}(\vec{x},t)}\rightarrow\ket{\vec{\Tilde{f}}(\vec{x},t+1)}$
\item Uncomputing $\vec{u}(\vec{x},t)$: $\ket{\vec{u}(\vec{x},t)}\rightarrow\ket{\vec{0}}$
\end{itemize}
where for simplicity we have assumed that the value at $t+1$ is only the value after the discrete densities have undergone collision, discussion in the context of streaming is left for later. Most importantly, in each step the variables being updated are updated based on the value of variables already updated. The unitaries corresponding to these three steps are detailed below:
\begin{itemize}
\item Computing $\vec{u}(\vec{x},t)$: $e^{-i\frac{1}{\Tilde{\rho}(t)}\Sigma_{i=1}^Q\Sigma_{d=1}^D\hat{\mu}_d \hat{q}_i e_{i,d}}$
\item Relaxing the discrete densities: $e^{-i\Tilde{\rho}(t+1)\frac{\Delta t}{\tau}\Sigma_{i=1}^Q\hat{p}_iw_i(1+ 3\frac{\Vec{e}_i\cdot\Vec{\hat{u}}}{c^2}+9\frac{(\Vec{e}_i\cdot\Vec{\hat{u}})^2}{2c^4}-3\frac{\Vec{\hat{u}}^2}{2c^2})}$
\item Uncomputing $\vec{u}(\vec{x},t)$: $e^{+i\frac{1}{\Tilde{\rho}(t+1)}\Sigma_{i=1}^Q\Sigma_{d=1}^D\hat{\mu}_d \hat{q}_i e_{i,d}}$
\end{itemize}
where $e_{i,d}$ is the $d^{th}$ component of the D-dimensional lattice vector along the $i^{th}$ direction. We note that rescaling the variables is likely to run into issues with numerical error, and normalization of the statevector not considered here.
\paragraph{Implementation without Rescaling}
An alternative to rescaling the econded discrete densities with a time-dependent factor would be to utilize the condition of incompressibility $\rho = 1$. We revisit the collision step:
\begin{equation}
f_i(\vec{x},t+1) -f_i(\vec{x},t) = -\frac{\Delta t }{\tau}f_i(\vec{x},t)+\frac{\Delta t}{\tau}f_i^{eq}[\vec{u}(\vec{x},t)]
\end{equation}
and replace $f_i(\vec{x},t)$ by $1-\Sigma_{j=1,\neq i}^Qf_j$ using $\rho = \Sigma_{i=1}^Q f_i = 1$ (Remember, in reality $\rho = 1 + O(Ma^2)$). Implementing the collision step then becomes described by the following unitaries:
\begin{itemize}
\item Computing $\vec{u}(\vec{x},t)$: $e^{-i\Sigma_{i=1}^Q\Sigma_{d=1}^D\hat{\mu}_d \hat{q}_i e_{i,d}}$
\item Relaxing the discrete densities: $e^{-i\Sigma_{i=1}^Q\hat{p}_i  \frac{\Delta t}{\tau} (\Sigma_{j =1, \neq i}^Q\hat{q}_j+f_i^{eq}[\vec{u}(\vec{x},t)]-1)}$
\item Uncomputing $\vec{u}(\vec{x},t)$: $e^{+i\Sigma_{i=1}^Q\Sigma_{d=1}^D\hat{\mu}_d \hat{q}_i e_{i,d}}$
\end{itemize}
Particular attention is to be paid to $j\neq i$ since this is what ensures that the operators commute, therefore their exponentiation being unitary. The price paid by this implementation is not solely a larger number of terms making up the Hamiltonian (therefore increasing the gate complexity as discussed later), but the additional $O(Q^2)$ operators acting on the statevectors encoding the discrete densities, introduced with the use of the incompressbility condition, each introduces an error upon the statevector as it is not an exact eigenstate, as discussed in the appendix. It is likely disadvantageous to repeatedly \textit{access} the set of variables $u_d \; \forall \;l d \in [1,D]$ given that each query introudces an error into the variable, whereas writing the equilibrium function in terms of the discrete densities would distribute such errors over a larger number of variables $Q = 3^D$ allows for the opportunity that they may cancel out \cite{itaniAnalysisCarlemanLinearization2022}.
\subsubsection{Choice of Unitary}

A brief survey of the methods described above for achieving a unitary evolution, the Hermitization approach appears to be best. First, it does not require introducing additional $D$ variables. Second, the anti-Hermitian part separated out could be performed exactly, which we expect to reduce the error, as opposed to the hydrodyanmic variables approach which introduces additional error-prone operations. For example, resetting the velocity to zero cannot be done exactly unitarily, mid-circuit measurement would be required. Third, the renormalization required in post-processing bodes better with the quantum algorithm as the values are \textit{deflated} in the quantum algorithm as opposed to the rescaling introduced as part of the hydrodynamic variables approach.

An additional advantage of the Hermitized Hamiltonian is the separability of the nonlinear and dissipative dynamics. This is particularly important when considering implementation on noisy intermediate scale quantum (NISQ) computers since the noise itself could be leveraged in achieving the dissipative behaviour of the system \cite{guimaraesNoiseassistedDigitalQuantum2023}. The complexity scaling developed in Sec.~\ref{sec:compan}, however, does not account for the limitations of NISQ devices.

\subsection{Unitary Streaming}

If we are to represent $Q$ variables over a total $G = \Pi_{d=1}^D N_d$  positions in space, where $N_d$ is the number of positions sampled across the $d^{th}$ dimension, then we are required to store a total of $QG$ values. Classically, this would require $QGb$ where $b$ is a number of classical bits required for storing a single value to a desired precision. On a quantum computer, ideally this is achieved by putting a quantum register into a superposition of $QG$ states, thus, requiring $\log_2{(QG)} = \log_2{(Q)}+\log_2{(G)}$ qubits, up to a normalization factor:
\begin{equation}
\Sigma_{i=1}^Q\Sigma_{\vec{x}\in G} \ket{f_i(\vec{x},t)} = \Sigma_{i=1}^Q\Sigma_{x_1,\dots,x_d,\dots,x_D}^{N_d} f_i(\vec{x},t)\ket{i}\bigotimes_{d=1}^D\ket{x_d}
\end{equation}
It is straightforward to see that the values stored could be streamed along the positive (negative) direction of a given $d$ dimension by incrementing (decrementing) the respective register $\ket{x_d}$. We note that $\ket{x_d} = \bigotimes_{b=0}^{\log_2{N_d}-1}\ket{b}$ where $b \in \{0,1\}$, and this incrementing (decrementing) operation could be achieved unitarily in parallel for all values \cite{todorovaQuantumAlgorithmCollisionless2020a,schalkersEfficientFailsafeCollisionless2022}. Thus, the lattice discretization happens in the quantum mechanical framework, as opposed to discretizing the classical problem as \cite{liuEfficientQuantumAlgorithm2020,akhalwayaEfficientQuantumComputation2022} have done.

\paragraph{Additional Register Encoding Lattice Direction}

In the case where $\ket{f_i(\vec{x},t)}$ uses binary representation of $f_i(\vec{x},t)$, qubits in the zero state have a zero contribution, making it possible to control the streaming operator on the qubits of a particular variable being in the $\ket{1}$ state to stream the variable. This is not the case for the position encoding we have chosen. As such an additional register $\ket{\mathbf{e}}$ must be introduced where to encode lattice vectors $\vec{e}_i$. Unitary evolution is then achieved by (de)incrementing the lattice position register in binary representation controlled on the correpsonding state of $\ket{\vec{e}_i}$.
\begin{equation}
\Sigma_{\vec{x}\in G}\Sigma_{i=1}^Q\ket{f_i(\vec{x},t)}\ket{\vec{e}_i}\ket{\vec{x}}
\end{equation}

\subsubsection{Streaming Step}

With the binary representation of the lattice position, the streaming step looks like:
\begin{equation}
    \bigotimes_{d}^D \ket{\vec{x}_d}
    \rightarrow \bigotimes_{d}^D \ket{\vec{x}_d+\Delta\vec{x}_d}
\end{equation}
Note that $\bigotimes_{d=1}^D {\Delta \vec{x}_d}$ has to be controlled by the discrete density register in binary representation, or the lattice direction register in the positoin embedding. This can be done independently of the lattice site. For example one can choose to implement the streaming step in real space as done in~\cite{todorovaQuantumAlgorithmCollisionless2020} or in Fourier space as \cite{schalkersEfficientFailsafeCollisionless2022}. We choose the latter as it is more illustrative.

\begin{figure}[htp]
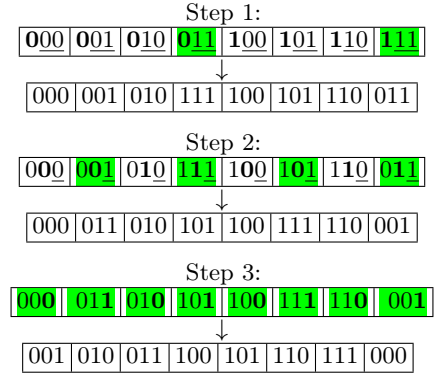

    \centering
    {\small
    \begin{tabular}{|c|c|c|c|c|c|c|c|}
  \hline
  0 & 1 & 2 & 3 & 4 & 5 & 6 & 7
  \\\hline
  \end{tabular} :
     \begin{tabular}{|c|c|c|c|c|c|c|c|}
     \hline
  000 & 001 & 010 & 011 & 100 & 101 & 110 & 111\\
  \hline
\end{tabular}
\\
\vspace{1.25cm}
Step 1:
\\
\begin{tabular}{|c|c|c|c|c|c|c|c|}
\hline
  \textbf{0}\underline{00} & \textbf{0}\underline{01} & \textbf{0}\underline{10} & \cellcolor{green}\textbf{0}\underline{11} & \textbf{1}\underline{00} & \textbf{1}\underline{01} & \textbf{1}\underline{10} & \cellcolor{green}\textbf{1}\underline{11}\\
\hline
\end{tabular}
\\
$\downarrow$
\\
\begin{tabular}{|c|c|c|c|c|c|c|c|}
\hline
  000 & 001 & 010 & 111 & 100 & 101 & 110 & 011\\
  \hline
\end{tabular}
\\
\vspace{0.25cm}
Step 2:
\\
\begin{tabular}{|c|c|c|c|c|c|c|c|}
\hline
  0\textbf{0}\underline{0} & \cellcolor{green}0\textbf{0}\underline{1} & 0\textbf{1}\underline{0} & \cellcolor{green}1\textbf{1}\underline{1} & 1\textbf{0}\underline{0} & \cellcolor{green}1\textbf{0}\underline{1} & 1\textbf{1}\underline{0} & \cellcolor{green}0\textbf{1}\underline{1}\\
\hline
\end{tabular}
\\
$\downarrow$
\\
\begin{tabular}{|c|c|c|c|c|c|c|c|}
\hline
  000 & 011 & 010 & 101 & 100 & 111 & 110 & 001\\
  \hline
\end{tabular}
\\ \vspace{0.25cm} Step 3: \\
\begin{tabular}{|c|c|c|c|c|c|c|c|}
\hline
  \cellcolor{green}00\textbf{0} &\cellcolor{green} 01\textbf{1} & \cellcolor{green}01\textbf{0} & \cellcolor{green}10\textbf{1}& \cellcolor{green}10\textbf{0} & \cellcolor{green}11\textbf{1} & \cellcolor{green}11\textbf{0} &\cellcolor{green} 00\textbf{1}\\
\hline
\end{tabular}
\\
$\downarrow$
\\
\begin{tabular}{|c|c|c|c|c|c|c|c|}
\hline
  001 & 010 & 011 & 100 & 101 & 110 & 111 & 000\\
  \hline
\end{tabular}
\vspace{1.25cm}

\begin{tabular}{|c|c|c|c|c|c|c|c|}
     \hline
  001 & 010 & 011 & 100 & 101 & 110 & 111 & 000\\
  \hline
\end{tabular}:
\begin{tabular}{|c|c|c|c|c|c|c|c|}
  \hline
  1 & 2 & 3 & 4 & 5 & 6 & 7 & 0
  \\\hline
  \end{tabular}
     }
    
\caption{A visualization of the index-shuffling streaming procedure with periodic boundaries introduced by \cite{todorovaQuantumAlgorithmCollisionless2020} for $2^3=8$ sites with a single variable. Contents of the cell refer to the corresponding index position. Green cells refer to the case where the condition of the control has been met and the digit being acted upon would change. Digits in binary representation being acted on by a controlled NOT gate (targets) are written in boldface whereas those controlling (controls) are underlined. The number of steps (3) is logarithmic in the number of sites (8).}
    \label{fig:streamingviz}
\end{figure}
Fig.~\ref{fig:streamingviz} visualizes the streaming procedure achieved by the binary representation of the lattice position for the case of a single variable with $8$  lattice sites. Reversing the direction of streaming could be achieved by reversing the control, as Fig.~\ref{fig:streamingCircuit} shows where streaming along the negative x-axis uses closed circle controls, conditioning on the control qubit being in the $\ket{1}$ state. Streaming along the positive direction, shown for the y-axis, utilizes an open-circle control, conditioning on the control qubit being in the $\ket{0}$ state. Streaming more than one variable in different directions would require controlling on the corresponding register of the variable in binary representation, or on the lattice direction register as Fig.~\ref{fig:streamingCircuit} shows.

\subsection*{Streaming Circuit in D2Q9}

\begin{figure}[!htbp]
\centering
\input{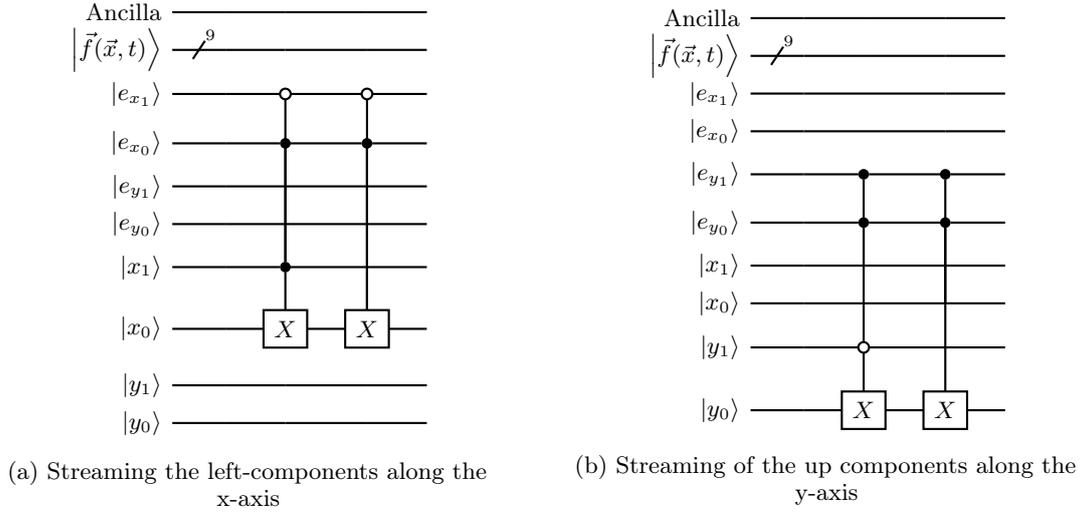}
\caption{A detailed representation of streaming discrete densities with a \ref{xStreaming} left component in their respective directions in D2Q9, where $\ket{e_{x_1}e_{x_0}}=\ket{01}$ corresponds to all directions which projection along the x-axis points in the negative direction, and $\ket{e_{y_1}e_{y_0}}=\ket{11}$ in \ref{yStreaming} respectively along the positive y-axis for streaming discrete densities with an upward component.}
    \label{fig:streamingCircuit}
\end{figure}

\FloatBarrier

We account for the need to control on the $2D$ lattice direction registers we introduced to stream variables in different directions, which contributes a an additional $2D$ factor to the gate complexity, making it:
\begin{equation}
O(TD\log_2^2{(G)})
\end{equation}

The lattice directions are an overcomplete (non-orthogonal) set. We need $\ceil{\log_2{(Q)}} = \ceil{\log_2{(3^D)}} = D\ceil{\log_2{(3)}} = 2D$ qubits to encode the lattice directions. For each of the $D$ dimensions the two qubits encode its positive, negative and null directions (a null direction corresponding to the lattice direction being perpendicular to the dimension under consideration):

\begin{table}[!htbp]
    \centering
        \caption{States encoding the different directions for a given dimension}
\begin{tabular}{|c|c|}
  \hline
  Direction & $\ket{\vec{e}_i} = \ket{e_{i,0} e_{i,1}}$
  \\\hline
Stationary & $\ket{10}$
\\ Positive Axis & $\ket{11}$
\\Negative Axis & $\ket{01}$
\\Constant Variable $(1)$ & $\ket{00}$
\\\hline
  \end{tabular}
\end{table}
\begin{table}[!htbp]
    \centering
        \caption{An example of lattice direction encoding in 2D qubits for D2Q9 lattice}
\begin{tabular}{|c|c|c|
}
  \hline
  Direction & $\ket{x_0x_1}$ & $\ket{y_0y_1}$
  \\\hline
Center & $\ket{10}$ & $\ket{10}$
\\ East & $\ket{11}$ & $\ket{10}$
\\Northeast & $\ket{11}$ & $\ket{11}$
\\North & $\ket{10}$ & $\ket{11}$
\\Northwest & $\ket{01}$ & $\ket{11}$
\\West & $\ket{01}$ & $\ket{10}$
\\Southwest & $\ket{01}$ & $\ket{01}$
\\South & $\ket{10}$ & $\ket{01}$
\\Southeast & $\ket{11}$ & $\ket{01}$
\\ \hline
  \end{tabular}

\end{table}

\begin{figure}[!htb]
\centering
\tdplotsetmaincoords{90}{90}
{\resizebox{5cm}{!}{\begin{tikzpicture}[tdplot_main_coords,
axis/.style={very thin, ->, >=stealth'}]

\coordinate (O) at (0.5,0.5,0.5);

\def \a {1}       
\def \b {1}       
\def \c {1}       
 \foreach \v in {0,1,...,\b}
      \foreach \w in {0,1,...,\c}
        \draw[very thin,gray] (0.5,\v,0) -- (0.5,\v,\w); 
 \foreach \v in {0,1,...,\b}
      \foreach \w in {0,1,...,\c}
        \draw[very thin,gray] (0.5,0,\w) -- (0.5,\v,\w); 

\draw plot [mark=*, mark size=0.1] coordinates{(O)};

    \foreach \v in {0,1,...,\b}
      \foreach \w in {0,1,...,\c}
        \draw[very thin,-latex,black](O) -- (0.5,\v,\w);
        
    \foreach \v in {0,1,...,\b}
        \draw[very thin,-latex,black](O) -- (0.5,\v,0.5); 
      \foreach \w in {0,1,...,\c}
        \draw[very thin,-latex,black](O) -- (0.5,0.5,\w); 
\node[scale = 0.25] at (0.5,0.725,0.55) {$\ket{10}\ket{10}$};
\node[scale = 0.25] at (0.5,1.25,0.5) {$\ket{11}\ket{10}$};
\node[scale = 0.25] at (0.5,1.25,1.15) {$\ket{11}\ket{11}$};
\node[scale = 0.25] at (0.5,-0.25,1.15) {$\ket{01}\ket{11}$};
\node[scale = 0.25] at (0.5,-0.25,0.5) {$\ket{01}\ket{10}$};
\node[scale = 0.25] at (0.5,-0.25,-0.15) {$\ket{01}\ket{01}$};
\node[scale = 0.25] at (0.5,0.5,-0.15) {$\ket{10}\ket{01}$};
\node[scale = 0.25] at (0.5,1.25,-0.15) {$\ket{11}\ket{01}$};
\node[scale = 0.25] at (0.5,0.5,1.15) {$\ket{10}\ket{11}$};
\end{tikzpicture}}}
\hspace{0.1\textwidth}
\caption{Lattice vector register states corresponding to the different directions for a D2Q9 lattice\label{fig:latticedirenc}}
\end{figure}
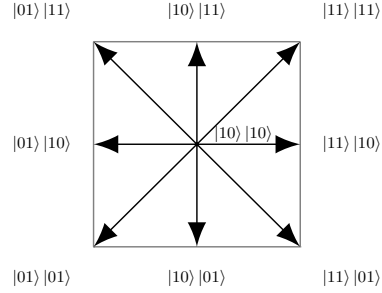

Another advantage of the procedure is that it enforces periodic boundary conditions in all dimensions at no extra computational cost. Specular, Neumann or other conditions, with the first being discussed in \cite{todorovaQuantumAlgorithmCollisionless2020}, would nevertheless bring up an additional coefficient of $log_2(N)$ into the complexity.

Having demonstrated how the streaming operator works by shifting indices in Fig.~\ref{fig:streamingviz}, we demonstrate how it works as a controlled operator with an example of D2Q9 lattice.

Starting from the state:
    \begin{align}
    \ket{\psi(t)} = &\frac{1}{\sqrt{G}}\sum_{\vec{x}\in G}\sum_{i=1}^Q \ket{f_i(\vec{x},t)}\ket{\vec{e}_i}\ket{\vec{x}}
\end{align}
where we assume that a $N=1$ such that a single qubit constitutes each discrete density register:
\begin{equation}
    \ket{f_i(\vec{x},t)} = C_{0,i}(\vec{x},t)\ket{0}+C_{1,i}(\vec{x},t)\ket{1}
\end{equation}
2 lattice directions registers, corresponding to each dimension, each consisting of 2 qubits:
\begin{equation}
    \ket{\vec{e}_i} = \frac{1}{\sqrt{9}}\ket{x_{1,i}x_{0,i}}\ket{y_{1,i}y_{0,i}}
\end{equation}
and $4$ lattice sites in each dimension indexed by the position register starting from index $0$, making $G=16$, such that the state corresponding to the site at $(1,3)$
\begin{equation}
    \ket{\vec{x}} = \ket{x}\ket{y} = \ket{01}\ket{11}
\end{equation}
As illustrated in Fig.~\ref{fig:streamingviz}, the action of the left streaming operator (negative x), followed by the up streaming operator (positive y) on the state encoding $(1,3)$ gives $(0,3)$ then $(0,0)$:
\begin{equation}
    \hat{S}_{\uparrow}\hat{S}_\leftarrow \ket{01}\ket{11} = \hat{S}_{\uparrow} \ket{00}\ket{11} = \ket{00}\ket{00}
\end{equation}
We now apply the controlled version on the full state of the system:
\begin{align}
    \ket{\psi(t)} = \frac{1}{\sqrt{9G}}\sum_{\vec{x}\in G} \ket{f_\cdot(\vec{x},t)}\ket{10}\ket{10}\ket{x,y}
    \nonumber\\+\ket{f_\rightarrow(\vec{x},t)}\ket{11}\ket{10}\ket{x,y}
    +\ket{f_\nearrow(\vec{x},t)}\ket{11}\ket{11}\ket{x,y}
    \nonumber\\+\ket{f_\uparrow(\vec{x},t)}\ket{10}\ket{11}\ket{x,y}
    +\ket{f_\nwarrow(\vec{x},t)}\ket{01}\ket{11}\ket{x,y}
    \nonumber\\+\ket{f_\leftarrow(\vec{x},t)}\ket{01}\ket{10}\ket{x,y}
    +\ket{f_\swarrow(\vec{x},t)}\ket{01}\ket{01}\ket{x,y}
    \nonumber\\+\ket{f_\downarrow(\vec{x},t)}\ket{10}\ket{01}\ket{x,y}
    +\ket{f_\searrow(\vec{x},t)}\ket{11}\ket{01}\ket{x,y}
\end{align}
where $\vec{x}:(x,y)$.
Then, the controlled application of the left-streaming operator acts as:
\begin{equation}
    \begin{aligned}
    C^2_{\bar{10}\;11,14\;15}\hat{S}_\leftarrow \ket{\psi(t)}= 
    \\\frac{1}{\sqrt{9G}}\sum_{\vec{x}\in G} (\ket{f_\cdot(\vec{x},t)}\ket{10}\ket{10}\ket{x,y}
    \\+\ket{f_\rightarrow(\vec{x},t)}\ket{11}\ket{10}\ket{x,y}
    \\+\ket{f_\nearrow(\vec{x},t)}\ket{11}\ket{11}\ket{x,y}
    \\+\ket{f_\uparrow(\vec{x},t)}\ket{10}\ket{11}\ket{x,y}
    \\+\ket{f_\nwarrow(\vec{x},t)}\ket{01}\ket{11}\ket{x-1[4],y}
    \\+\ket{f_\leftarrow(\vec{x},t)}\ket{01}\ket{10}\ket{x-1[4],y}
    \\+\ket{f_\swarrow(\vec{x},t)}\ket{01}\ket{01}\ket{x-1[4],y}
    \\+\ket{f_\downarrow(\vec{x},t)}\ket{10}\ket{01}\ket{x,y}
    \\+\ket{f_\searrow(\vec{x},t)}\ket{11}\ket{01}\ket{x,y})
    \label{eq:2DstreamingExample}
\end{aligned}
\end{equation}
where the operator is controlled on the tenth and eleventh qubits (the first 9 qubits represent the discrete density registers) representing the lattice directions along the x-axis, and acts on the fourteenth and fifteenth qubits encoding the lattice site index along the x-axis. The bar above the qubit number in the control means the action of the operator is conditioned on the zero control-state. We note that the coordinates of position contain a $\mod(4)$ corresponding to the periodic boundaries enforced by the streaming operator.

We note that the streaming operation for any dimension $d$ is written as an operator acting on the $\ket{x_d}$ register:
\begin{equation}
    \hat{S}_{d+} = \hat{X}_0\cdot {[CX]}^1_{\bar{1},0} \cdots {[CX]}^{\log_2{(N_{x_d})}-1}_{\bar{1}\dots\bar{\log_2{(N_{x_d})}-1},0}
    \end{equation}
where streaming in the opposite direction could be achieved by complementing the control state:
\begin{equation}
    \hat{S}_{d-} = \hat{X}_0\cdot {[CX]}^1_{1,0} \cdots {[CX]}^{\log_2{(N_{x_d})}-1}_{1\dots\bar{\log_2{(N_{x_d})}-1},0}
\end{equation}
The notation of the controlled gate used specifies the control qubits first, then the target qubit, with the superscript indicating the number of control qubits. A bar over the number indicating a control qubit indicates a control state of $0$, whereas its absence refers to the default control state of $1$. For example, ${[CX]}^1_{1,0}$ applies a NOT-gate on the $0^{th}$ qubit assuming the $1^{st}$ qubit is in the $1$ state. To stream only certain discrete densities in the chosen direction, one needs to control on the register encoding the lattice direction. We demonstrate this by considering a detailed circuit 
of the 2D example considered in Eq.~(\ref{eq:2DstreamingExample}), in Fig.~\ref{fig:streamingCircuit}.

\section{Unitary Evolution}
\subsection{Limitations of the Unitary Collision \& Streaming}

We compare the encoding in the ideal scaling discussed above, to the one considered above in linear embedding discussed above. In the ideal scaling, the statevectors for different positions, and different variables, are all in superposition:
\begin{equation}
\Sigma_{\vec{x}\in G} \ket{\vec{x}}\Sigma_{i=1}^Q \ket{f_i(\vec{x},t)}
\end{equation}
, whereas the linear embedding requires the $Q$ values to be tensored:
\begin{equation}
\bigotimes_{i=1}^Q \ket{f_i(\vec{x},t)}
\label{eq:tensored}
\end{equation}
requiring a number of qubits that scales as $Q$ instead of $\log_2{(Q)}$. This is due to the evolution operator $\Omega(\vec{f})$ possibly containing nonlinear and coupled terms $f_1^2,f_1f_2,\dots$, such that the evolution operator needs to access all terms to update each one of them. The values in superposition could be streamed separately. Considering only two variables at $\vec{x}=(x_1,x_2,x_3)$ for example:
\begin{equation}
\ket{f_1(\vec{x},t)}\ket{\vec{x}} + \ket{f_2(\vec{x},t)}\ket{\vec{x}} \rightarrow \ket{f_1(\vec{x},t)}\ket{\vec{x}+\Delta \vec{x}_1} + \ket{f_2(\vec{x},t)}\ket{\vec{x}+\Delta \vec{x}_2}
\end{equation}
it is possible to stream them separately in different directions, in a unitary fashion. 

It is not possible to stream the variables in different directions unitarily when it is tensored by acting on the register encoding the lattice position. During the preparation of this manuscript, a preprint \cite{schalkersImportanceDataEncoding2023} has surfaced discussing this issue specifically, but claims that it could only be resolved by linear scaling in time of the qubits. We improve their linear scaling, and show that it is a tradeoff between linear scaling of the qubits and the complexity, before showing how logarithmic scaling could be maintained. 

It is not immediately clear how tensored variables could be streamed in different directions otherwise, until at least two sites are considered:
\begin{equation}
\ket{\vec{x}_1} \ket{f_1(\vec{x}_1,t)}\ket{f_2(\vec{x}_1,t)}+\ket{\vec{x}_2} \ket{f_1(\vec{x}_2,t)}\ket{f_2(\vec{x}_2,t)}
\end{equation}
whereas the two variables are tensored at each site, they are in superposition with the values at the other site. Acting upon the register encoding the position forces us to stream all variables of the site together, at once, in the same direction.
\begin{equation}
\begin{aligned}
&\ket{\vec{x}_1} \ket{f_1(\vec{x}_1,t)}\ket{f_2(\vec{x}_1,t)}+\ket{\vec{x}_2} \ket{f_1(\vec{x}_2,t)}\ket{f_2(\vec{x}_2,t)}+\ket{\vec{x}_3} \ket{f_1(\vec{x}_3,t)}\ket{f_2(\vec{x}_3,t)}+\ket{\vec{x}_4} \ket{f_1(\vec{x}_4,t)}\ket{f_2(\vec{x}_4,t)}
\\&\rightarrow \ket{\vec{x}_1} \ket{f_1(\vec{x}_2,t)}\ket{f_2(\vec{x}_4,t)}+\ket{\vec{x}_2} \ket{f_1(\vec{x}_3,t)}\ket{f_2(\vec{x}_1,t)}+\ket{\vec{x}_3} \ket{f_1(\vec{x}_4,t)}\ket{f_2(\vec{x}_2,t)}+\ket{\vec{x}_4} \ket{f_1(\vec{x}_1,t)}\ket{f_2(\vec{x}_3,t)}
\end{aligned}
\label{eq:streaming}
\end{equation}
where it is assumed that sites position vectors denoted with consecutive numbers are neighbors, and $f_1$ and $f_2$ are variables associated with opposite lattice directions (e.g. left and right respectively) along a single dimension dictated. Moreover, the boundaries are assumed to be periodic ($\vec{x}_1$ denotes a neighbor of the cell at $\vec{x}_4$).

Three remarks are due. First, the incompressible assumption $\rho(\vec{x},t)  = \Sigma_{i=1}^Q f_i(\vec{x},t) = 1$ reduces the number of independent variables from $Q$ to $Q-1$ at each site. Second, the variable at the center of lattice site corresponds to zero lattice direction vector (rest particles) and, thus, is not streamed. Third, Upon further inspection of the streaming procedure in lattice Boltzmann, we see that each pair of neighboring lattice sites exchange a pair of discrete densities along opposite lattice directions, such that, with $Q$ variables, we can divide the streaming step into $Q-2$ non-unitary, multiplexed entangelment forging \cite{eddinsDoublingSizeQuantum2022}, step. We speak of entanglement forging in the sense that classical resources are used to capture the pair of variables being streamed across neighboring cells and entangled with the rest of the variables at each respective site. In simpler terms, we may simply measure each respective variable of the pair, calculate the difference, and "translate" each variable, e.g. $f_1$ and $f_2$, at neighboring cites, by the addition/subtraction of the difference to emulate the streaming process. A final unitary step where only values of other variables are used to adjust the value of the variable yet to be streamed. For example, if at a given site $f_1$ is determined, knowledge thereof is enough to perform the final unitary step calculate $f_2 = 1 - f_1$ without "streaming" the value from the neighboring cell, as $\hat{p}_2$ commutes with $\hat{q}_1$. Recall, however, that in standard form, the incompressible assumption holds up to an error, scaling with the square of the Mach number, during evolution $\rho = 1 + O(Ma^2)$. It is optional to perform a $Q^{th}$ unitary step involving $\frac{Q-1}{2}$ swaps of variables to avoid the bookkeeping resulting from change the order the variables are tensored in. 

In short, unitary streaming by acting on the lattice position register as described in \cite{todorovaQuantumAlgorithmCollisionless2020a,schalkersEfficientFailsafeCollisionless2022} is incompatible with the linear embedding without the latter mixing states in time. The latter means that unitary streaming requires the collision not only to be non-unitary, but involving first-order terms only, as we would discuss hereafter. Such a nonunitary collision is described in further sections. 

For the time being, we note that sacrificing quantum parallelism (either by tensoring values at different lattice sites/timesteps, or by measurement, or a combination of the two) lifts this restriction. The qubit or gate complexity would have to scale linearly in the lattice volume $G$.
\begin{equation}
\bigotimes_{\vec{x}\in G} \ket{\vec{f}(\vec{x},t)}
\end{equation}
It is straightforward to see the intuition behind this choice. If we had encoded the discrete densities into tensored states to resolve the interdependency in their evolution, doing the same, representing them at different sites in a tensored fashion, would also allow for intersite-dependency in the evolution (swapping pairs of discrete densities pointing along opposite directions, each to a neighboring cell, among neighboring cells). We note of another possiblity described by of using a number of qubit registers linear in time, rather than volume. With the mapping at hand, one would have the state of each site in superposition with the states of all other sites, so a logarithmic number of qubits is still needed to represent the lattice positions. However, at each site, we still require $T$ copies of each discrete density, and we would need to perform $T-t$ collisions (where $t$ is the number of timesteps taken thus far) every timestep since they are not in superposition. The basic idea is that as we are solving over the lattice, at a given site, to evolve the time, the lattice site at hand must know the solution which is $t$ steps away in time and space. The limitation of this for simulating fluids is immediately observed. It is only useful if one is useful in evolution of some local properties, given an appropriate initial state. In the case where $G$ copies are used, a uniform state could be used a the cost of further timesteps. With $T$ copies, further attention should be given to the initial conditions. If you initialize a pipe flow with zero velocity for example, one can only reach the solution where the water reached $T$ cells into the pipe, no further. When simulating idealized, homogeneous isotropic, turbulence (HIT), one might indeed be interested in a few timesteps after the initial condition (which would cost $O(G)$ gates to initialize). In short, there is a tradeoff, but achieving unitary collision and streaming as described above would requrie a qubit count of:
\begin{equation}
O(min(G,T+\log_2{(G)})Q\log_2{(N+1)})
\end{equation}
\subsubsection{Numerical Results}

\begin{figure}[!htb]
    \centering
    \includegraphics[width = \columnwidth]{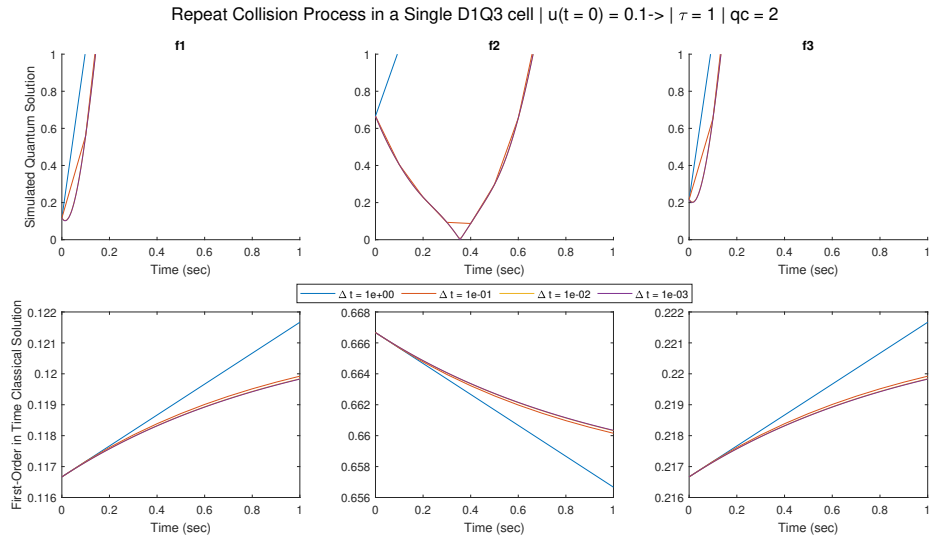}
    \caption{A comparison of the solution of the quantum algorithm (top) with the non-Hermitian Hamiltonian and the classical results (bottom) for a D1Q3 single-site lattice cell.}
    \label{fig:timefig}
\end{figure}

\begin{figure}[!htb]
    \centering
    \includegraphics[width = \columnwidth]{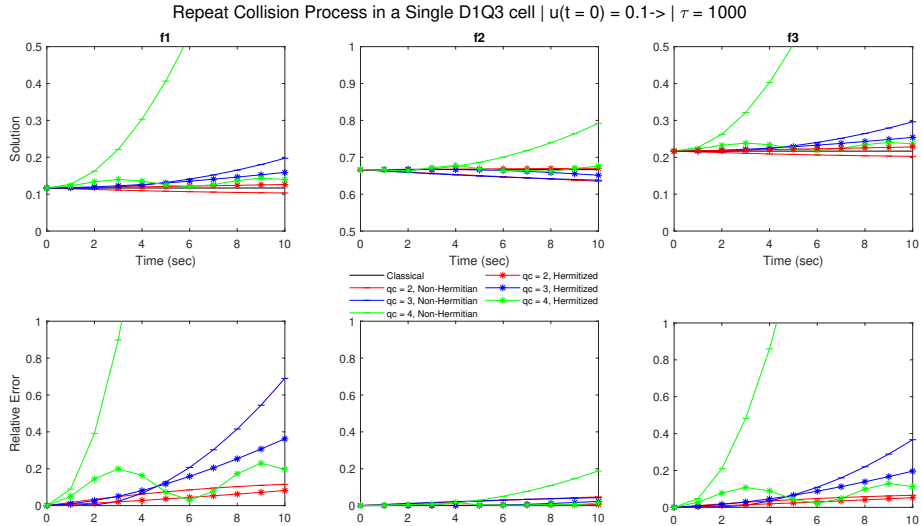}
    \caption{A comparison between the quantum solution solution (top) of the non-Hermitian Hamiltonian (dashes) and the Hermitian Hamiltonian with post-processed dissipation (stars), with the relative error with respect to the classical solution shown in the bottom.}
    \label{fig:Hermitfig}
\end{figure}

\paragraph{Single Site D1Q3 Collision (Effective 0D)}
We start by considering successive collisions in a single-cell lattice in Figs.~(\ref{fig:timefig},\ref{fig:Hermitfig}). We use direct matrix exponentiation to directly compare the inherent linearization error, without the implementation error arising from the linear combination of unitaries method used for either the non-Hermitian Hamiltonian, nor the truncated Taylor series of for the exponentiation of the Hermitian Hamiltonian. In Fig.~(\ref{fig:timefig}), we observe the solution of the three discrete densities $f_1$,$f_2$ and $f_3$ of a D1Q3 cell. The  quantum solution with $qc = 2$, $N = 3$, shown in the top of the figure is seen to diverge. The classical solution, which is first-order in time, is observed to converge to acceptable limits for $\Delta t = 10^{-3} $ time units.

In Fig.~(\ref{fig:Hermitfig}), we compare the solution for the non-Hermitian Hamiltonian with that of the Hermitian Hamiltonian for a different number of qubits per discrete density $qc$. Here, in the top figure, the solution for the different parameters is shown, whereas the bottom figure shows the relative error defined as:
\begin{equation}
    \abs{\frac{\text{quantum solution}(t) -\text{classical solution}(t)}{\text{classical solution(t)}}}
\end{equation}
In the non-Hermitian solution, the error scales better with a \textit{decreasing} number of qubits of discrete density, remaining within $20\%$ with the time of the simulation for $qc = 2$, whereas that of $qc = 4$ diverges within two timesteps, for all three discrete densities. In the case of the Hermitized Hamiltonian, the larger amplitude of oscillations observed for the higher number of qubits $qc = 4$ causes the latter's error to be larger than that of $qc = 3$. However, in the long run, the error for $qc = 3$ overtakes that of $qc = 4$ due to its larger growth rate. $qc = 2$ still outperforms the other two, however. This might be due to the fact that lattice Boltzmann for a single-phase fluid has no truncation error when linearized to second order \cite{itaniAnalysisCarlemanLinearization2022}.
One thing to note here is that part of the error is due to applying the initialization translation operator to a zero in a computational basis $\ket{0}^{\otimes qc}$ instead of a zero position state $\ket{q = 0}$. We do not discuss initialization here, as we do with the collision and streaming steps in Sec.~\ref{sec:compan} since data loading is a hurdle yet to be solved for quantum computing. This simplification is justified in Appendix~\ref{sec:eigvec}. Moreover, we note that boundary value problems typical of fluids could generally benefit from a reduced initialization cost of $O(\Pi_{d=1}^D-1(N_d))= O(G^{\frac{D-1}{D}})$ at the cost of additional timesteps $O(N_d)$. While this works outs to be $O(G)$ if no efficient initialization method is found, if an appropriate data loading method is found, flow in highly skewed domains have an advantage. We mention here that approximate data loading methods exist \cite{griffinInvestigationQuantumAlgorithms2019,araujoApproximatedQuantumstatePreparation2022}.
The difference in error scaling shown between the three discrete densities tells us that the error is not dependent on the initial value as conjenctured in Appendix~\ref{sec:erroran}, since the rest particle discrete density has the largest initial value of $\frac{2}{3}$, but on the relative distance of the initial value on the equilibrium value. 

This means that the error scaling might be expected to improve with increasing the number of discrete densities $Q$, up to a point where the increase in the number of terms in the equilibrium function cancels out any advantage for the discrete densities being closer to their equilibrium value. As such, for a given lattice, the hydrodynamic variables, introduced in Sec.~\ref{sec:hydrovar}, approach might prove beneficial, as it has two advantages to mitigate the error. First, it decreases the number of terms in the equilbrium function, especially for larger $Q$. it allows one to to make use of mid-circuit reset of the qubits for near-exact reset of velocity to zero. The hydrodynamic variables approach also opens up a new modeling paradigm in which partial spectral distribution of the velocity is supplemented with an \textit{guessed} distribution to reduce the scaling of the mid-circuit reset from $O(GD)$ to $O((GD)^{p})$ for $p<1$. Further description of the method requires a discussion on fluid dynamics not suitable for a pedagogical paper. As such, the hydrodynamic variables approach is not further discussed in Sec.~\ref{sec:compan}, after its introduction in Sec.~\ref{sec:hydrovar}.

\subsection{Lattice Position as a Mapped Variable}

One way to achieve an overall unitary evolution is to also introduce the position as a (normalized) variable mapped to a position operator:
\begin{equation}
\begin{aligned}
\vec{x} &\rightarrow \vec{\hat{x}}
\\ \hat{x}_d \ket{\vec{x}} &= \frac{x_d}{N_d} \ket{\vec{x}}
\end{aligned}
\end{equation}
where $N_d$ is the number of cells along the $d^{th}$ dimension. There would be a corresponding momentum operator $\hat{v}_d$. Then, the streaming step, stemming from the convection term, for the $i^{th}$ variable along the the $d^{th}$ dimension could be implemented as:
\begin{equation}
e^{-i\hat{v}_d\frac{c}{N_d}\hat{q}_i}
\end{equation}
We note that this is not exact when the truncated Fock space is considered. Therefore, we have sacrificed exact streaming for unitarity. Moreover, since the truncated bosonic position operator for a finite Fock space with $N$ excitation levels has $\frac{2^N}{2}$ linearly indepedent eigenvectors (for each of the $2^N$ eigenvectors, its opposite (multiplication by $-1$) is also an eigenvector) then we see why we need $N_d+1$ excitation levels considered when we have $N_d$ lattice sites along a given dimension, such that overall number of qubits used for the lattice position encoding is:
\begin{equation}
O(\log_2{(G+D)}) = O(\log_2{(G)}\log_2{(D)}) 
\end{equation}
such that it scales as $O(D\log_2{(D)})$ in the number of dimensions $D$, and $O(\log_2{(N_d)})$ in $N_d$. 	We note that this mapping of the of the lattice position:
\begin{equation}
\bigotimes_{d=1}^D \bigotimes_{k = 0}^{\log_2{(N_d+1)}-1} \ket{q} \; q \in \{0,1\}
\end{equation}
which yields inxeact streaming would require errorneous mixed state arising from the collision step to an at most equal magnitude of error, such that $N = N_d$, and the total number of qubits required for lattice position and discrete densities becomes:
\begin{equation}
O((D+Q)\log_2{(N_d+1)})
\end{equation}
A further disadvantage of this method of streaming is that now the lattice position register requires a more complicated initialization consisting of putting the register for each dimension in superposition of states consisting of a translation operator acting on the zero position state:
\begin{equation}
\ket{\vec{x}=\vec{0}} \rightarrow \frac{1}{\sqrt{G}}\Sigma_{\vec{x}\in G} e^{-i\Sigma_{d=1}^D \hat{v}_d \frac{x_d}{N_d}}
\end{equation}
where $\vec{x} = (x_1,x_2,x_3)$, e.g. for a two-dimensional problem $\ket{\vec{x}} = \ket{x_1x_20}$, rather than superposition of binary number representation achievable with Hadamard gates.

\section{Complexity Analysis}
\label{sec:compan}
\subsection{Hamiltonian Simulation}

We use the method outlined by \cite{berrySimulatingHamiltonianDynamics2015} for Hamiltonian simulation. The error that appears in the discussion is solely the error arising when simulating the Hamiltonian, which itself models the underlying problem up to an error to be determined. In practice, the model error, the Hamiltonian simulation error, and the error attributed to physical realization on the hardware need to be put in context of one another.
\subsubsection{Linear Combination of Unitaries from  a Truncated Taylor Series Expansion of a Unitary}

 To simulate a Hamiltonian $\hat{H} = \Sigma_{l = 1}^L \alpha_l \hat{H}_l$, where $\hat{H}_l$ are unitaries for which an efficient implementation method is known, for a total simulation time $T$, up to an error $\varepsilon$, \cite{berrySimulatingHamiltonianDynamics2015} utilize:
\begin{equation}
    O(\frac{\log_2{(L)}\log_2{(\frac{ST}{\varepsilon})}}{\log_2{\log_2{(\frac{ST}{\varepsilon})}}})
\end{equation}
ancillas, where we denote $\Sigma_{l=1}^L\alpha_l = S$.

The associated gate complexity is:
\begin{equation}
    O(STL(n+\log_2L)\frac{\log_2{(\frac{ST}{\varepsilon})}}{\log_2\log_2{(\frac{ST}{\varepsilon})}})
\end{equation}
where $n$ is the number of qubits the Hamiltonian acts on.
\subsubsection{Unitary Collision}

 For a single-phase fluid with a BGK equilibrium function, the Hamiltonian we have derived is written as a linear combination of $m$ monomials of momentum and position operators
\begin{equation}
    m = (Q^2+2Q+1+1) = Q^2+2Q+2
\end{equation}
where each is mapped to a sum of $\ceil{\log_2(N+1)}^2$ Pauli words. Hence:
\begin{equation}
    L = m\ceil{\log_2{(N+1)}}^2
\end{equation}

We also differentiate between the sum of coefficients for monomials of different order:
\begin{equation}
\begin{gathered}
    \\S_{0} = \frac{1}{\tau}
    \\S_1 = 2Q(\frac{1}{\tau})
    \\S_2 = -\frac{1}{\tau} Q(\frac{9}{2}-\frac{3}{2}Q)
    \end{gathered}
\end{equation}
where they are multiplied by the sum of coefficients of a single operator
\begin{equation}
    \sqrt{2(N+1)}\ceil{\log_2{(N+1)}}^2
\end{equation}
to the power of their order:
\begin{equation}
    \begin{gathered}
        \\Q(\sqrt{2(N+1)}\ceil{\log_2{(N+1)}}^2)
        \\Q(\sqrt{2(N+1)}\ceil{\log_2{(N+1)}}^2)^2
        \\Q(\sqrt{2(N+1)}\ceil{\log_2{(N+1)}}^2)^3
    \end{gathered}
\end{equation}
respectively, such that the overall sum of coefficients is:
\begin{equation}
\begin{aligned}
    S = O(\frac{1}{\tau}(\sqrt{2(N+1)}\ceil{\log_2{(N+1)}}^2)
    \\+ 2Q(\frac{1}{\tau})(\sqrt{2(N+1)}\ceil{\log_2{(N+1)}}^2)^2
    \\-\frac{1}{\tau} Q(\frac{9}{2}-\frac{3}{2}Q)(\sqrt{2(N+1)}\ceil{\log_2{(N+1)}})^2)^3)
    \\=O(\frac{1}{\tau}Q^2poly(\sqrt{2(N+1)}\ceil{\log_2{(N+1)}})^2))
\end{aligned}
\end{equation}

In regards to $\frac{1}{\tau}Q^2poly(\sqrt{2(N+1)}\ceil{\log_2{(N+1)}})^2)T$ which would appear in $\frac{log_2{x}}{\log_2\log_2{x}}$, we note:
\begin{align}
    \tau >\;& \frac{1}{2} &\;& \text{for a finite Reynold's number \cite{itaniAnalysisCarlemanLinearization2022}}
    \\ Q \;\in\;& [3,9,27] & \; & \text{by choice of lattice}
    \\ N+1 =& 2^{qc} \; \text{for} \; qc \; \in \; \mathcal{N}^+
    \\ T >\;& 1
\end{align}
Therefore, apart from a discontinuity around $2$, $\frac{log_2{x}}{\log_2\log_2{x}} = 100$, for $x = O(10^{300})$. We also note that to match the first-order in time accuracy of the classical method, it is $1$. Therefore, we drop the term.
The collision step then requires:
    \begin{equation}
            O(\log_2{(L)}\frac{\log_2{(\frac{ST}{\varepsilon})}}{\log_2{\log_2{(\frac{ST}{\varepsilon})}}})
            =O(\log_2{((Q^2+2Q+2)\ceil{\log_2{(N+1)}}^2)})
    \end{equation}
ancillas, and has a gate complexity:
\begin{equation}
\begin{aligned}
&O(STL(n+\log_2L)\frac{\log_2{(\frac{ST}{\varepsilon})}}{\log_2\log_2{(\frac{ST}{\varepsilon})}})
\\=&O(\frac{\Delta t}{\tau}Q^2poly(\sqrt{2(N+1)}\ceil{\log_2{(N+1)}}^2)T(Q^2+2Q+2)\ceil{\log_2{(N+1)}}^2
\\&(Q\ceil{\log_2{(N+1)}}+\log_2{((Q^2+2Q+2)\ceil{\log_2{(N+1)}}^2)}))
\end{aligned}
\end{equation}
\subsubsection{Unitary Streaming with Embedded Lattice 
Position}

In this case, $L = QD\ceil{\log_2{(N+1)}}\ceil{\log_2{(G)}}$, $S = {QD}$, and $n = \ceil{\log_2{G}}+Q\ceil{\log_2{(N+1)}}$, such that it requires a number of ancillas:
\begin{equation}
    O(\log_2{(L)}) = O(\log_2{(QD\ceil{\log_2{(N+1)}}\ceil{\log_2{(G)}})})
\end{equation}
and has a gate complexity of:
\begin{equation}
\begin{aligned}
    &O(STL(n+\log_2{(L)})) 
    \\=& O({QD}TQD\ceil{\log_2{(N+1)}}\ceil{\log_2{(G)}}(\ceil{\log_2{G}}+Q\ceil{\log_2{(N+1)}}+\log_2{(QD\ceil{\log_2{(N+1)}}\ceil{\log_2{(G)}})}))
\end{aligned}
\end{equation}
\subsection{Overall Complexity}

\begin{table}[!htb]
    \centering
    \caption{Summary of the qubit and gate complexities where $G$ is the lattice volume, $D$ the number of dimensions,  $T$ total number of timesteps taken, $Q$ number of discrete densities, and $b$ desired decimal precision of the results.
    X\textsuperscript{*}: Unitary Streaming by Embedding of Lattice Position
    X\textsuperscript{**}: Unitary Streaming with Binary Representation of Lattice Position
    X\textsuperscript{***}: Unitary Streaming by Swapping Discrete Density Registers}
    \resizebox{16cm}{!}{
    \begin{tabular}{|c|c|c|c|c|c|c|}
         \hline
         \multicolumn{2}{|c|}{Collision}&\multicolumn{2}{c|}{Streaming}& Qubit Complexity $(O)$ & LCU Ancillas $(O)$ &Gate Complexity $(O)$  
         \\\hline &&X\textsuperscript{*}&&$\log_2{(G)}+Q$&$\log_2{(QD\ceil{\log_2{(G)}})}$&$T{Q^2D^2}\ceil{\log_2^2{(G)}}$
         \\\hline &&X\textsuperscript{**}&&$\log_2{(QG)}+2D$&-&$TD\log_2^2{(G)}$
         \\\hline &&X\textsuperscript{***}&&$GQ$&-&$3T\frac{Q-1}{2}G$
         \\\hline         X&&&&$Q\log_2{(T+b)}$&$\log_2{(Q\log_2{(T+b)})}$&$(T^5\frac{1}{\tau}Q^5)$
         \\\hline         X&&X\textsuperscript{*}&&$\log_2{(G)}+Q(max(\log_2{(G)},\log_2{(T+b)}))$&$\log_2{(QD\ceil{\log_2{(G)}})}$&$T{Q^2D^2}\ceil{\log_2^2{(G)}}+T^5\frac{1}{\tau}Q^5$
         \\\hline X&&X\textsuperscript{***}&&$min(G,T+\log_2{(G)})Q\log_2{(T+b)}$&$\log_2{(Q\log_2{(T+b)})}$&$T^5\frac{1}{\tau}Q^5+TQG$
         \\\hline
X&&&X&$Q\log_2{(T+b)}+\log_2{(G)}+2D$&$\log_2{(Q\log_2{(T+b)})}$&$G(T^5\frac{1}{\tau}Q^5)$
         \\\hline
         Unitary & Non-Unitary & Unitary & Non-Unitary & && \\
         \hline
    \end{tabular}}
    \label{tab:detailedcomp}
\end{table}
As Table~\ref{tab:detailedcomp} shows, the binary representation of lattice position achieves the best scaling, logarithmic in $Q$ and $G$ for both, the qubit and the gate, complexities. However, the collision step scales as a power of the total number of timesteps $T$ and the number of discrete densities $Q$. Overall, the combination of the collision with the position embedding of the lattice position is the only method which does not show linear scaling in the lattice volume $G$. As such, we focus on it in our  discussion in terms of Reynolds number. For a given fluid viscosity, the lattice volume would increase with $Re^D$, whereas the total number of timesteps can be expected to increase with $Re$. The suboptimal, polynomial, dependence on the number of timesteps for achieving the collision then means that our algorithm has not yet achieved a quantum advantage. We might need to relax our requirements for unitarity and consider a nonunitary collision step achieved probabilisticly on a quantum computer to mitigate the undesirable scaling. This is left for future work. Moreover, we wish to observe that
the physics of fluids is populated with interesting problems
at low Reynolds numbers, especially in soft matter and biological flows \cite{bernaschiMesoscopicSimulationsPhysicschemistrybiology2019}.
For instance, it would be of great interest to devise a
{\it quantum multiscale} application, coupling quantum algorithms for 
biomolecules swimming in a water solvent described by a quantum 
algorithm for low-Reynolds-number flow.   
\section{Conclusion \& Outlook}
We have presented a prospective quantum computing algorithm 
for the solution of classical fluid dynamics based on the quantum linearization of the lattice Boltzmann equation.
The main result is that collision and streaming could both be achieved by unitary evolution. The quantum advantage for high-Reynolds-number flows remains entirely open at this stage. On philosophical grounds, it would not be surprising if high Reynolds
numbers would prove too hard for quantum computing. 
Indeed, coming back to Feynman, while it is true that Nature is not classical, it is
equally true that Nature has a very strong innate tendency to {\it become}
classical at macroscopic scales. In this respect, the prominence of non-locality
in exchange for the release of nonlinearity, may well represent yet another 
signature of this tendency. Figuring this out is a worthy enterprise, regardless of the practical 
outcome. 

\begin{acknowledgments}
One of the authors, S. Succi, is grateful to the SISSA program on 
"Collaborations of Excellence" that allowed him to visit and focus on the 
work presented in this paper. 
He also wishes to acknowledge support from National Centre for HPC, Big Data 
and  Quantum Computing (Spoke 10, CN00000013). 

Illuminating discussions with S. S. Bharadwaj (NYU), D. Buaria (NYU), V. Kumar (IBM), A. Mezzacapo (IBM), H. Ouerfelli (TUM), S. Ruffo (SISSA), L. Schleeper (IBM), K. Sharma (IBM), A. Solfanelli (SISSA), R. Steijl (UoGlas), and M. C. Tran (UMD) are kindly acknowledged. 
\end{acknowledgments}

\section*{Data Availability}
The code used in this study is openly available in GitHub at: \url{https://github.com/waelitani/Quantum-Carleman-Lattice-Boltzmann-Simulation-of-Fluids/}.

\section*{Conflict of Interest}
The authors have no conflicts to disclose.

\newpage
\appendix
\section*{Appendix}
\section{Error Analysis}
\label{sec:erroran}
The linearization order for a general nonlinear system is generally intractable for infinite time $N \propto e^{O(T)}$ \cite{gellerUniverseNonlinearQuantum2021}, and Carleman linearization is usually discussed in the context of prediction time within a chosen level of accuracy. One may conjencture that for a $k^{th}$ order in time accurate simulation of a system $\vec{f}$ driven by the function $\Omega$ with a polynomial type nonlinearity of order $p$, and with number of terms $m$, the error scales as $\norm{\vec{f}(t = 0)}^{N-kpm\norm{\Omega}T}$. The power-law scaling is chosen after \cite{foretsExplicitErrorBounds2017}. For the case of the single-phase lattice Boltzmann, for which we know is first-order in time, and is exactly linear in second-order linearization, a decimal precision of $b$ requires $N = O(T)+b$. Having set our expectations, we go on for a more detailed analysis. We analyze the error resulting from truncating the number of excitation levels in the bosonic Fock space considered to $N$. 
\vspace{-1cm}
\subsection{Error in the Action of Operators on a Truncated Position Eigenstate}
Starting with a position eigenstate encoding a discrete density $f_i(\vec{x},t)$:
\begin{equation}
    \ket{f_i(\vec{x},t)} = \frac{e^{-\frac{f_i(\vec{x},t)^2}{2}}}{\pi^{\frac{1}{4}}}\Sigma_{n=0}^\infty 2^{-\frac{n}{2}} (n!)^{-\frac{1}{2}} H_n(f_i(\vec{x},t)) \ket{n},
\end{equation}
consider the action of the corresponding position operator $\hat{q}_i$:
\begin{equation}
    \begin{aligned}
        \hat{q}_i \ket{f_i(\vec{x},t)} =& \hat{q}_i \frac{e^{-\frac{f_i(\vec{x},t)^2}{2}}}{\pi^{\frac{1}{4}}}\Sigma_{n=0}^\infty 2^{-\frac{n}{2}} (n!)^{-\frac{1}{2}} H_n(f_i(\vec{x},t)) \ket{n}
        \\ =& \frac{1}{\sqrt{2}}(\hat{a}_i+\hat{a}^\dagger_i) \frac{e^{-\frac{f_i(\vec{x},t)^2}{2}}}{\pi^{\frac{1}{4}}}\Sigma_{n=0}^\infty 2^{-\frac{n}{2}} (n!)^{-\frac{1}{2}} H_n(f_i(\vec{x},t)) \ket{n}
        \\ =& \frac{1}{\sqrt{2}} \frac{e^{-\frac{f_i(\vec{x},t)^2}{2}}}{\pi^{\frac{1}{4}}}\Sigma_{n=0}^\infty 2^{-\frac{n}{2}} (n!)^{-\frac{1}{2}} n^{\frac{1}{2}} H_n(f_i(\vec{x},t)) \ket{n-1}
        \\ &+\frac{1}{\sqrt{2}}\frac{e^{-\frac{f_i(\vec{x},t)^2}{2}}}{\pi^{\frac{1}{4}}}\Sigma_{n=0}^\infty 2^{-\frac{n}{2}} (n!)^{-\frac{1}{2}}(n+1)^{\frac{1}{2}} H_n(f_i(\vec{x},t)) \ket{n+1}
    \end{aligned}
\end{equation}
Clearly $n^{\frac{1}{2}} = 0$ for $n = 0$. We shift the indices to combine the terms:
\begin{equation}
    \begin{aligned}
        \hat{q}_i \ket{f_i(\vec{x},t)} =& \frac{1}{\sqrt{2}} \frac{e^{-\frac{f_i(\vec{x},t)^2}{2}}}{\pi^{\frac{1}{4}}}\Sigma_{n=1}^\infty 2^{-\frac{n}{2}} (n!)^{-\frac{1}{2}} n^{\frac{1}{2}} H_n(f_i(\vec{x},t)) \ket{n-1}
        \\ &+\frac{1}{\sqrt{2}}\frac{e^{-\frac{f_i(\vec{x},t)^2}{2}}}{\pi^{\frac{1}{4}}}\Sigma_{n=1}^\infty 2^{-\frac{n-1}{2}} ((n-1)!)^{-\frac{1}{2}}(n)^{\frac{1}{2}} H_{n-1}(f_i(\vec{x},t)) \ket{n}
        \\ =& \frac{1}{\sqrt{2}} \frac{e^{-\frac{f_i(\vec{x},t)^2}{2}}}{\pi^{\frac{1}{4}}}\Sigma_{n=0}^\infty 2^{-\frac{n+1}{2}} ((n+1)!)^{-\frac{1}{2}} (n+1)^{\frac{1}{2}} H_{n+1}(f_i(\vec{x},t)) \ket{n}
        \\ &+\frac{1}{\sqrt{2}}\frac{e^{-\frac{f_i(\vec{x},t)^2}{2}}}{\pi^{\frac{1}{4}}}\Sigma_{n=1}^\infty 2^{-\frac{n-1}{2}} ((n-1)!)^{-\frac{1}{2}}(n)^{\frac{1}{2}} H_{n-1}(f_i(\vec{x},t)) \ket{n}
        \\ =& \frac{e^{-\frac{f_i(\vec{x},t)^2}{2}}}{\pi^{\frac{1}{4}}}\Sigma_{n=0}^\infty 2^{-\frac{n}{2}} (n!)^{-\frac{1}{2}} \frac{1}{2} H_{n+1}(f_i(\vec{x},t)) \ket{n}
        \\ &+\frac{e^{-\frac{f_i(\vec{x},t)^2}{2}}}{\pi^{\frac{1}{4}}}\Sigma_{n=1}^\infty 2^{-\frac{n}{2}} (n!)^{-\frac{1}{2}} n H_{n-1}(f_i(\vec{x},t)) \ket{n}
        \\ =& \frac{e^{-\frac{f_i(\vec{x},t)^2}{2}}}{\pi^{\frac{1}{4}}}\Sigma_{n=0}^\infty 2^{-\frac{n}{2}} (n!)^{-\frac{1}{2}} \frac{1}{2} H_{n+1}(f_i(\vec{x},t)) \ket{n}
        \\ &+\frac{e^{-\frac{f_i(\vec{x},t)^2}{2}}}{\pi^{\frac{1}{4}}}\Sigma_{n=0}^\infty 2^{-\frac{n}{2}} (n!)^{-\frac{1}{2}} n H_{n-1}(f_i(\vec{x},t)) \ket{n}
        \\ =& \frac{e^{-\frac{f_i(\vec{x},t)^2}{2}}}{\pi^{\frac{1}{4}}}\Sigma_{n=0}^\infty 2^{-\frac{n}{2}} (n!)^{-\frac{1}{2}} (\frac{1}{2} H_{n+1}(f_i(\vec{x},t))+n H_{n-1}(f_i(\vec{x},t))) \ket{n}
    \end{aligned}
\end{equation}
We now use the recurrence relation for the Hermite polynomials:
\begin{equation}
    H_{n+1}(f_i(\vec{x},t)) = 2f_i(\vec{x},t)H_n(f_i(\vec{x},t))-2nH_{n-1}(f_i(\vec{x},t))
\end{equation}
which, when rearranged, gives the expression for $f_i(\vec{x},t)H_n(f_i(\vec{x},t))$ in terms of one-order higher and lower Hermite polynomials:
\begin{equation}
    \frac{1}{2}H_{n+1}(f_i(\vec{x},t))+nH_{n-1}(f_i(\vec{x},t)) = f_i(\vec{x},t)H_n(f_i(\vec{x},t))
\end{equation}
such that:
\begin{equation}
    \begin{aligned}
        \hat{q}_i \ket{f_i(\vec{x},t)} =& \frac{e^{-\frac{f_i(\vec{x},t)^2}{2}}}{\pi^{\frac{1}{4}}}\Sigma_{n=0}^\infty 2^{-\frac{n}{2}} (n!)^{-\frac{1}{2}} (\frac{1}{2} H_{n+1}(f_i(\vec{x},t))+n H_{n-1}(f_i(\vec{x},t))) \ket{n}
        \\=& \frac{e^{-\frac{f_i(\vec{x},t)^2}{2}}}{\pi^{\frac{1}{4}}}\Sigma_{n=0}^\infty 2^{-\frac{n}{2}} (n!)^{-\frac{1}{2}} f_i(\vec{x},t) H_n(f_i(\vec{x},t) ) \ket{n} = f_i(\vec{x},t) \ket{f_i(\vec{x},t)}
        \label{eq:poswithrecur}
    \end{aligned}
\end{equation}
We see how the truncation error arises due to the dependence on higher-order Hermite polynomials in the expression to bring out the variable of the Hermite polynomial, $f_i(\vec{x},t)$ in this case, out as a coefficient.

Now consider, a truncated position eigenstate:
\begin{equation}
\begin{aligned}
    \ket{f_i(\vec{x},t)} = C_i(\vec{x},t) \Sigma_{n=0}^N 2^{-\frac{n}{2}} (n!)^{-\frac{1}{2}} H_n(f_i(\vec{x},t)) \ket{n}
\end{aligned}
\label{eq:trucpos}
\end{equation}
where $C_i(\vec{x},t)$ corresponds to the appropriate statevector normalization accounting for the truncated series:
\begin{equation}
    C_i(\vec{x},t) = (\Sigma_{n=0}^N 2^{-n} (n!)^{-1} H_n^2(f_i(\vec{x},t)))^{-1}
\end{equation}
From Eqs.~\ref{eq:poswithrecur}~and~\ref{eq:trucpos}, we see that:
\begin{equation}
    \hat{q}_i \ket{f_i(\vec{x},t)} = f_i(\vec{x},t) \ket{f_i(\vec{x},t)} + O(C_i(\vec{x},t) 2^{-\frac{N}{2}} (N!)^{-\frac{1}{2}} \frac{1}{2} H_{N+1}(f_i(\vec{x},t))) \ket{N}
\end{equation}
We recall that $\abs{f_i(\vec{x},t)} \le 1 $, and the monic Hermite polynomials are used such that we may write:
\begin{equation}
    \abs{\braket{n+1}{f_i(\vec{x},t)}}\le\abs{\braket{n}{f_i(\vec{x},t)}} \; \forall n
    \label{eq:hercomp}
\end{equation}

With no abuse of the big $O$ notation, we may write:
\begin{equation}
    \hat{q}_i \ket{f_i(\vec{x},t)} = (f_i(\vec{x},t)+O(2^{-\frac{N}{2}} (N!)^{-\frac{1}{2}} \frac{1}{2} H_{N+1}(f_i(\vec{x},t)))) \ket{f_i(\vec{x},t)}
\end{equation}
Having invoked Eq.~(\ref{eq:hercomp}) implies any error expression must now be derived as an upper bound.
We now shift our attention to the momentum operator:
\begin{equation}
    \begin{aligned}
        \hat{p} = (\frac{i}{\sqrt{2}})(\hat{a}^\dagger-\hat{a}) = i (\sqrt{2}\hat{a}^\dagger-\hat{q})
    \end{aligned}
\end{equation}

where we have kept the raising operator and replaced the lowering operator since the truncation means the raising operator is causing loss of information when raising lower order polynomials into excitation levels beyond the truncation. On the other hand, the lowering operator brings in a dependency on the unknown (unencoded) higher order Hermite polynomials. Then the truncation introduces an error in the momentum operator as:
\begin{equation}
    \hat{p} = -i (\frac{\partial}{\partial \hat{q}} +O(2^{-\frac{N}{2}} (N!)^{-\frac{1}{2}} \frac{1}{2} H_{N+1}(f_i(\vec{x},t))))
\end{equation}
This is to say that the error introduced by the momentum operator could be accounted for since we have differed the dissipative aspect of the algorithm. While this argument is more obvious for the simplified and bounding expression used by introducing the error to all levels directly, attention to the overall state norm, and relative scaling of amplitude of the different excitation levels and ground is sufficient for the correction. An exact differential operator corresponding to the truncated position operator could be constructed. We see that we can define an exact canonical conjugate of truncated position:
\begin{equation}
    [\hat{q},i\hat{A}] = i(\hat{q}\hat{A}-\hat{A}\hat{q}) = i\hat{I}
\end{equation}
such that:
\begin{align}
    (\hat{A}-\hat{q}^{-1}\hat{A}\hat{q}) = \hat{q}^{-1}
    \\\hat{A} = \hat{q}^{-1}(\hat{I}+\hat{A}\hat{q})
\end{align}
and:
\begin{equation}
    \hat{A}=-\hat{A}^\dagger
\end{equation}

\subsection{Linear Combination under the Incompressible Nonlinear Collision Term}
For a single-phase incompressible fluid, recall the definition of the collision operator under the BGK appromixation:
\begin{equation}
    \Omega_i(\vec{f}) = -\frac{1}{\tau} (f_i -f_i^{eq})
\end{equation}
in terms of the equilibrium function defined  as:
\begin{equation}
        f_i^{eq}(\vec{x}, t) ={w_i}(1+3\vec{e}_i\cdot f_j\vec{e}_j+\frac{9}{2}(\vec{e}_i\cdot f_j\vec{e}_j)^2-\frac{3}{2}f_jf_k\vec{e}_j\cdot\vec{e}_k))
\end{equation}
such that:
\begin{equation}
    \Omega_i(\vec{f}+\vec{\gamma}) = -\frac{1}{\tau} (f_i+\gamma_i -{w_i}(1+3\vec{e}_i\cdot (f_j+\gamma_j)\vec{e}_j+\frac{9}{2}(\vec{e}_i\cdot (f_j+\gamma_j)\vec{e}_j)^2-\frac{3}{2}(f_j+\gamma_j)(f_k+\gamma_k)\vec{e}_j\cdot\vec{e}_k)))
\end{equation}
Let $i,j = 0$ always denote the rest particles discrete density. The zero basis vector is orthogonal to every other vector by construction.
\begin{equation}
    \vec{e}_0\cdot \vec{e}_i = 0 \; \forall i
\end{equation}
Additionally, for an any given dimension, there exists two lattice basis vectors for each different dimension completely orthogonal to the given dimension, such that the inner product of a given lattice basis vector $\vec{e_i}$ is nonorthogonal with at most $(Q-2(D-1)-1) = Q-2D+1$ other basis vectors. Then, we may write:

\begin{equation}
\begin{aligned}
    &\Omega_i(\vec{f}+\vec{\gamma}) 
    \\=& -\frac{1}{\tau} (f_i+\gamma_i -{w_i}(1+3\vec{e}_i\cdot (f_j+\gamma_j)\vec{e}_j+\frac{9}{2}(\vec{e}_i\cdot (f_j+\gamma_j)\vec{e}_j)^2-\frac{3}{2}(f_j+\gamma_j)(f_k+\gamma_k)\vec{e}_j\cdot\vec{e}_k)))
    \\=& -\frac{1}{\tau} (f_i -{w_i}(1+3\vec{e}_i\cdot f_j\vec{e}_j+\frac{9}{2}(\vec{e}_i\cdot (f_j)\vec{e}_j)^2-\frac{3}{2}f_jf_k\vec{e}_j\cdot\vec{e}_k)))
    \\&+O(-\frac{1}{\tau}\norm{\vec{\gamma}}
    \\&+\frac{1}{\tau}w_i
    (3(Q-2D+1)\norm{\vec{\gamma}})
    \\&+\frac{1}{\tau}w_i(\frac{9}{2}((Q-2D+1)^2(2\norm{\vec{f}}\norm{\vec{\gamma}}+\norm{\vec{\gamma}}^2)))
    \\&-\frac{1}{\tau}w_i\frac{3}{2}(Q-1)(Q-2D+1)(\norm{\vec{\gamma}}^2+2\norm{\vec{\gamma}}\norm{\vec{f}})
    )))
\end{aligned}
\end{equation}

By considering the bound and using the big $O$ notation we are by default sacrificing any symmetries unique to the lattice Boltzmann formulation.
We note that invoking the bound for the discrete density removes the dependence of the error expression derived on the initial conditions, which renders a direct compraison to the results of \cite{liuEfficientQuantumAlgorithm2020,anEfficientQuantumAlgorithm2022} harder.

When considering the bound for the discrete densities, the bound for the expression in the big $O$ may be alternatively simplified as:
\begin{equation}
    O(\abs{\frac{1}{\tau}(6Q^2\norm{\vec{\gamma}}^2+(12Q^2+3Q+1)\norm{\vec{\gamma}}+1)})
    \label{eq:modbound}
\end{equation}
for which we define:
\begin{align}
    a  =& 6Q^2
    \\ b =& 12Q^2+3Q+1
    \\ c =& 1
\end{align}
such that:
\begin{equation}
    b^2-4ac = (12Q^2 + 3Q + 1)^2 - 24Q^2 = 144Q^4 + 72Q^3 + 9Q^2 + 6Q + 1
    \label{eq:delta}
\end{equation}
Our aim is to be able to factor the recurrence relation in Eq.\ref{eq:recerror} such that it is a complete square. Then we can rearrange the equality in Eq.\ref{eq:delta}:
\begin{equation}
    b^2-4ac -4a(6Q^2 + 3Q + \frac{3}{8} + \frac{1}{4Q} + \frac{1}{24Q^2}) = 0
    \label{eq:deltarear}
\end{equation}
such that we may introduce a modified polynomial in $\norm{\vec{\gamma}}$ with coefficients:
\begin{align}
    a' =& a
    \\b'=& b
    \\ c' =& c+6Q^2 + 3Q + \frac{3}{8} + \frac{1}{4Q} + \frac{1}{24Q^2}
\end{align}
bounding the polynomial at hand in Eq.~(\ref{eq:modbound}). We have chosen to modify the $c$ coefficient since it is the coefficient for the zero order term. An alternative strategy would be to have:
\begin{equation}
    a' = a+36Q^4 + 18Q^3 + \frac{9}{4}Q^2 + \frac{3}{2}Q + \frac{1}{4}
\end{equation}
which would be decided upon once the error bounds are computed. The  bounds for the expression in Eq.~(\ref{eq:modbound}) would then be of the form:
\begin{equation}
    \\O(\abs{\frac{1}{\tau}(C_1\norm{\vec{\gamma}}+C_0)^2}) \; \text{for} \; C_1 \ge 0 \;,\; C_0 \ge 0
    \label{eq:genbound}
\end{equation}
We note that there are infinitely many choices for a complete square polynomial that bounds that appearing in Eq.\ref{eq:modbound}. The two presented here are associated with either $\norm{\Vec{\gamma}}$ approaching zero in which case we are interested in introduced the additional terms to the coefficient of its highest power, or otherwise approaching infinite in which case we introduce the additional terms to the constant rather than coefficient terms. An optimal choice of is made in Sec.\ref{sec:optimal} after the role of the coefficients in the dynamical evolution of the error bound is clarified.

\subsection{Error Definition}
The monic Hermite polynomial for $n=1$ is given by:
\begin{equation}
    H_1(x) = x
\end{equation}
such that the truncated position eigenestate defined in Eq.\ref{eq:trucpos} allows us to retrieve the encoded variable as:
\begin{equation}
    \braket{1}{f_i(\vec{x},t)} = C_i(\vec{x},t) f_i(\vec{x},t)
\end{equation}
We defer the discussion of $C_i(\vec{x},t)$ and suffice with the assumption that $\Pi_{i=1}^Q C_i = cst = C_N$, which bodes well with the fact that the collision term is linear in a rescaling factor. This is to say there is a way to retrieve $f_i(\vec{x},t)$ encoded in a truncated position eigenstate $\ket{f_i(\vec{x},t)}$ free of error apart from the error arising in the evolution. For the same reason, the initial values could be retrieved exactly given the initial state; thus, we say that the initial state of the system is free of error, as expected since no nonlinear evolution has yet been approximated.

To differentiate between the exact and the solution approximated by evolving over a finite Hilbert space, hereafter in this subsection, we denote the former as $f_i(\vec{x},t)$ and the latter $\Tilde{f}_i(\vec{x},t)$, and define the error $\varepsilon_i(\vec{x},t)$:
\begin{equation}
    \varepsilon_i(\vec{x},t) = \abs{f_i(\vec{x},t)-\Tilde{f}_i(\vec{x},t)}
\end{equation}
and:
\begin{equation}
\label{eq:errordef}
    \varepsilon(t) = O(\norm{\vec{\varepsilon}(\vec{x},t)}_{\infty,\forall \vec{x}})
\end{equation}
where $\vec{\varepsilon}(\vec{x},t) = (\varepsilon_1(\vec{x},t),\dots,\varepsilon_i(\vec{x},t),\dots,\varepsilon_Q(\vec{x},t))$.
Finally, we define the expression:
\begin{equation}
    \varepsilon_N = O(\norm{2^{-\frac{N}{2}} (N!)^{-\frac{1}{2}} \frac{1}{2} H_{N+1}(f_i(\vec{x},t))}_{\infty,\forall i \& \forall \vec{x} \& \forall t})
\end{equation}
noting that $\abs{f_i(\vec{x},t)} \le 1$ such that $\varepsilon_N \le 1 \forall N$.

\subsection{Error Recurrence Relation}
Consider the state:
\begin{equation}
    \ket{\vec{\Tilde{f}}(\vec{x},t)} = \bigotimes_{i=1}^Q \ket{f_i(\vec{x},t)+\varepsilon_i(t)}
\end{equation}
Performing a collision and streaming step with the Hamiltonian defined in Eq.~(\ref{orgexp}):
\begin{equation}
    \hat{H} = \vec{\hat{p}} \cdot \Omega(\vec{\hat{q}})
\end{equation}
updates the error embedded in the approximate solution as well as the solution itself:
\begin{equation}
    \bigotimes_{i=1}^Q \ket{f_i(\vec{x}+\vec{c}_i\Delta t,t+\Delta t)+\varepsilon(t+\Delta t)} = e^{-i\Delta t \vec{\hat{p}} \cdot \Omega(\vec{\hat{q}})} \bigotimes_{i=1}^Q \ket{f_i(\vec{x},t)+\varepsilon(t)}
\end{equation}
This is further justified by the fact that the mapping is exact for $N \rightarrow \infty $.
By using the triangular inequality:
\begin{equation}
    \norm{\varepsilon_N+\varepsilon(t)} \le \varepsilon_N+\varepsilon(t)
\end{equation}
combined with Eq.~(\ref{eq:modbound}), we arrive at the recurrent relation:
\begin{equation}
\begin{aligned}
\label{eq:recerror}
    \varepsilon(t+1) =& \frac{\Delta t}{\tau}(C_1(\varepsilon_N+\varepsilon(t))+C_0)^2
\end{aligned}
\end{equation}
\subsection{Error Logistic Map}
We now make the substitution:
\begin{equation}
\label{eq:epsdef}
    \epsilon(t) = \sqrt{\frac{\Delta t}{\tau}}(C_1(\varepsilon(t)+\varepsilon_N)+C_0)
\end{equation}
with:
\begin{equation}
        \epsilon(t = 0) = \sqrt{\frac{\Delta t}{\tau}}(C_1\varepsilon_N+C_0)
\end{equation}
into Eq.\ref{eq:recerror}, giving:
\begin{equation}
\begin{aligned}
\label{eq:recerrorsubs}
    \\ \epsilon(t) =& C_1\sqrt{\frac{\Delta t}{\tau}}(\epsilon^2(t)-\frac{C_0}{C_1}-\varepsilon_N)
\end{aligned}
\end{equation}
we define $\kappa$ to satisfy:
\begin{equation}
    \kappa^2+\kappa-\frac{C_0}{C_1}-\varepsilon_N = 0
\end{equation}
which admits solutions:
\begin{equation}
\label{eq:kappadef}
    \kappa = \frac{-1\pm \sqrt{1+4(\frac{C_0}{C_1}+\varepsilon_N)}}{2}
\end{equation}

so that we define:
\begin{align}
\label{eq:zetaeps}
    \zeta(t) =& \epsilon(t) + \kappa
\end{align}
with:
\begin{equation}
\begin{aligned}
    \zeta(t = 0) =& \sqrt{\frac{\Delta t}{\tau}}(C_1\varepsilon_N+C_0) + \kappa
    \\=& \sqrt{\frac{\Delta t}{\tau}}C_1(\varepsilon_N+\frac{C_0}{C_1}) + \kappa
    \\=& \frac{1}{4}\sqrt{\frac{\Delta t}{\tau}}C_1((2\kappa+1)^2-1) + \kappa
\end{aligned}
\end{equation}
which evolves according to:
\begin{equation}
    \begin{aligned}
        \zeta(t+1) =& C_1\sqrt{\frac{\Delta t}{\tau}}((\zeta(t)-\kappa)^2-\frac{C_0}{C_1}-\varepsilon_N) 
        \\ =& C_1\sqrt{\frac{\Delta t}{\tau}}\zeta(t)(\zeta(t)-2\kappa)
        \\ =& -2\kappa C_1\sqrt{\frac{\Delta t}{\tau}}\zeta(t)(1-\frac{\zeta(t)}{2\kappa})
    \end{aligned}
\end{equation}
We bring this down to a logistic map by dividing both sides by making a change of variable

\begin{equation}
\label{eq:zdef}
    Z(t) = \frac{\zeta(t)}{2\kappa}
\end{equation}
giving:
\begin{equation}
    \begin{aligned}
        Z(t+1) =& -2\kappa C_1\sqrt{\frac{\Delta t}{\tau}}Z(t)(1-Z(t))
    \end{aligned}
\end{equation}
with:
\begin{equation}
    Z(t=0) = \frac{1}{4}\sqrt{\frac{\Delta t}{\tau}}C_1((\sqrt{2\kappa}+\frac{1}{\sqrt{2\kappa}})^2-\frac{1}{2\kappa}) + \frac{1}{2}
\end{equation}
where we note that the initial value could not be imaginary because $\sqrt{2\kappa}$ appears such that any imaginary unit is squared.
\begin{equation}
    Z(t=0) = \frac{1}{8\kappa}\sqrt{\frac{\Delta t}{\tau}}C_1((2\kappa+1)^2-1) + \frac{1}{2}
\end{equation}

\subsection{Impossibility of a General Time-Independent Bound}
\label{sec:optimal}
We can rearrange Eq.~(\ref{eq:zetaeps}) as:
\begin{equation}
    \epsilon(t) = \zeta(t)-\kappa
\end{equation}
As we need to relate the error $\varepsilon(t)$ defined in Eq.~(\ref{eq:errordef}) to the variable showing the logistic map $Z(t)$, we replace the expressions of $\epsilon(t)$ and $\zeta(t)$ from Eqs.~(\ref{eq:epsdef})~and~(\ref{eq:zdef}) respectively.
\begin{equation}
   \sqrt{\frac{\Delta t}{\tau}}(C_1(\varepsilon(t)+\varepsilon_N)+C_0)  = 2\kappa Z(t) - \kappa
\end{equation}
    
We also use the defintion of $\kappa$ from Eq.~(\ref{eq:kappadef}):
\begin{equation}
\frac{1}{4}((2\kappa+1)^2-1)-\varepsilon_N = \frac{C_0}{C_1}    
\end{equation}
to arrive at:
\begin{align}
   \sqrt{\frac{\Delta t}{\tau}}C_1((\varepsilon(t)+\varepsilon_N)+\frac{C_0}{C_1}) =& 2\kappa (Z(t) - \frac{1}{2})
   \\\sqrt{\frac{\Delta t}{\tau}}C_1((\varepsilon(t)+\frac{1}{4}((2\kappa+1)^2-1)) =& 2\kappa (Z(t) - \frac{1}{2})
   \\\varepsilon(t) =& \frac{1}{C_1} \sqrt{\frac{\tau}{\Delta t}}2\kappa (Z(t) - \frac{1}{2})-(\kappa^2+\kappa)\ge0
   \\\varepsilon(t) =& \frac{1}{C_1} \sqrt{\frac{\tau}{\Delta t}}2\kappa (Z(t) - \frac{1}{2})-(\frac{C_0}{C_1}+\varepsilon_N)\ge0
\end{align}

\subsubsection{Logistic Map with Coefficient in [-2,0]}
\begin{equation}
    2\kappa = -1\pm{\sqrt{1+4(\frac{C_0}{C_1}+\varepsilon_N)}}
\end{equation}
$\kappa$ could only be bounded by unity $\abs{\kappa}<1$ when it is positive. 
Observing the evolution of $Z(t)$:
\begin{equation}
    \begin{aligned}
        Z(t+1) =& -2\kappa C_1\sqrt{\frac{\Delta t}{\tau}}Z(t)(1-Z(t))
    \end{aligned}
\end{equation}

We further require:
\begin{equation}
    \kappa C_1 \sqrt{\frac{\Delta t}{\tau}} < 1
\end{equation}
and know \cite{itaniAnalysisCarlemanLinearization2022}:
\begin{equation}
    \frac{\Delta t}{\tau} < 2
\end{equation}
Therefore:
\begin{equation}
    \kappa C_1 < \frac{1}{\sqrt{2}}
\end{equation}
The constraint on the initial condition:
\begin{equation}
\begin{aligned}    
    -\frac{1}{2}\le Z(t=0) \le \frac{3}{2}
\end{aligned}
\end{equation}
yields the condition:
\begin{align}
-\frac{1}{2}\le& \frac{1}{2}(1+\sqrt{\frac{\Delta t}{\tau}}C_1(\kappa+1)) \le& \frac{3}{2}
\\-1\le& 0 \le \frac{1}{2}(\sqrt{\frac{\Delta t}{\tau}}C_1(\kappa+1)) \le& 1
\\C_1(\kappa+1)< & \sqrt{2} <& 2\sqrt{\frac{\tau}{\Delta t}}
\end{align}
Going back to $\varepsilon(t)\ge0$:
\begin{equation}
    \varepsilon(t) = \frac{\kappa}{C_1} \sqrt{\frac{\tau}{\Delta t}}2(Z(t) - \frac{1}{2})-(\kappa+1)
\end{equation}
\begin{align}
   0  \le& -\frac{\kappa}{C_1} \sqrt{2}-(\kappa+1) \le& -\frac{\kappa}{C_1} \sqrt{\frac{\tau}{\Delta t}}2-(\kappa+1) \le& \frac{\kappa}{C_1} \sqrt{\frac{\tau}{\Delta t}}2(Z(t) - \frac{1}{2})-(\kappa+1) \le& \frac{\kappa}{C_1} \sqrt{\frac{\tau}{\Delta t}}2-(\kappa+1)
\end{align}
where the leftmost inequality contradicts with our assumption that $\kappa$ is positive.
\subsubsection{Logisitic Map with Coefficient in [0,4]}
Alternatively, we may seek:
\begin{equation}
    0< -2\kappa C_1\sqrt{\frac{\Delta t}{\tau}} < 4
\end{equation}
\begin{align}
    0< -\kappa C_1 < \sqrt{2}
\end{align}
which is only achievable with $\kappa < 0$. The constraint on the initial condition:
\begin{equation}
    0 < Z(t = 0) < 1
\end{equation}
yields:
\begin{align}
0 <& \frac{1}{2}(1+\sqrt{\frac{\Delta t}{\tau}}C_1(\kappa+1)) <& 1
\\-1 <& \sqrt{\frac{\Delta t}{\tau}}C_1(\kappa+1) <& 1
\end{align}
\begin{align}
\abs{\sqrt{\frac{\Delta t}{\tau}}C_1(\kappa+1)} <& 1
\\\abs{\sqrt{\frac{\Delta t}{\tau}}C_1(\kappa+1)} <& 1
\\\abs{\kappa} <& \sqrt{\frac{\tau}{\Delta t}}\frac{1}{C_1}-1
\end{align}
The requirement that $\varepsilon(t)\ge 0$ gives:
\begin{align}
    -\frac{1}{C_1} \sqrt{\frac{\tau}{\Delta t}}\kappa -(\frac{C_0}{C_1}+\varepsilon_N)\ge&0
    \\\sqrt{\frac{\Delta t}{\tau}}(C_0+C_1\varepsilon_N) \le& \abs{\kappa}
\end{align}
To summarize, our choice of $C_0$, $C_1$ and $\Delta t$ is determined by the inequality:
\begin{equation}
\sqrt{\frac{\Delta t}{\tau}}(C_0+C_1\varepsilon_N) \le \frac{1+ \sqrt{1+4(\frac{C_0}{C_1}+\varepsilon_N)}}{2} < \sqrt{\frac{\tau}{\Delta t}}\frac{1}{C_1}-1
\end{equation}

To confirm the physicality of the constraints, we note that for $\Delta t \rightarrow 0$ and/or $\tau \rightarrow +\infty $, or in other words $Re \rightarrow 0$, the inequality is satisfied for any choice of $C_0$ and $C_1$, and the error term maintains its null initial value. However, for $\frac{\Delta t}{\tau} = O(1)$, the inequality is impossible
\section{Binary Representation of Fock Space}
\subsection{Mapping of Bosonic Operators to Qubit Gates}
For a scalable implementation, we seek to utilize the binary number representation for the occupation number basis. In addition, it is desirable to have the bosonic operators written in terms of standard qubit gates for implementation. As such, we make use of the expression of \cite{liScalableBosonEncoding2020} writing the truncated operators in terms of linear combination of Pauli words (tensored matrices) based on the Pauli matrices \ standard qubit gates:
\begin{equation}
    \hat{X}_{k,j} = \begin{pmatrix}
        0 & 1 \\ 1 & 0
    \end{pmatrix}
    \\ \hat{Y}_{k,j} = \begin{pmatrix}
        0 & -i \\ i & 0
    \end{pmatrix}
    \\ \hat{Z}_{k,j} = \begin{pmatrix}
        1 & 0 \\ 0 & -1
    \end{pmatrix}
\end{equation}
where the subindex $(k,j)$ refers to the gate acting on the $k^{th}$ qubit of the $j^{th}$ register. 
such that:
\begin{equation}
    \hat{\sigma}^\pm_{k,j} = \frac{1}{2}(\hat{X}_{k,j}\pm i\hat{Y}_{k,j})
\end{equation}
In the binary number representation, each register, representing a bosonic mode, would need to have a qubit count $qc$ logarithmic in the decimal number it encodes, the number of excitation levels $N$:
\begin{equation}
    qc = \ceil{\log_2{(N+1)}}
\end{equation}
where $N+1$ appears rather than $N$ to account for the ground level corresponding to decimal zero. Throughout the paper, we assume that $N$ is adequately chosen to make full use of the number of qubits it would require, that is:
\begin{equation}
\begin{aligned}
    qc &= \ceil{\log_2{(N+1)}} = \log_2{(N+1)}
    \\N &= 2^{qc}-1
    \end{aligned}
\end{equation}
\begin{equation}
    \begin{aligned}
    \label{a_map}
    \hat{a}_i^\dagger &= \sqrt{\frac{N+1}{2}}
    &{\Sigma}_{j=0}^{q_c-1} (\sqrt{\frac{\hat{I}}{2^j}+{\Sigma}_{k=0}^{j-1}(\frac{\hat{n}_{k,i}}{2^k})}
    \hat{\sigma}_{j,i}^+\bigotimes_{k=j+1}^{q_c-1}(\hat{\sigma}_{k,i}^-))
    \end{aligned}
\end{equation}
\begin{equation}
\label{adag_map}
    \begin{aligned}
    \hat{a}_i &= \sqrt{\frac{N+1}{2}} {\Sigma}_{j=0}^{q_c-1} (\sqrt{\frac{\hat{I}}{2^j}+{\Sigma}_{k=0}^{j-1}(\frac{\hat{n}_{k,i}}{2^k})}
    \hat{\sigma}_{j,i}^-\bigotimes_{k=j+1}^{q_c-1}(\hat{\sigma}_{k,i}^+))
    \end{aligned}
\end{equation}

The position and momentum operators can then be represented as:
\begin{equation}
\begin{aligned}
    \hat{q}_i &=\frac{\sqrt{\ceil{N+1}}}{2}
    {\Sigma}_{j=0}^{q_c-1} (\sqrt{\frac{\hat{I}}{2^j}+{\Sigma}_{k=0}^{j-1}(\frac{\hat{n}_{k,i}}{2^k})} \hat{X}_{j,i} \bigotimes_{k=j+1}^{q_c-1}(\hat{X}_{k,i}))
\end{aligned}
\end{equation}

and:
\begin{equation}
\begin{aligned}
    \hat{p}_i &=\frac{\sqrt{\ceil{N+1}}}{2}
    {\Sigma}_{j=0}^{q_c-1} (\sqrt{\frac{\hat{I}}{2^j}+{\Sigma}_{k=0}^{j-1}(\frac{\hat{n}_{k,i}}{2^k})} \hat{Y}_{j,i} \bigotimes_{k=j+1}^{q_c-1} (i\hat{Y}_{k,i}))
    \end{aligned}
\end{equation}
written as a linear combination of unitary gates. This representation achieves truncated operators:
\begin{equation}
    \hat{a} = \begin{pmatrix}
              0 & 0 & 0 & 0 & \dots & 0 & 0
        \\ \sqrt{1} & 0 & 0 & 0 & \dots & 0 & 0
        \\ 0 & \sqrt{2} & 0 & 0 & \dots & 0 & 0
        \\ 0 & 0 & \sqrt{3} & 0 & \dots & 0 & 0
        \\ \vdots & \vdots & 0 & 0 & \ddots & \dots & 0
        \\ 0 & 0 & \dots & 0 & \sqrt{N-1} & 0 & 0
        \\ 0 & 0 & 0 & \dots & 0 & \sqrt{N} & 0
    \end{pmatrix}
    \label{amatrix}
\end{equation}

The representation is not complete without noting:
\begin{equation}
\begin{aligned}
    \\ \hat{n} &= \begin{pmatrix}
        0 & 0 \\ 0 & 1
    \end{pmatrix}
    \\ \hat{n}_{k,i} &= \frac{1}{2}(\hat{I}-\hat{Z}_{k,i})
\end{aligned}
\end{equation}
such that it is real-valued and diagonal, and their linear combinations with the identity operator remains so under the diagonal. Since we are using them for Hamiltonian simulation, we may do without expressing the term under the square root as an operator, and use explicit coefficients instead.

The position and momentum operators are approximated as a linear combination of $O(qc) = O(\ceil{\log_2{(N+1)}}^2)$ Pauli words, consisting of $O(qc)= O(\ceil{\log_2{(N+1)}})$ tensored Pauli operators, with coefficients $O(\sqrt{2(N+1)})$. The Pauli words have a $2$-sparsity (The Hamiltonian for the lattice Boltzmann formulation, which has second-order nonlinearity, then has a sparsity (or lack thereof) of $(2^{(2+1)})^{qc} = 8^{qc}$.

\subsection{Hermiticity}
The truncated position and momentum operator are still Hermitian:
\begin{equation}
    \hat{q}^\dagger=\hat{q}
\end{equation}
and
\begin{equation}
    \hat{p}^\dagger=\hat{p}
\end{equation}

\subsection{Commutation Relations}
It follows from the definition of the truncated ladder operators that the commutation relations are:
\begin{equation}
\begin{aligned}
    [\hat{q},\hat{p}] &= i[\hat{a},\hat{a}^\dagger] = i\begin{pmatrix}
        1 & 0 & 0 & 0 & \dots & 0 & 0
        \\ 0 & 1 & 0 & 0 & \dots & 0 & 0
        \\ 0 & 0 & 1 & 0 & \dots& 0 & 0
        \\ 0 & 0 & 0 & 1 & 0 & \dots & 0
        \\ \vdots & \vdots & \vdots & \ddots & \ddots & \ddots & \vdots
        \\ 0 & 0 & 0 & \dots & 0 & 1 & 0
        \\ 0 & 0 & 0 & \dots & 0 & 0 & -N
    \end{pmatrix}
    \\&= i(\hat{I}-(N+1)\ket{N}\bra{N})
    \end{aligned}
\end{equation}
More explicitly, we can write:
\begin{equation}
    \begin{aligned}
        [\hat{q},\hat{p}]&= i(\hat{I}-(N+1)\bigotimes_{k=0}^{qc-1}\hat{n}_{k,i})
    \end{aligned}
\end{equation}
\subsection{Momentum Operator in Position Space}
The truncated operators warrant revisiting the momentum operator in position space. For a general state $\ket{\psi}$
\begin{equation}
    \begin{aligned}
\bra{q'}[\hat{q},\hat{p}]\ket{\psi} &= \int_{-\infty}^\infty dq \bra{q'}(i-(N+1)\ket{N}\bra{N})\ket{q}\bra{q}\ket{\psi} 
\\&= \int_{-\infty}^{\infty} dq (q'-q)\bra{q'}\hat{p}\ket{q}\bra{q}\ket{\psi}
\end{aligned}
\end{equation}

\begin{equation}
    \begin{aligned}
        \\(q'-q)\bra{q'}\hat{p}\ket{q} &= (i\delta(q-q')-(N+1)\bra{q'}\ket{N}\bra{N})\ket{q}
    \end{aligned}
\end{equation}

\begin{equation}
    \begin{aligned}
-i\delta(q-q')/(q'-q) &= \bra{q'}(\hat{p}+(N+1)\ket{N}\bra{N})\ket{q}
\\i \frac{\partial\delta(q-q')}{\partial \hat{q}} &= \bra{q'}(\hat{p}+(N+1)\ket{N}\bra{N})\ket{q}
    \end{aligned}
\end{equation}
\begin{equation}
    \begin{aligned}
\\\bra{q'}(\hat{p}+(N+1)\ket{N}\bra{N})\ket{q}\ket{\psi} &= \int_{-\infty}^{\infty}dq \frac{\partial\delta(q-q')}{\partial \hat{q}} \bra{q}\ket{\psi}
\\&= -i \frac{\partial \psi(q)}{\partial \hat{q}}|_{q=q'}
    \end{aligned}
\end{equation}
\subsection{Eigenvalues \& Eigenvectors}
\subsubsection{Lowering \& Raising Operators}
Following the definition of the truncated ladder operators, the operators, $\hat{a}$ and $\hat{a}^\dagger$, are nilpotent:
\begin{equation}
    \hat{a}^k=(\hat{a}^\dagger)^k = 0 \; \text{for} \; k = 2^{qc}
\end{equation}
\subsubsection{Position Operator}
$\hat{q}$ and $\hat{p}$, on the other hand, are symmetric tridiagonal matrices with zero diagonal entries, and non-zero off-diagonal terms. They could be diagonalized in an orthonormal basis, which forms our computational basis.

\paragraph{Eigenvalues}
Since $\hat{q}$ is Hermitian, it follows that there exists an operator $\hat{O}$
\begin{equation}
    \hat{O}^{-1} \hat{q} \hat{O} = \hat{\lambda}
\end{equation}
where $\lambda$ is formed by a diagonal containing the eigenvalues of $\hat{q}$,such that:
\begin{equation}
    \hat{O}^\dagger = \hat{O}^{-1}
\end{equation}
\subparagraph{Eigenvalues of Untruncated Operator}
The untruncated `position operator corresponds to a Jacobi matrix, with its characteristic polynomial being the monic Hermite polynomial, such that the eigenvalues are the roots of the Hermite polynomial:
\begin{equation}
    H_n(\lambda) = (-1)^ne^{\frac{1}{2}\lambda^2}\frac{d^n}{d\lambda^n}e^{-\frac{1}{2}\lambda^2}
\end{equation}
We are interested in seeking the eigenvalues of the truncated operator. 
\subparagraph{Three-Term Recurrence Relation for Characteristic Polynomial}
For a general symmetric tridiagonal matrix:
\begin{equation}
A = 
    \begin{pmatrix}
        a_1 & b_1 & 0 & 0 & 0 & 0 & \dots & 0
        \\b_1 & a_2 & b_2 & 0 & 0 & 0 & \dots & 0
        \\ 0 & b_2 & a_3 & b_3 & 0 & 0 & \dots & 0
        \\ 0 & 0 & b_3 & a_4 & b_4 & 0 & \dots & 0
        \\ 0& 0 & 0 & b_4 & a_5 & b_5  & \dots & 0
        \\ 0 & 0 & 0 & \ddots & \ddots & \ddots & \ddots & \vdots
        \\0 & 0 & \dots & 0 &0 & b_{n-2} & a_{n-1} & b_{n-1}
        \\0 & 0 & 0 & 0 & 0 & 0 & b_{n-1} & a_n
    \end{pmatrix}
\end{equation}
the eigenvalues are roots of the characteristic polynomial $P_n(\lambda)=det(A_n-\lambda I_n)$ obeying the recurrence relation:
\begin{equation}
    P_j(\lambda) = (a_j-\lambda)P_{j-1}(\lambda)-b^2_{j-1}P_{j-2}(\lambda) \; 2\le j \le n
\end{equation}
with
\begin{align}
    b_0 &= 0
    \\P_0(\lambda) &= 1
    \\P_1(\lambda) &= (a_1-\lambda)
\end{align}
where $P_j(\lambda)$ is the charactersitic polynomial of the $j^{th}$ principal minor of $A$.

Since we are interested in normalized eigenvectors, we can study the eigenvalues of $\sqrt{2}\hat{q}$. In this case, $a_j = 0$, and $b_j^2 = j-1$, such that we have:
\begin{equation}
    P_j(\lambda) = (-\lambda)P_{j-1}(\lambda)-(j-1)P_{j-2}(\lambda) \; 2\le j \le n
\end{equation}
with
\begin{align}
    b_0 &= 0
    \\P_0(\lambda) &= 1
    \\P_1(\lambda) &= -\lambda
\end{align}
\subparagraph{Generating Function of Characteristic Polynomial}
Let $g_{\lambda}(x)$ be the generating function of $P_n(\lambda)$:
\begin{equation}
    g_{\lambda}(x) = \Sigma_{n=0}^\infty P_n(\lambda)x^n
\end{equation}
\begin{align}
        g_{\lambda}(x) =& 1-\lambda x+
        \Sigma_{n=2}^\infty ((-\lambda)P_{n-1}(\lambda)-(n-1)P_{n-2}(\lambda))x^n
        \\=& 1-\lambda x
        +\Sigma_{n=2}^\infty (-\lambda)P_{n-1}(\lambda)x^n-\Sigma_{n=2}^\infty(n-1)P_{n-2}(\lambda)x^n
        \\=& 1-\lambda x-\lambda x(-1+1+\Sigma_{n=1}^\infty P_{n}(\lambda)x^n)
        +-\Sigma_{n=2}^\infty(n-1)P_{n-2}(\lambda)x^n
        \\=& 1-\lambda x-\lambda x(g_{\lambda}(x)-1)
        -x^2\Sigma_{n=0}^\infty(n+1)P_{n}(\lambda)x^{n}
        \\=& 1-\lambda x-\lambda x(g_{\lambda}(x)-1)
        -x^2(g_{\lambda}(x)+x\Sigma_{n=0}^\infty(n)P_{n}(\lambda)x^{n-1})
        \\=& 1-\lambda x-\lambda x(g_{\lambda}(x)-1)-x^2(g_{\lambda}(x)+xg'(x))
\end{align}
\vspace{-1cm}
\begin{align}
g(x)=& 1-((\lambda x+x^2)g_{\lambda}(x)+x^3g'(x))
\\x^3 g'(x) =& -(1+\lambda x + x^2)g(x)+1
\end{align}
The nonlinear ordinary differential equation yields a solution:
\begin{equation}
\begin{aligned}
    g_{\lambda}(x) =& e^{-2x^2-\frac{x}{\lambda}}\frac{1}{x}(c+\int_1^x e^{- \frac{2\lambda z+1}{2z^2}}z^{-2}dz)
    \\=& e^{-2x^2-\frac{x}{\lambda}}\frac{1}{x}(c-\sqrt{\frac{\pi}{2}}e^{\frac{1}{2}\lambda^2}erf(\frac{1}{\sqrt{2}}(\lambda+\frac{1}{z}))|_{z=1}^{z=x})
    \\=& e^{-2x^2-\frac{x}{\lambda}}\frac{1}{x}\sqrt{\frac{\pi}{2}}e^{\frac{1}{2}\lambda^2}(C-erf(u)|_{u=\frac{\lambda+1}{\sqrt{2}}}^{u=\frac{\lambda+\frac{1}{x}}{\sqrt{2}}})
    \\=& e^{1/2(\lambda^2+2\frac{\lambda}{x}+\frac{1}{x^2})}\frac{1}{x}\sqrt{\frac{\pi}{2}}(C-erf(u)|_{u=\frac{\lambda+1}{\sqrt{2}}}^{u=\frac{\lambda+\frac{1}{x}}{\sqrt{2}}})
    \\=& e^{\frac{1}{2}(\lambda+\frac{1}{x})^2}\frac{1}{x}\sqrt{\frac{\pi}{2}}(C-erf(\frac{\lambda+\frac{1}{x}}{\sqrt{2}})+erf(\frac{\lambda+1}{\sqrt{2}})))
    \\=& \sqrt{\frac{\pi}{2}}e^{\frac{1}{2}(\lambda+\frac{1}{x})^2}\frac{1}{x}(C-erf(\frac{\lambda+\frac{1}{x}}{\sqrt{2}}))
    \\+&\sqrt{\frac{\pi}{2}}e^{\frac{1}{2}(\lambda+\frac{1}{x})^2}\frac{1}{x}erf(\frac{\lambda+1}{\sqrt{2}})
\end{aligned}
\end{equation}

We evaluate its derivative to determine the constant:
\begin{equation}
    \begin{aligned}
        g'(x)=& \sqrt{\frac{\pi}{2}}e^{\frac{1}{2}(\lambda+\frac{1}{x})^2}\frac{1}{x^3} (-(\lambda+\frac{1}{x}+x)(C-erf(\frac{\lambda+\frac{1}{x}}{\sqrt{2}}))
        \\-&(\lambda+\frac{1}{x}+x)erf(\frac{\lambda+1}{\sqrt{2}})
        +\frac{1}{x^3}
    \end{aligned}
\end{equation}
where we want to put it into the form of the underlying nonlinear differential equation:
\begin{widetext}
\begin{equation}
    \begin{aligned}
        x^3g'(x)=& (-(\lambda x+1+x^2)\sqrt{\frac{\pi}{2}}e^{\frac{1}{2}(\lambda+\frac{1}{x})^2}\frac{1}{x}(C-erf(\frac{\lambda+\frac{1}{x}}{\sqrt{2}})+erf(\frac{\lambda+1}{\sqrt{2}}))+1
        \\=&-(\lambda x+1+x^2)g(x)+1
    \end{aligned}
\end{equation}
\end{widetext}

with $g_\lambda(x\rightarrow0) \rightarrow P_0(\lambda) = 1$, and $g'_\lambda(x\rightarrow0) \rightarrow P_1(\lambda) = -\lambda$, such that:
\begin{equation}
    C = \pm1 -erf(\frac{\lambda+1}{\sqrt{2}})
\end{equation}
yielding:
\begin{equation}
    \begin{aligned}
        g_\lambda(x)=& \sqrt{\frac{\pi}{2}}e^{\frac{1}{2}(\lambda+\frac{1}{x})^2}\frac{1}{x}(\pm1-erf(\frac{\lambda+\frac{1}{x}}{\sqrt{2}}))
    \end{aligned}
\end{equation}
We now perform a change of variables, $u = \frac{1}{\sqrt{2}}(\lambda+\frac{1}{x})$:
\begin{equation}
    \begin{aligned}
        g_\lambda(u)=& \sqrt{\frac{\pi}{2}}e^{u^2}(\sqrt{2}u-\lambda)(\pm1-erf(u))
    \end{aligned}
\end{equation}
Using the Taylor series of the error function:
\begin{widetext}
\begin{equation}
    \begin{aligned}
        erf(u) =& \pm 1 +\Sigma_{n=1}^\infty \frac{d^n erf(u)}{du} \frac{u^n}{n!}
        \\=& \pm 1 +\Sigma_{n=0}^\infty \frac{d^{n+1} erf(u)}{du} \frac{u^{n+1}}{{n+1}!}
        \\=& \pm 1 +\Sigma_{n=0}^\infty \frac{2}{\sqrt{\pi}}(-1)^n H_{n}(u)e^{-u^2} \frac{u^{n+1}}{{n+1}!}
    \end{aligned}
\end{equation}
\end{widetext}
we write:
\begin{widetext}
\begin{equation}
    \begin{aligned}
        g_\lambda(u)=& \sqrt{2}(\sqrt{2}u-\lambda)(-\Sigma_{n=0}^\infty (-1)^n H_{n}(u) \frac{u^{n+1}}{{n+1}!})
        \\=& \sqrt{2}(\sqrt{2}u-\lambda)(-\Sigma_{n=0}^\infty H_{n}(-u) \frac{u^{n+1}}{{n+1}!})
    \end{aligned}
\end{equation}
\end{widetext}
We now reverse the change of variables and expand the Hermite polynomials and the binomial expression:
\begin{widetext}
\begin{equation}
    \begin{aligned}
        g_\lambda(x)=& \sqrt{2}\frac{1}{x}\Sigma_{n=0}^\infty \frac{(-1)^{n+1}}{(n+1)!} (\Sigma_{m=0}^n{\binom{n}{m}}H_{n}(\frac{\lambda}{\sqrt{2}})2^{\frac{1}{2}(n-m)}x^{m-n})2^{-\frac{1}{2}(n+1)} \Sigma_{k=0}^{n+1}\binom{n+1}{k}x^{k-n-1}\lambda^{k}
        \\=& \Sigma_{n=0}^\infty\Sigma_{m=0}^n \Sigma_{r=0}^{\lfloor \frac{n}{2}\rfloor} \Sigma_{k=0}^{n+1} \frac{(-1)^{n+1}}{(n+1)!} \binom{n}{m}\binom{n}{2r}\binom{n+1}{k} \frac{(2r)!}{r!} 2^{\frac{1}{2}(n-m-2r)}H_{n-2i}(\lambda)\lambda^{k}x^{k+m-2(n+1)}
    \end{aligned}
\end{equation}
\end{widetext}
As a result, we see that the eigenvalues of the position operator no longer form roots of Hermite polynomials, and the characterstic polynomial of the truncated position eigenvector only corresponds to the untruncated case for $N \rightarrow \infty$. 

\paragraph{Eigenvectors}
\label{sec:eigvec}
We use the recurrence relations for eigenvectors of symmetric tridiagonal matrices \cite{godunovComputationEigenvectorSymmetric1986} to determine the components of the eigenvector $\ket{q_\lambda}$, corresponding to an eigenvalue $\lambda$ of the operator $\hat{q}$, in the computational basis:
\begin{equation}
\begin{aligned}
    0 &= -\lambda \bra{0}\ket{q_\lambda}+\frac{1}{\sqrt{2}}\bra{1}\ket{q_\lambda}
    \\ 0 &= \sqrt{\frac{n-1}{2}}\bra{n-2}\ket{q_\lambda}-\lambda \bra{n-1}\ket{q_\lambda} + \sqrt{\frac{n}{2}}\bra{n}\ket{q_\lambda}
    \;,\\&\; \forall \; n \in [2,N]
    \\ 0 &= \sqrt{\frac{N}{2}}\bra{N-1}\ket{q_\lambda}-\lambda \bra{N}\ket{q_\lambda}
    \end{aligned}
\end{equation}

which can 
where the last equation is automatically satisfied for $\lambda$ being an eigenvalue, and is not used, leaving a degree of freedom in choosing one of the components, typically taken to be the first. The resulting eigenvectors do not necessarily satisfy the normalization condition, requiring to be appropriately rescaled by their L2 norms a posteriori.

We may write:
    \begin{equation}
\begin{aligned}
    0 &= \sqrt{\frac{n-1}{2}}\bra{n-2}\ket{q_\lambda}-\lambda \bra{n-1}\ket{q_\lambda} + \sqrt{\frac{n}{2}}\bra{n}\ket{q_\lambda}
    \;,\\&\; \forall \; n \in [1,N]
    \end{aligned}
\end{equation}
noting that $n-1 =0 \; \text{for} \; n = 1$.

\subsection{Zero Vectors}
To initialize the system, one might be interested to start from a definite zero occupation number or position, before applying a displacement or translation operator respectively. For the lowering operator, this is straightforward, as the desired vector corresponds to the zero in computational basis:
\begin{equation}
    \hat{a}^n \ket{0} = 0 \; \forall \; n \in [1,\infty)
\end{equation}

However, for the position operator, it is more tricky to construct such a vector not corresponding to the trivial solution. However, one may choose a vector where the only occupied level corresponds to the excitation level which number is the truncation number:
\begin{equation}
    \ket{N}
\end{equation}
such that it is asymptotically a zero vector:
\begin{equation}
    \ket{N} \rightarrow 0 \; \text{for} \; N \rightarrow \infty
\end{equation}
This yeilds:
\begin{equation}
    \hat{q}^n \ket{N} \neq 0\ket{m} \; \forall \; m \in [N-n,N]
\end{equation}

Consider the fact that the eigenstates of the untruncated position operator correspond to a Dirac delta in position space. Then, constructing eigenstates for the truncated position operator correspond to finding a finite sum representation of a Dirac delta function. In other words, we ought to use polynomials orthogonal to a linear functional on a finite domain, which we consider in Appendix~\ref{sec:finite}.

\section{Finite Position Embedding}
\label{sec:finite}
\subsection{Truncated Hermite Polynomials}
We survey the definition and properties of the truncated Hermite polynomials as introduced by \cite{dominiciTruncatedHermitePolynomials2022} before using them in our linear embedding scheme.
\subsubsection{Linear Functional}
 Consider the truncated Hermite polynomials $P_n(x,z)$ \cite{dominiciTruncatedHermitePolynomials2022} orthogonal with respect to the linear functional:
\begin{equation}
    L[p] = \int_{-z}^z p(x)e^{-x^2}dx \; p \in \mathcal{R}[x], \; z>0
\end{equation}

The linear functional satisifies the Pearson equation:
\begin{equation}
    L[\partial_X((x^2-z^2)p)] = L[2x(x^2-z^2)p]
\end{equation}
which is equivalent to:
\begin{equation}
    L[(x^2-z^2)\partial_xp] = L[2x(x^2-z^2-1)p]
\end{equation}
\subsubsection{Recurrence Relation}
The truncated Hermite polynomials satisfy the recurrence relation:
\begin{equation}
    P_{z,n+1}(x) = xP_{z,n}-\gamma_{n}(z)P_{z,n-1}(x) \; n \ge 0 \label{eq:rec}
\end{equation}
with $P_{z,-1} = 0$.

\subsubsection{Decay Coefficient}
The coefficient $\gamma_n(z)$ satisfies:
\begin{equation}
    z \partial_z ln(\gamma_n) = 2(\gamma_{n-1}-\gamma_{n+1}+1)
\end{equation}

Alternatively, they may be defined by the Laguerre-Freud equation they satisfy:
\begin{eqnarray}
    \frac{z^2}{2} = &\gamma_n (\gamma_{n-1}+\gamma_n-z^2+\frac{1}{2}-n)
    \\&-\gamma_{n+1}(\gamma_{n+1}+\gamma_{n+2}-z^2-n-\frac{3}{2})
\end{eqnarray}
or the equation:
\begin{equation}
    \gamma_n (n+\frac{1}{2}-\gamma_n-\gamma_{n+1})(n-\frac{1}{2}-\gamma_n-\gamma_{n-1}) = z^2(\frac{n}{2}-\gamma_n)^2
\end{equation}
Additionally, one may define:
\begin{equation}
    g_n(z) = \frac{n}{2} - \gamma_n(z)
\end{equation}
satisfying:
\begin{equation}
    (\frac{n}{2}-g_n)(g_n+g_{n+1})(g_n+g_{n-1}) = z^2g_n^2
\end{equation}
\subsubsection{Symmetry}
Since the $n^{th}$ polynomial only appears weighed by $x$ in the recurrence relation, then:
\begin{equation}
    P_{z,n}(-x) = (-1)^nP_{z,n}(x)
\end{equation}
\subsubsection{Differential-Recurrence Relation}
The derivative of $P_{z,n}(x)$ with respect to $x$ is related to the polynomial through:
\begin{equation}
    (x^2-z^2)\partial_x P_{z,n+1} = (n+1)P_{z,n+2}+\lambda_n P_{z,n}+\tau_nP_{z,n-2}\label{eq:dif-rec}
\end{equation}
for:
\begin{equation}
    \lambda_n(z) = (2(\gamma_n+\gamma_{n+1}+\gamma_{n+2}-z^2-1)-n)\gamma_{n+1}
\end{equation}
and:
\begin{equation}
    \tau_n(z) = 2\gamma_{n+1}\gamma_n \gamma_{n-1}
\end{equation}
\subsubsection{Ladder Operators}
\paragraph{Lowering Operator}
Let:
\begin{eqnarray}
    A_{z,n}(x) =& \frac{x^2-z^2}{2\gamma_n(z) C_{z,n}(x)}
    \\B_{z,n}(x) =& \frac{n-2\gamma_n(z)}{2\gamma_n(z) C_{z,n}(x)} x
    \\C_{z,n}(x) =& x^2-z^2+\gamma_n(z)+\gamma_{n+1}(z)-n-\frac{1}{2}
\end{eqnarray}
Then, one may speak of a lowering "operator":
\begin{equation}
\label{eq:lower}
    U_{z,n} = A_{z,n}(x) \partial_x - B_{z,n}(x)
\end{equation}
acting on the polynomial $P_{z,n}(x)$ as:
\begin{equation}
    U_{z,n}P_{z,n}(x) = P_{z,n-1} \; \forall n \in \mathcal{N}
\end{equation}

\subsection{Linear Embedding in Infinite Hilbert Space}
In the following section, we set $z = 1$ to simplify the issue of normalizing the quantum statevector down the road.
\subsubsection{Finite Position State}
For a state $\ket{x}$:
\begin{equation}
    \braket{n}{x} = P_n(x) \; \forall n \in \mathcal{N} \label{eq:def}
\end{equation}
with:
\begin{equation}
    P_0(x) = 1,
\end{equation}
we define an operator:
\begin{equation}
    \hat{q} \ket{x} = x \ket{x} \; \forall x \in [-1,1]
\end{equation}
We say that the state $\ket{x}$ is a finite position eigenstate with an eigenvalue $x$ with respect to the finite position operator $\hat{q}$. 
\subsubsection{Position Operator}
Using the definition of the finite position state, Eq.\ref{eq:def}, and the recurrence relation, Eq.~\ref{eq:rec}, we may write:
\begin{eqnarray}
    \hat{q} \ket{x} = &x \ket{x} 
    \\= &x \Sigma_{n=0}^\infty P_n(x) \ket{n} 
    \\= &\Sigma_{n=0}^\infty (P_{n+1}(x)+\gamma_{n}P_{n-1}(x))\ket{n}
    \\= &\Sigma_{n=0}^\infty P_{n+1}(x)\ket{n}+\Sigma_{n=0}^\infty\gamma_{n}P_{n-1}(x)\ket{n}
    \\= &\Sigma_{n=0}^\infty P_{n+1}(x)\ket{n}+\Sigma_{n=1}^\infty\gamma_{n}P_{n-1}(x)\ket{n}
    \\= &\Sigma_{n=0}^\infty P_{n+1}(x)\ket{n}
    \\&+\Sigma_{n=0}^\infty\gamma_{n+1}P_{n}(x)\ket{n+1}
\end{eqnarray}
where we have used the fact that $P_{-1}(x) = 0$. Now consider two operators:
\begin{eqnarray}
    \hat{a}^{-} =& \Sigma_{n=1}^\infty \ket{n-1}\bra{n}
    \\ \hat{a}^{+} =& \Sigma_{n=0}^\infty \ket{n+1}\bra{n}
    \\ \hat{\gamma} =& \Sigma_{n=0}^\infty \gamma_{n}\ket{n}\bra{n}
\end{eqnarray}
such that:
\begin{eqnarray}
    \hat{a}^{-} \ket{n} =& \ket{n-1}
    \\ \hat{a}^{+} \ket{n} =& \ket{n+1}
    \\ \hat{\gamma} \ket{n} =& \gamma_n \ket{n}
\end{eqnarray}
Therefore, the action of the position operator on a position eigenstate may be described in terms of those two operators:
\begin{eqnarray}
    \hat{q} \ket{x} = &x \ket{x}
    \\= &\Sigma_{n=0}^\infty P_{n+1}(x)\ket{n}
    \\&+\Sigma_{n=0}^\infty\gamma_{n+1}P_{n}(x)\ket{n+1}
    \\= &(\hat{a}^{-}+\hat{\gamma}\hat{a}^{+})\ket{x}
\end{eqnarray}
\paragraph{Hermiticity}
Unlike the \textit{classical} quantum mechanical position operator, $\hat{q}$ here is not Hermitian.
Consider the conjugate transpose of the operators:
\begin{eqnarray}
    (\hat{a}^{-})^\dagger = &\hat{a}^+
    \\(\hat{a}^{+})^\dagger = &\hat{a}^-
    \\ \hat{\gamma}^\dagger = &\hat{\gamma}
\end{eqnarray}
The conjugate transpose of the finite position operator could be written as:
\begin{equation}
    \hat{q}^\dagger = \hat{a}^+ + \hat{a}^-\hat{\gamma}
\end{equation}
\paragraph{Inversion}
We define the inverses:
\begin{eqnarray}
    (\hat{a}^{-}) ^{-1} =& \hat{a}^{+}
    \\(\hat{a}^{+}) ^{-1} =& \hat{a}^{-}
    \\\hat{\gamma}^{-1} =& \Sigma_{n=0}^\infty \gamma_n^{-1}\ket{n}\bra{n}
\end{eqnarray}
the last one is straightforward since $\hat{\gamma}$ is diagonal, while the first two are intuitive, and we verify:
\begin{eqnarray}
    \hat{a}^+\hat{a}^- = &(\Sigma_{m=0}^\infty \ket{m+1}\bra{m})(\Sigma_{n=1}^\infty \ket{n-1}\bra{n}
    \\= &\Sigma_{n=0}^\infty \ket{n}\bra{n} = \hat{I}
\end{eqnarray}
We argue that for an operator $\hat{q}$:
\begin{equation}
    \hat{q}\ket{x} = x \ket{x}
\end{equation}
if there exists $\hat{q}^{-1}$, then:
\begin{equation}
    \hat{q}^{-1}\hat{q} \ket{x} = \hat{I}\ket{x} = \hat{q}^{-1}x\ket{x}
\end{equation}
such that:
\begin{equation}
    \hat{q}^{-1}\ket{x} = \frac{1}{x}\ket{x}
\end{equation}
\subsubsection{Translation}
In order to use the said polynomials for embedding a dynamical system into the Hilbert space, we must be able to initialize a state $\ket{x(t=0)}$ from $\ket{0}$, then further \textit{translate} it by $\Delta t \Omega(x)$ where:
\begin{equation}
    \partial_t f(x) = \Omega(x)
\end{equation}
We start by restating the differential-recurrence relation, Eq.\ref{eq:dif-rec}:
\begin{equation}
    (x^2-1)\partial_x P_{n}(x) = nP_{n+1}(x)+\lambda_{n-1} P_{n-1}(x)+\tau_{n-1}P_{n-3}(x)
\end{equation}
with:
\begin{equation}
    \lambda_{n-1} = (2(\gamma_{n-1}+\gamma_{n}+\gamma_{n+1}-2)-n)\gamma_{n}
\end{equation}
and:
\begin{equation}
    \tau_{n-1} = 2\gamma_{n}\gamma_{n-1} \gamma_{n-2}
\end{equation}
\begin{widetext}
\begin{eqnarray}
    (x^2-1)\partial_x P_{n}(x) = &nP_{n+1}(x)
    \\&+(2(\gamma_{n-1}+\gamma_{n}+\gamma_{n+1}-2)-n)\gamma_{n} P_{n-1}(x)
    \\&+2\gamma_{n}\gamma_{n-1} \gamma_{n-2}P_{n-3}(x)
\end{eqnarray}
\end{widetext}
In a more straightforward manner, we may utilize the lowering operator defined in Eq.~\ref{eq:lower} in terms of:
\begin{eqnarray}
    A_{n}(x) =& \frac{x^2-1}{2\gamma_n(1) C_{n}(x)}
    \\B_{n}(x) =& \frac{n-2\gamma_n(1)}{2\gamma_n(1) C_{n}(x)} x
    \\C_{n}(x) =& x^2-1+\gamma_n(1)+\gamma_{n+1}(1)-n-\frac{1}{2}
\end{eqnarray}
such that the "lowering operator":
\begin{equation}
    U_{n}(x) = A_{n}(x) \partial_x - B_{n}(x)
\end{equation}
acts on the polynomial $P_n(x)$:
\begin{equation}
    U_{n}(x) P_n(x) = P_{n-1}(x)
\end{equation}
Rearranging the expression, we have:
\begin{equation}
    \partial_x P_n(x) = \frac{B_n(x)}{A_n(x)}P_n(x)+\frac{1}{A_n(x)}P_{n-1}(x)
\end{equation}
As such, the modified embedding is likely to exacerbate errors rather than provide a closure due to the appearance of $x$ in the coefficients $A_n$ and $B_n$.

\newpage
\bibliography{references.bib}

\end{document}